\documentclass{ar-1col-S2O}
\usepackage[comma]{natbib}
\usepackage{url}
\usepackage[bottom=1.6in]{geometry}
\usepackage{aas_macros}
\usepackage{amsmath,amssymb, graphics}
\usepackage{bm}
\usepackage{soul}
\usepackage{graphicx}
\usepackage{subfigure}
\usepackage{epstopdf}
\usepackage{float}
\usepackage{mathrsfs}
\usepackage{booktabs}
\usepackage{multirow}
\usepackage{tabularx}
\usepackage{makecell}
\usepackage{etoolbox}
\usepackage{comment}
\usepackage[version=4]{mhchem}

\newcommand{\mb}{\mathbf}

\newcommand{\bgeq}{\begin{equation}}
\newcommand{\edeq}{\end{equation}}
\newcommand{\bgsp}{\begin{split}}
\newcommand{\edsp}{\end{split}}
\newcommand{\pa}{\partial}

\newbool{supplement}
\setbool{supplement}{false}

\jname{Annu. Rev. Astron. Astrophys.}
\jvol{AA}
\jyear{2026}
\doi{10.1146/((please add article doi))}

\def\gtrsim{\mathrel{\raise.3ex\hbox{$>$}\mkern-14mu
             \lower0.6ex\hbox{$\sim$}}}
\def\lesssim{\mathrel{\raise.3ex\hbox{$<$}\mkern-14mu
             \lower0.6ex\hbox{$\sim$}}}

\begin{document}

\markboth{X.-N. Bai}{Angular Momentum Transport in Protoplanetary Disks}

\title{Angular Momentum Transport in Protoplanetary Disks}
\author{Xue-Ning Bai$^{1,2}$
\affil{$^1$Institute for Advanced Study, Tsinghua University, Beijing, China; email: xbai@tsinghua.edu.cn}
\affil{$^2$Department of Astronomy, Tsinghua University, Beijing, China}
}


\begin{abstract}
We review our current understanding on the physical processes that govern angular momentum transport and evolution of protoplanetary disks. Extremely rich in physics, these processes are intimately connected to disk gas dynamics, with profound implications for planet formation. We organize them into a three-level hierarchical framework:

\vspace{0.1in}

\begin{minipage}[l]{0.66\textwidth}
\begin{itemize}
\item[{\scriptsize$\blacksquare$}] The coupling of gas with magnetic fields and radiation sets the microphysical foundation for understanding protoplanetary disk dynamics. Key ingredients include non-ideal magnetohydrodynamic effects (requiring ionization chemistry), along with heating and cooling processes. The disk can be divided into three radial sectors governed by distinct microphysics.
\item[{\scriptsize$\blacksquare$}] Protoplanetary disks host diverse gas dynamical processes, including hydrodynamic, magnetic and gravitational instabilities, along with thermally and magnetically-driven disk winds. Many of these {\it individual} processes are reasonably well understood, while others still require detailed investigation.
\item[{\scriptsize$\blacksquare$}] Protoplanetary disks are highly complex ecosystems where multiple processes interact. It is recognized that the bulk disk exhibits weak turbulence, with magnetically-driven wind likely serving as the primary transport mechanism. However, our knowledge remains highly limited regarding the disk's innermost region, early stages, long-term evolution, and environmental effects.
\end{itemize}
\end{minipage}

\vspace{0.1in}
\end{abstract}
\maketitle

\vspace{-0.1in}
\tableofcontents

\section{INTRODUCTION}
\label{sec:1}

Protoplanetary disks are gaseous and dusty disks orbiting young stars. They are the byproduct of on-going star formation following the gravitational collapse of an overdense core in a molecular cloud, and serve as a buffer channeling the dusty gas from the parent cloud/core to the central (proto)star. The physical processes mediating the flow of gas are complex, yet they control the formation, structure and evolution of protoplanetary disks, with profound astrophysical consequences, particularly on the formation of planets. Such processes can be most effectively interpreted from the principle of angular momentum conservation, more specifically the transport of angular momentum, usually associated with turbulent, magnetic and/or gravitational stresses that allow neighboring materials to exchange momentum and energy.

Angular momentum transport is the most fundamental question behind the theory of accretion disks. Compared to other types of accretion disks surrounding compact objects, protoplanetary disks are distinct in two aspects. First, protoplanetary disks are weakly ionized, leading to poor coupling between gas and magnetic fields. This greatly complicates the role magnetic fields play which is dependent upon the ionization chemistry. Second, it is expected that except for the very inner regions, protoplanetary disks are primarily passively heated by the central (proto)star known as stellar irradiation, instead of by accretion heating itself. This may allow for simplified prescriptions of disk temperature, but more realistically, the disk thermal properties in the inner and outer regions become coupled through the global disk geometry. Both aspects imply that the microphysics mediating angular momentum transport in protoplanetary disks is spatially inhomogeneous, and generally much richer than other types of accretion systems.

\begin{textbox}
\section{Evolutionary Stages of Protoplanetary Disks and Terminologies}
The evolution of protoplanetary disks can be characterized by two physical stages \cite[see][for a recent review on the observational classification scheme]{TobinSheehan24}:

-- Protoplanetary disks initially form embedded within an infalling envelope as the central protostar actively builds up its mass. For low-mass stars ($M_*\lesssim 2M_\odot$), this embedded phase corresponds to Class 0/I and typically lasts a few dynamical times ($\lesssim1$Myrs) \citep{Evans_etal09,Bell_etal13}.

-- The star lately enters the pre-main-sequence upon approaching its final mass and the disk subsequently becomes more visible as the envelope is largely accreted. This phase is known as the Class II or Classical T Tauri phase for low-mass stars, typically lasting a few Myrs \cite[e.g.][]{Haisch_etal01,Bell_etal13}. 
Disks are also found around brown dwarfs, as well as more massive stars (Herbig Ae/Be stars, $M_*\sim2-8M_\odot$). 

Protoplanetary disks are also known as protostellar disks, with the latter emphasizing that the central star is still in the protostellar phase (i.e., during the Class 0/I stage). However, the two terms are often used interchangeably in the literature and we do not make this distinction in this review.
\end{textbox}

The primary objective of studying disk angular momentum transport is to understand, at microscopic level, detailed gas dynamics across different regions in protoplanetary disks. This will ultimately enable us to piece together a coherent macroscopic picture of global disk evolution. On top of this, protoplanetary disks host a variety of physical and chemical processes thanks to the interplay among gas, dust, radiation, magnetic fields and nascent planets \cite[see recent reviews by][]{KleyNelson12,Oberg23,Birnstiel24}, which establishes a dynamic ``ecosystem" and sets the foundation for understanding almost all stages of planet formation \cite[e.g.][]{Armitage20}.

Testing the theory of angular momentum transport rests on disk observations.
The most direct observational constraint on disk angular momentum transport is the stellar accretion rate. This can be measured based on the accretion luminosity to a precision within a factor of 2-3 for Class II disks (see the sidebar titled Evolutionary Stages of Protoplanetary Disks and Terminologies), with a typical value of $10^{-8}M_\odot$ yr$^{-1}$ for Sun-like stars with large scatter \cite[e.g.][]{Hartmann_etal16}. Accretion rates are generally higher ($10^{-7}$-$10^{-6}M_\odot$ yr$^{-1}$) and more variable in the Class 0/I disks, reaching up to $10^{-4}M_\odot$ yr$^{-1}$ during FU Orionis outburst (FUor) \citep{Fischer_etal23}. Clues to angular momentum transport and disk evolution also reside in fundamental disk properties, such as disk mass and size, evaluated over a large disk population across different evolutionary stages
\citep{Manara_etal23,Zhang_etalAGEPRO}. 
For individual systems, it has recently become possible to directly measure (small-scale) disk substructures \citep{Andrews20,Bae_etal23,Benisty_etal23}, kinematics \citep{Pinte_etal23,Teague_etal25b}, and potentially magnetic field \citep{Vlemmings_etal19,Harrison_etal21,Teague_etal25}. All these observables are intimately linked to disk gas dynamics and the underlying angular momentum transport processes.

In the solar system context, protoplanetary disks represent analogs of the hypothesized solar nebula. A major historical issue of the ``nebular hypothesis" for the formation of solar system planets was how to reconcile the fact that the Sun contains $>99.8\%$ of the system's mass but $<1\%$ of its angular momentum. This problem is naturally resolved by recognizing that angular momentum transport occurs in the disk. Interdisciplinary study between astrophysics and planetary science over the past decade has provided new insights into planet assembly and its physical environment \citep{Lichtenberg_etal23}. Notably, it has become possible to extract information on nebula magnetic field from meteorites, offering vital complementary constraints to disk physics and evolution \citep{Weiss_etal21}.

\subsection{Scope}

This review is primarily theoretically-oriented, serving as a successor of the last annual review on this topic by \citet{Armitage11}. 
The field has almost been completely transformed over the past 15 years thanks to major advances in theory, simulations, observations, and solar system studies.
These advances have also revealed that the dynamics and evolution of protoplanetary disks are far more complex than previously thought.
There are a few recent reviews covering similar grounds, notably the pedagogical lecture notes by \citet{Armitage19} and \citet{Lesur21JPP} detailing gas dynamical processes, and the Protostars and Planets VII review by \citet{Lesur_etal23} with comprehensive coverage on the recent literature in both gas and dust dynamics. 

Our approach is to embed this review within the broader protoplanetary disk ``ecosystem", by organizing the relevant physical processes into a hierarchical framework across three levels:
\begin{itemize}
\item {\bf Microphysics} (Section \ref{sec:micro}): processes acting at atomic/molecular level that set the coupling among gas, magnetic fields, and radiation. These include ionization chemistry and non-ideal magnetohydrodynamic (MHD) effects, together with heating/cooling processes and radiation transport.
\item {\bf Gas dynamical processes} (Section \ref{sec:process}): isolated fluid-level processes as a result of specific microphysics. We focus on ``clean" physical processes such as individual MHD and hydrodynamic instabilities, and thermal/MHD disk winds.
\item {\bf Full disk applications} (Section \ref{sec:fulldisk}): embrace the complexity toward realistic scenarios, where multiple processes act concurrently. This is made convenient by dividing the disk into distinct regions.
\end{itemize}
We provide a systematic review in each level, highlighting recent advances. The microphysical and gas dynamical processes serve as foundations for full disk applications, where our understanding remains far from complete. Such ever-improving understandings can be built into models of global disk evolution (Sections \ref{sec:diskevol}, \ref{sec:frontier}), and tested against observational and experimental constraints
\ifbool{supplement}
{(Supplemental Text Section 2)}
{(Section \ref{sec:obs})}.
While additional processes such as dust dynamics, chemistry and planet formation are beyond the scope of this review, we believe they can be naturally integrated into this hierarchical framework.

\subsection{A Base Disk Model}\label{ssec:basemodel}

For theoretical purposes, protoplanetary disks are usually treated as isolated and axisymmetric. Fundamental disk properties are characterized by radial and vertical profiles of density ($\rho$), pressure ($P$), velocity (${\mb v}$) and magnetic field (${\mb B}$). Disk temperature ($T\propto P/\rho$) can be conveniently expressed via the isothermal sound speed $c_s\equiv\sqrt{P/\rho}$. We adopt either cylindrical coordinates $(R, \phi, z)$ or spherical polar coordinates $(r, \theta, \phi)$ as appropriate.
Protoplanetary disks are observed to be geometrically thin, thus disk rotation is close to Keplerian $v_\phi\approx v_K=(GM_*/R)^{1/2}$ with $M_*$ being the stellar mass, and we have angular velocity $\Omega\approx v_K/R\equiv\Omega_K$ and specific angular momentum $j\approx v_KR$. The disk aspect ratio is given by $H/R\approx c_s/v_K\ll1$, with $H$ being the disk scale height. Disk evolution is usually described through the surface density profile $\Sigma(R)$, and the physics mediating such evolution requires detailed understanding of disk vertical structure, which leads to diverse physical processes taking place in different disk regions. 

To facilitate discussion, we consider a simple base disk model around a solar-mass star, as follows
\bgeq\label{eq:basedisk}
\Sigma=500R_{\rm AU}^{-1}\exp(-R/R_t)\ \textrm{g cm}^{-2}\ ,\quad T=230R_{\rm AU}^{-1/2}\ K\ ,
\edeq
with a vertically isothermal structure, and default truncation radius $R_t=50$AU. This model resembles the minimum-mass solar nebula \citep{Hayashi81} with slightly adjusted scaling and normalizations, and aligns with typical Class II disk observations \cite[e.g.][]{Andrews_etal10}, yielding a total disk mass of about $0.018M_\odot$. 
For this model, we have the sound speed $c_s\approx9\times10^4R_{\rm AU}^{-1/4}$ cm s$^{-1}$ (for mean molecular weight $\mu=2.34$), $H/R\approx0.03R_{\rm AU}^{1/4}$, and midplane gas density $\rho_{\rm mid}\approx\Sigma/\sqrt{2\pi}H\approx4.4\times10^{-10}R_{\rm AU}^{-9/4}$g cm$^{-3}$. The gas number density (largely neutral molecules) can be found by definition $n_n=\rho/\mu m_H$, but it is also convenient to use the number density of \ce{H} atoms given by $n_H=\rho/1.40m_H$ for abundance calculations.

We keep in mind that protoplanetary disks are observed encompassing a highly diverse range of properties. This base model will serve as a reference point for initial discussion. As we progressively incorporate more complex disk physics, necessary adjustments to this model will be introduced. Additionally, we list in Table \ref{tab:symbols} the main symbols used in this review.

\begin{table}
\caption{List of commonly used symbols in this review and their meanings.}
\label{tab:symbols}
\begin{center}
\begin{tabularx}{\textwidth}{@{}c|c|c@{}}
\hline
{\bf Symbol} & {\bf Meaning} & {\bf Reference} \\\hline
$\mathcal{B}$ & Bernoulli constant along the wind & Eq. (\ref{eq:bernoulli})   \\\hline
$c_s$ & sound speed &   \\\hline
$H$ & disk scale height &  \S\ref{ssec:basemodel}   \\\hline
$j, l$ & disk and wind specific angular momentum &  \S\ref{ssec:basemodel}, Eq. (\ref{eq:lever})  \\\hline
$L_X, T_X$ & stellar X-ray luminosity and temperature & textbox in \S\ref{sssec:ionfid}   \\\hline
$\dot{M}_{\rm acc}$, $\dot{M}_{\rm acc}^{R, z}$ & mass accretion rate, and that due to radial/vertical transport &  \S\ref{sec:diskevol}  \\\hline
$\dot{M}_{\rm wind}(R) $ & cumulative mass loss rate within a given radius $R$ & Eq. (\ref{eq:mdotwind})  \\\hline
$n_H, N_H$ & number density/column density of \ce{H} atoms    \\\hline
$n_e, n_n, n_{\rm gr}$ & number density of electrons, neutrals and dust grains & before Eq. (\ref{eq:xe0})  \\\hline
$N^2, N_{R,z}^2$ & Brunt-V\"ais\"al\"a frequency squared & Eq.(\ref{eq:buoancy})   \\\hline
$p_{T,\rho,\Sigma}$ & radial power-law index of $T$, $\rho_{\rm mid}$, and $\Sigma$ in the disk & Before Eq.(\ref{eq:buoancy})  \\\hline
$q_{\rm Joule}, q_{\rm irr}, q_{\rm vis}$ & Joule, irradiation and effective viscous heating rates & Eq.(\ref{eq:Joule}), \S\ref{sssec:diskTemp}  \\\hline
$Q$ & Toomre $Q$ parameter for the onset of GI & Eq.(\ref{eq:TmrQ}) \\\hline
$R_A$ & Alfv\'en radius of disk winds & \S\ref{sssec:windbasics}  \\\hline
$R_m, R_{\rm co}$ & Magnetospheric radius, corotation radius & Eqs. (\ref{eq:Rm}), (\ref{eq:Rco})  \\\hline
$R_t$ & disk outer truncation radius &  \S\ref{ssec:basemodel}  \\\hline
$s$ & specific entropy & Before Eq.(\ref{eq:buoancy})  \\\hline
$t_{\rm acc}, t_{\rm loss}$ & accretion and mass loss timescale & After Eq.(\ref{eq:master})  \\\hline
$T_{R\phi}, T_{z\phi}$ & radial and vertical stress & Eq.(\ref{eq:Trphi}), (\ref{eq:Tzphi})  \\\hline
$v_A, v_{Ap}$ & Alfv\'en speed and its poloidal component & Eq. (\ref{eq:Elsasser}), \S\ref{sssec:windbasics}  \\\hline
$x_e, x_{\rm gr}$ & $n_e/n_n$, $\langle Z\rangle n_{\rm gr}/n_n$ & Eq.(\ref{eq:xe0}), \S\ref{sssec:nimhd}  \\\hline
$z_s$ & $z$-coordinate of disk surface & After Eq.(\ref{eq:amt})  \\\hline 
$\alpha_S, \alpha_W$ & Shakura-Sunyaev $\alpha$ and its counterpart for disk wind & Eqs. (\ref{eq:macc_R}), (\ref{eq:alphaw}) \\\hline
$\beta, \beta_z$ & Plasma $\beta$ and its $z-$component & After Eq.(\ref{eq:Elsasser})  \\\hline
$\beta_{\rm cool}$ & dimensionless thermal relaxation time & Eq.(\ref{eq:betacool})  \\\hline
$\beta_{c, {\rm VSI}}$ & critical $\beta_{\rm cool}$ for the onset of VSI & Eq.(\ref{eq:betacvsi})  \\\hline
$\beta_{c, {\rm GI}}$ & critical $\beta_{\rm cool}$ for GI fragmentation & After Eq.(\ref{eq:TmrQ}) \\\hline
$\beta_s$ & The Hall parameter for species $s$ & Eq.(\ref{eq:betaj})  \\\hline
$\gamma$ & adiabatic index & Before Eq.(\ref{eq:buoancy}) \\\hline
$\gamma_s$ & $\langle\sigma v\rangle_s/(m_n+m_s)$ & Eq.(\ref{eq:balance}) \\\hline
$\zeta$ & ionization rate per \ce{H} nucleus & After Eq.(\ref{eq:OD74})  \\\hline
$\zeta_{{\rm CR}, X, R}$ & $\zeta$ from CR, X-ray and radionuclides & \S\ref{sssec:ionfid}  \\\hline
$\eta_{O,H,A}$ & Ohmic/Hall/Ambipolar diffusivities & Eq.(\ref{eq:etas})  \\\hline
$\kappa$ & Epicyclic frequency & After Eq.(\ref{eq:buoancy}) \\\hline
$\kappa_R, \kappa_P$ & Rosseland mean and Planck mean opacities & Eq. (\ref{eq:meanopacity}) \\\hline
$\lambda$ & wind lever arm & \S\ref{sssec:windbasics}  \\\hline
$\Lambda, Ha, Am$ & Ohmic/Hall/Ambipolar Elsasser numbers & Eq.(\ref{eq:Elsasser})  \\\hline
$\mu$ & mean molecular weight &  \S\ref{ssec:basemodel} \\\hline
$\xi$ & ejection index of disk wind & After Eq.(\ref{eq:lever})  \\\hline
$\dot{\Sigma}_w$ & local mass loss rate & \S\ref{sec:diskevol}  \\\hline
$\Sigma_{\rm FUV}$ & threshold FUV penetration column & \S\ref{sssec:ionfid}  \\\hline
$\langle\sigma v\rangle_s$ & momentum exchange rate for species $s$ with neutrals & \S\ref{sssec:nimhd}  \\\hline
$\omega_s$ & fastness parameter for magnetospheric accretion & After Eq. (\ref{eq:Rco})  \\\hline
\end{tabularx}
\end{center}
\end{table}

\section{GENERAL FRAMEWORK FOR DISK EVOLUTION}\label{sec:diskevol}

In this section, we provide an overview on the governing equations of angular momentum transport and disk evolution. This sets the stage for the rest of the review, which primarily aims at revealing the underlying physics behind the controlling parameters.

\subsection{The Master Equation}

Working with vertically-integrated disk quantities in cylindrical coordinates, disk evolution is governed by the radial profile of the mass accretion rate $\dot{M}_{\rm acc}(R)\equiv-2\pi R v_R\Sigma$ (positive for accretion) and 
the local wind mass loss rate $\dot{\Sigma}_w$ (positive for mass loss):
\bgeq
2\pi R\frac{\pa\Sigma}{\pa t}-\frac{\pa\dot{M}_{\rm acc}}{\pa R}+2\pi R\dot{\Sigma}_w=0\ .\label{eq:cont}
\edeq
We define the cumulative mass loss rate from an inner disk radius $R_{\rm in}$ as
\bgeq\label{eq:mdotwind}
\dot{M}_{\rm wind}(R)\equiv\int_{R_{\rm in}}^R2\pi R'\dot{\Sigma}_wdR'\ ,
\edeq
so that the last term in (\ref{eq:cont}) can be rewritten as $\pa\dot{M}_{\rm wind}/\pa R$.

The accretion rate follows from angular momentum conservation, given by 
\bgeq
2\pi R\frac{\pa(\Sigma j)}{\pa t}+\frac{\pa}{\pa R}\underbrace{\bigg(-\dot{M}_{\rm acc}j+2\pi R^2\int_{-z_s}^{z_s}T_{R\phi}dz\bigg)}_{\textrm{Radial angular momentum flux}}+\underbrace{2\pi R\dot{\Sigma}_wj+2\pi R^2T_{z\phi}\bigg|_{-z_s}^{z_s}}_{\textrm{Vertical angular momentum loss}}=0\ ,\label{eq:amt}
\edeq
where $\pm z_s$ denotes the upper/lower vertical location of the disk surface (note $\Sigma$ integrates $\rho$ between $\pm z_s$),
\bgeq\label{eq:Trphi}
T_{R\phi}\equiv\overline{\rho\delta v_R\delta v_\phi}-\frac{1}{4\pi}\overline{B_RB_\phi}+\frac{1}{4\pi G}\overline{g_Rg_\phi}
\edeq
is the radial stress which is associated with the radial angular momentum flux, 
\bgeq\label{eq:Tzphi}
T_{z\phi}\equiv\overline{\rho v_z\delta v_\phi}-\frac{1}{4\pi}\overline{B_zB_\phi}\approx-\frac{1}{4\pi}\overline{B_zB_\phi}
\edeq
is the vertical stress to be evaluated at disk surfaces, associated with additional angular momentum flux carried by the outflow. Overlines denote azimuthal averages. The terms $\delta v_R$ and $\delta v_\phi$ represent local velocity deviations from the mean flow, with $\overline{\rho\delta v_R\delta v_\phi}$ being the Reynolds stress, usually resulting from turbulence\footnote{The angular momentum flux can also be carried by waves, which leads to non-local transport of angular momentum. This process can be oscillatory, and the outcome depends on wave dissipation (see Sections \ref{ssec:nonturb} and \ref{sssec:BL}).}. The magnetic part of $T_{R\phi}$ is the Maxwell stress, while ${\mb g}$ is the self-gravitational acceleration with the $\overline{g_Rg_\phi}$ as gravitational stress \citep{LyndenBellKalnajs72}. Moreover, for nearly Keplerian disk rotation ($\delta v_\phi\ll v_K$),  only magnetic stress contributes significantly to $T_{z\phi}$.

Combining Equations (\ref{eq:cont}) and (\ref{eq:amt}), we obtain the fundamental equation expressing angular momentum transport mechanisms
\bgeq\label{eq:mdotacc}
\dot{M}_{\rm acc}\frac{dj}{dR}\approx\frac{\pa}{\pa R}\bigg(2\pi R^2\int_{-z_s}^{z_s}T_{R\phi}dz\bigg)+ 2\pi R^2T_{z\phi}\bigg|_{-z_s}^{z_s}\ .
\edeq
For thin disks, $j\approx v_KR\propto R^{1/2}$ and hence $dj/dR\approx v_K/2$. At a given radius, the left hand side represents the angular momentum loss rate per unit radius due to accretion. This is compensated by the radial gradient of angular momentum flux and the extraction of angular momentum through disk wind, set by $T_{R\phi}$ and $T_{z\phi}$ on the right hand side, respectively.

A positive $T_{R\phi}$ transports angular momentum radially outward, but its effect on accretion depends on its radial gradient. It is customary to parameterize $T_{R\phi}\equiv\alpha_S P$, with $\alpha_S$ being dimensionless, leading to the $\alpha-$disk model \citep{ShakuraSunyaev73}. The $T_{R\phi}$ term may be seen as equivalent to a viscous stress $\sigma_{R\phi}=\rho\nu Rd\Omega/dR\approx-(3/2)\rho\nu\Omega$, with $\nu$ being the kinematic viscosity. The $\alpha$-prescription corresponds to $\nu=(2/3)\alpha_S c_sH$. In reality, molecular viscosity in protoplanetary disks is orders of magnitude too small to account for disk accretion. Radial angular momentum transport therefore must be mediated by $T_{R\phi}$ through specific physical processes (Section \ref{sec:process}), especially turbulence (known as ``turbulent viscosity"). Using the $\alpha$-prescription, the accretion rate driven by radial (viscous) transport is
\begin{marginnote}
\entry{Molecular viscosity}{Microscopic viscosity due to molecular collisions.} 
\entry{Turbulent viscosity}{An effective viscosity from turbulent mixing of bulk fluid.}
\end{marginnote}
\bgeq
\dot{M}_{\rm acc}^{R}\approx\frac{4\pi}{v_K}\frac{\pa}{\pa R}(\alpha_S\Sigma c_s^2R^2)\ .\label{eq:macc_R}
\edeq
Applying this to our base disk model with pure radial transport, we obtain $\dot{M}_{\rm acc}\approx2\pi\alpha_S\Sigma Hc_s(1-2R/R_t)$, where the factor in the parenthesis arises from the radial gradient of an exponentially truncated surface density profile. Clearly, the disk accretes within $R_t/2$ but spreads out at larger radii. This inner disk accretion and outer disk expansion (viscous spreading) is the typical outcome for radial angular momentum transport. For $\dot{M}_{\rm acc}\sim10^{-8}M_{\odot}$ yr$^{-1}$, we find $\alpha_S\sim5\times10^{-3}$ prior to disk truncation in our base model.

The vertical stress is generally attributed to magnetized disk winds. For a vertical field threading a thin disk, $B_z\approx$ constant across the disk. With a typical hourglass field geometry, the radial field $B_r$ changes sign across the disk, and Keplerian shear then produces $B_\phi$ of opposite signs. By symmetry and using Equations (\ref{eq:Tzphi}) and (\ref{eq:mdotacc}), we can express $\dot{M}_{\rm acc}\approx 2R^2|B_zB_\phi|_{z_s}/v_K$, which always drives accretion. 
One can attempt for a similar $\alpha$-like prescription for vertical angular momentum transport \citep{Tabone_etal22}. In doing so, it is desirable that $RT_{z\phi}|^{z_s}_{-z_s}$ and $\int_{-z_s}^{z_s}T_{R\phi}dz$ share similar forms, and we choose
\bgeq\label{eq:alphaw}
|T_{z\phi}|_{z_s}\equiv\frac{\alpha_W\Sigma c_s^2}{4R}
\approx\alpha_WP_{\rm mid}\cdot\bigg(\frac{\sqrt{2\pi}H}{4R}\bigg)\approx\alpha_WP_{H}\cdot\bigg(\frac{H}{R}\bigg)\ ,
\edeq
where $P_{\rm mid}\approx\Sigma c_s^2/\sqrt{2\pi}H$ is the midplane pressure and $P_H=P_{\rm mid}e^{-1/2}$ is the pressure at one disk scale height. The resulting wind-driven accretion rate is
\bgeq
\dot{M}_{\rm acc}^{z}\approx\frac{8\pi}{v_K}|T_{z\phi}|_{z_s}R^2=\frac{2\pi}{v_K}\alpha_W\Sigma c_s^2R=2\pi\alpha_W\Sigma Hc_s\ .\label{eq:Mdotz}
\edeq
Compared to the viscous $\alpha$-prescription, $\alpha_W$ includes an additional $\sim H/R$ factor in normalization. Physically, if $T_{R\phi}$ and $|T_{z\phi}|$ share similar values, especially if magnetic fields dominate angular momentum transport, then $\alpha_W$ is about a factor $\sim R/H$ larger than $\alpha_S$. This suggests that wind-driven accretion is about $\sim R/H$ times more efficient than viscously-driven accretion. The reason lies in the fact that the vertical stress exerted by winds operates with a long lever arm ($R$), whereas radial stress is integrated only over the disk thickness ($H$).

As we will discuss in Section \ref{ssec:magwind}, the wind flow carries higher specific angular momentum (denoted by $l$) relative to the disk gas at the wind streamline's footpoint. Wind-driven accretion results from this {\it excess} angular momentum extraction:
\bgeq\label{eq:lever}
\dot{M}_{\rm acc}^{z}\frac{dj}{dR}=\frac{\pa\dot{M}_{\rm wind}}{\pa R}(l-j)\equiv\frac{\pa\dot{M}_{\rm wind}}{\pa R}(\lambda-1)j\ ,\quad{\rm or}\quad\frac{d\dot{M}_{\rm wind}/d\ln R}{\dot{M}_{\rm acc}}\bigg|_{R=R_0}=\frac{1}{2}\frac{1}{\lambda-1}
\edeq
where we have defined the ratio $l/j$ to be $\lambda>1$, known as the magnetic lever arm parameter. This relation thus connects the wind mass loss rate to the wind-driven accretion rate via the $\lambda$ parameter, and the ratio $(d\dot{M}_{\rm wind}/d\ln R)/\dot{M}_{\rm acc}$ is sometimes called ``ejection index", denoted by $\xi$.

Combining Equations (\ref{eq:cont}) and (\ref{eq:amt}) with the above parameterizations, we arrive at the master equation governing global disk evolution:
\bgeq\label{eq:master}
\frac{\pa\Sigma}{\pa t}=\underbrace{\frac{2}{R}\frac{\pa}{\pa R}\bigg[\frac{1}{\Omega R}\frac{\pa}{\pa R}(\alpha_S\Sigma c_s^2R^2)\bigg]}_{\textrm{Radial (viscous) transport}}+\underbrace{\frac{1}{R}\frac{\pa}{\pa R}\bigg(\frac{\alpha_W\Sigma c_s^2}{\Omega}\bigg)}_{\textrm{Vertical (wind) transport}}-\underbrace{\frac{\alpha_W\Sigma c_s^2}{\Omega R^2[2(\lambda-1)]}}_{\textrm{Wind mass loss}}=0\ .
\edeq
The physical meaning of each term is clear, with a diffusive term from viscous transport, an advective term from wind-driven accretion, and a sink term due to wind mass loss. These further define two characteristic timescales of disk evolution: an accretion timescale $t_{\rm acc}(R)=R^2/(\alpha c_sH)$ with $\alpha=\alpha_S+\alpha_W$, and a mass loss timescale $t_{\rm loss}(R)=2(\lambda-1)R^2/(\alpha_Wc_sH)$.
It should also be noted that unless disk temperature can be prescribed, this equation should also be combined with an energy equation to solve for $c_s^2$ (Section \ref{ssec:thermo}).

Finally, we make a statement on the usage of viscosity-related terms. 
It is customary to use ``viscously-driven" accretion and ``viscous heating" to refer to radial transport by $T_{R\phi}$ and the associated dissipation. We retain the use of ``viscosity" in pure mathematical models. However, as the gas is physically inviscid, we advocate to use ``effective viscosity" and associated terminology in the physical context for scientific rigor.

\subsection{Representative Solutions}\label{ssec:diskevolsol}

\begin{figure*}[h]
\centering
\includegraphics[width=\textwidth]{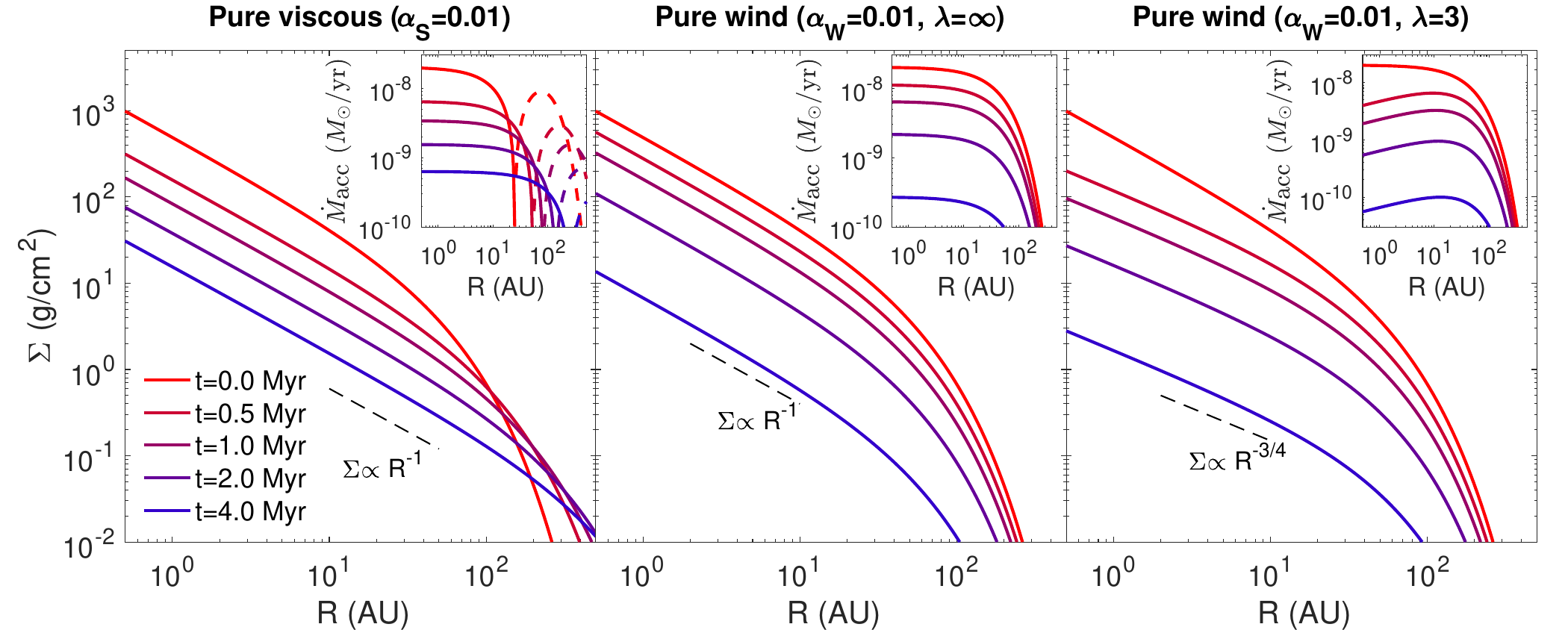}
\caption{Evolution of disk surface density (main panels) and accretion rate profiles (insets) from our base disk model (\ref{eq:basedisk}) assuming constant ($\alpha_S, \alpha_W, \lambda$). Left: pure viscous disk model with $\alpha_S=0.01$. Middle: pure wind model with $\alpha_W=0.01$ and $\lambda=\infty$. Right: pure wind model with $\alpha_W=0.01$ and $\lambda=3$. Parameters are chosen so that the initial accretion rate profiles within the disk truncation radius are identical among all models. Dashed lines in insets indicate disk spreading (negative accretion). Power-law indices of late-time surface density profiles are also marked.
\label{fig:alphadisk}}
\end{figure*}

This master equation generalizes the viscous evolution model \cite[e.g.][]{Pringle81} to incorporate disk winds, and has been adopted by several authors to study disk evolution \citep{Armitage_etal13,Bai16,Suzuki_etal16,Tabone_etal22}. Here we show representative solutions with constant ($\alpha_S, \alpha_W, \lambda$), akin to the results of \citet{Tabone_etal22}.

Assuming $c_s^2\propto R^{-1/2}$ (as in our base model), a steady state solution can be obtained by setting $\Sigma\propto R^{-1+\xi}$ ($\xi\geq0$)
in Equation (\ref{eq:master}), which requires $\xi$ to satisfy
$2\xi^2+(1+\alpha_W/\alpha_S)\xi-(\alpha_W/\alpha_S)/[2(\lambda-1)]=0$.
For $\alpha_W=0$, the case reduces to pure viscous disk solution with $\xi=0$,
and a constant accretion rate $\dot{M}_{\rm acc}=2\pi\alpha_S\Sigma Hc_s$ (note $Hc_s\propto R$ in the model). For $\alpha_S=0$, we have pure wind-driven accretion with ejection index $\xi=[2(\lambda-1)]^{-1}$ and $\dot{M}_{\rm acc}=2\pi\alpha_W\Sigma Hc_s\propto R^{\xi}$. In the limit of negligible mass loss ($\lambda\rightarrow\infty$), a steady state accretion maintains the same surface density profile as the pure viscous case. However, accounting for mass loss, the surface density profile becomes shallower, with a radially increasing accretion rate to compensate for the mass loss.

Figure \ref{fig:alphadisk} shows the representative outcomes of disk evolution starting from our base disk model. The parameters are chosen so that the initial accretion rates are identical. All models exhibit self-similar evolution, characterized by a power-law surface density profile matching the steady state solution in the inner region, together with an exponential truncation \citep{Tabone_etal22}. The truncation radius $R_t$ expands linearly in time in the pure viscous disk as $R_t(t)=R_t(0)[1+t/t_{\rm acc}(R_t(0)/2)]$, following the \citet{LyndenBellPringle74} solution, while it stays constant in pure wind models. In the hybrid case with both radial and vertical transport, the disk expands at the rate solely set by $t_{\rm acc}$ evaluated at viscous $\alpha_S$. With our choice of $\lambda=3$, wind mass loss modifies the disk surface density profile, and significantly accelerates disk evolution.

Finally, it is important to bear in mind that in real disks, ($\alpha_S, \alpha_W, \lambda$) can be inhomogeneous, and likely evolve over time. Without considering detailed disk physics, even such simple parameterization has infinite degrees of freedom that can produce diverse evolutionary paths. A primary goal of studying disk gas dynamics is to reveal disk physical processes across space and time, which we focus on in this review, and this will eventually allow us to reduce uncertainties toward a comprehensive understanding of global disk evolution.

\section{GOVERNING MICROPHYSICS}\label{sec:micro}

In this section, we introduce the important microphysical processes that describe the coupling between gas and magnetic field, known as the non-ideal magnetohydrodynamic (MHD) effects set by disk ionization, and the coupling between gas and radiation, which is exhibited as thermodynamical effects.

\subsection{Magnetic Coupling: Ionization and Non-ideal MHD Effects}\label{ssec:mag}

Protoplanetary disks are sufficiently cold to be almost completely neutral, and electric conductivity is just set by a trace amount of charged particles. However, magnetic coupling is not measured by ionization fraction, but by magnetic diffusivities that also depend on local environment, as we elaborate in this subsection.

\subsubsection{Ionization-recombination chemistry: standard model}\label{sssec:ionfid}

In weakly ionized protoplanetary disks, magnetic coupling is set by the disk ionization level, which is the outcome of an ionization-recombination chemical reaction network. For this purpose, the complexity of the adopted chemical network varies but is usually (much) simpler than those used in the disk chemistry community \cite[e.g.,][]{Oberg23}.\footnote{Simple networks can well reproduce the ionization fraction in the dense midplane region and the disk atmosphere unshielded to UV radiation, but tend to overpredict ionization fraction by a factor of a few in between \citep{Semenov_etal04,Xu_etal19}. The situation may be alleviated with AI-assisted network reduction \citep{Grassi_etal22}.} The main elements of gas-phase reactions can be encapsulated by a simple network by \citet{OD74} with free electrons e$^-$, neutral molecules $\textrm{m}$, molecular ions $\textrm{m}^+$ (e.g., \ce{HCO+}), metal atoms $\textrm{M}$ (e.g., \ce{Mg}) and their ions $\textrm{M}^+$:
\bgeq\label{eq:OD74}
\ce{m ->[\zeta] m+ + e-},\quad
\ce{m+ + e- ->[k_{dr}] m}, \quad 
\ce{M+ + e- ->[k_{rr}] M}, \quad
\ce{m+ + M ->[k_{ce}] m + M+}\ .
\edeq
Respectively, they correspond to non-thermal ionization reactions with effective ionization rate $\zeta$ (often quoted as the rate per hydrogen nucleus), dissociative recombination with rate $k_{\rm dr}\approx3\times10^{-6}T^{-1/2}$cm$^3$s$^{-1}$, radiative recombination with much slower rate $k_{\rm rr}\approx3\times10^{-11}T^{-1/2}$cm$^3$s$^{-1}$, and charge exchange with a rate $k_{\rm ce}\approx3\times10^{-9}$cm$^3$s$^{-1}$, where the adopted rates follow \citet{Fromang_etal02}, and $T$ is the gas temperature in Kelvin. The non-thermal ionization source here includes cosmic-rays (CRs), stellar X-rays, and short-lived radionuclides. These ionization processes effectively act on the hydrogen and helium as the most abundant species, thereby activating the reaction network in the bulk (cold) disk regions. 

\begin{textbox}
\section{Non-thermal ionization: standard prescriptions}
Cosmic-rays (CRs) are energetic charged particles that pervade the Galaxy, with typical energy $\sim$GeV, but lower-energy particles are responsible for most ionization \citep{Grenier_etal15}.
The widely adopted CR ionization rate is  based on the flux of cosmic-rays arriving on Earth, given by \citep{UN81}
\bgeq
\zeta_{\rm CR}\approx\frac{\zeta_{\rm CR,0}}{2}\bigg[\exp{(-\Sigma^{\rm top}(z)/\Sigma_{\rm CR})}+\exp{(-\Sigma^{\rm bot}(z)/\Sigma_{\rm CR})}\bigg]\ ,\label{eq:zetaCR}
\edeq
where the attenuation depth is $\Sigma_{\rm CR}=96$ g cm$^{-2}$, and $\Sigma^{\rm top/bot}$ represents the column density to the disk's top/bottom surface. At large column densities ($\Sigma^{\rm top/bot}\gtrsim\Sigma_{\rm CR}$), the ionization rate can be further reduced by a factor of $\sim\Sigma^{\rm top/bot}/\Sigma_{\rm CR}$ assuming CRs impinge the disk isotropically \citep{UmebayashiNakano09}. Models typically adopt unattenuated interstellar value of $\zeta_{\rm CR,0}\approx10^{-17}$s$^{-1}$ \cite[e.g.][]{SpitzerTomasko68,BlackDalgarno77}. Observations of interstellar diffuse clouds suggest higher values \cite[e.g.][]{McCall_etal03,Indriodo_etal15}, with recent quote of $\zeta_{\rm CR,0}\approx(3-10)\times10^{-17}$s$^{-1}$ \citep{Obolentseva_etal24}. The actual flux reaching the disk is very uncertain depending on the poorly constrained CR transport models.

Young protostars are strong X-ray emitters due to coronal activities \citep{Feigelson_etal07}. The typical X-ray luminosity of a T Tauri star is $L_X\sim10^{29-32}$erg s$^{-1}$ \citep{Preibisch_etal05,Gudel_etal07}, with characteristic X-ray temperature $T_X$ of 1 to 8 keV \citep{Wolk_etal05}. The X-rays impinging the disk undergo photoionization loss (direct absorption) and Compton scattering, and the latter deflects part of the X-ray photons into the disk and enhances their penetration. Based on Monte-Carlo calculations of \citet{IG99},
for $L_X=10^{30}$\,erg\,s$^{-1}$, the ionization rate from direct absorption is $\zeta_{X0, {\rm abs}}\approx(4\text{--}6)\times10^{-11}$\,s$^{-1}$ at 1\,AU, attenuating over a column of $\sim5\times10^{-3}$\,g\,cm$^{-2}$. In contrast, the scattered component provides a lower rate, $\zeta_{X0, {\rm sca}}\approx(1\text{--}2)\times10^{-14}$\,s$^{-1}$ at 1\,AU, but penetrates much deeper ($\sim2$\,g\,cm$^{-2}$).
A fitting formula may be adopted from \citet{BaiGoodman09} (their Equation 21), showing that $\zeta_X$ decreases with radius approximately as $R^{-2.2}$, exhibiting an attenuation profile that is slower than exponential.

In addition, a base level of ionization is provided by the decay of short-lived radionuclides, primarily $^{26}$Al. With solar abundances, the resulting ionization rate is about $\zeta_{\rm R}\approx(3\text{--}10)\times10^{-19}$\,s$^{-1}$ \cite[e.g.,][]{UmebayashiNakano09,TurnerDrake09}, which decays with a half-life of 0.717\,Myr.
\end{textbox}

Another major source of recombination is dust grains \citep{UmebayashiNakano80,Sano_etal00,IlgnerNelson06}. For spherical grains of size $a$, the reaction rate coefficient is approximately 
$k_{\rm gr}\sim s\pi a^2 v_{\rm th,e/i}\tilde{J}$, 
where $v_{\rm th,e/i}=\sqrt{8kT/\pi m_{e/i}}$ is the thermal speed of the electrons/ions, and $\tilde{J}$ is a correction factor \cite[see Section IIIa of][]{DraineSutin87}. The sticking coefficient $s$ is the probability of recombination following a collision of a charged particle with the dust grain. It is typically assumed to be unity for ions and $0.1$ (but highly uncertain) for electrons, with the latter expected to decrease with temperature \citep{Nishi_etal91,IlgnerNelson06,Bai11a}.
To leading order (ignoring the $\tilde{J}$ factor), 
the rate coefficients are $k_{\rm gr}\sim2\times10^{-3}(a/\mu{\rm m})^2T^{1/2}$ and $\sim10^{-4}(a/\mu{\rm m})^2T^{1/2}$ cm$^3$s$^{-1}$ for electrons and ions, respectively. These values are orders of magnitude higher than gas-phase recombination rates ($k_{\rm dr}$). With dust number density $n_{\rm gr}\propto a^{-3}$, the total dust recombination rate $n_{\rm gr}k_{\rm gr}$ is primarily proportional to the total dust surface area. 

Let $n_e$ and $n_n$ be the number density of the free electrons and neutrals. In the dust-free case, balancing ionization and recombination rates, the ionization fraction $x_e\equiv n_e/n_n$ can be crudely estimated to be
\bgeq
x_e\approx\sqrt{\frac{\zeta}{k_{\rm dr}n_n}}\approx6\times10^{-10}\bigg(\frac{\zeta}{10^{-17}{\rm s}^{-1}}\bigg)^{1/2}\bigg(\frac{n_n}{10^{8}{\rm cm}^{-3}}\bigg)^{-1/2}\bigg(\frac{T}{100{\rm K}}\bigg)^{1/4}\ .\label{eq:xe0}
\edeq
This can be boosted to $x_e\approx\sqrt{\zeta/k_{\rm rr} n_n}$ when free metals are abundant. In contrast,
assuming single-sized dust with dust-to-gas mass ratio $f_{d2g}$, grain-phase recombination alone leads to an ionization fraction of $x_e\approx\sqrt{\zeta/(n_{\rm gr}k_{\rm gr})}$.
Therefore, grain-phase recombination dominates over gas-phase recombination when 
\bgeq
x_e\lesssim3\times10^{-10}\bigg(\frac{f_{d2g}}{10^{-2}}\bigg)\bigg(\frac{a}{\mu{\rm m}}\bigg)^{-1}\bigg(\frac{T}{100{\rm K}}\bigg)\ .
\edeq
A rough estimate based on grain size distribution from grain-growth calculations \citep{Birnstiel_etal11} gives $x_e\lesssim3\times10^{-11}(T/100{\rm K})$. This is generally satisfied in the dense midplane regions of the inner $\sim20$AU where $n_n\gtrsim10^{11}$cm$^{-3}$ in our base disk model. It should also be noted that grains can carry multiple charges \cite[e.g.][]{Wardle07,Okuzumi09}, and a more detailed discussion can be found in \citet{Ivlev_etal16}.

In addition, stellar far-UV (FUV) photons penetrate the warmer surface layers of the disk. These photons cannot ionize hydrogen/helium, but can fully ionize atomic carbon and sulfur, bringing the ionization fraction to $x_e\gtrsim10^{-5}$ over a thin layer \citep{PerezBeckerChiang11b}. 
This UV radiation also triggers a rich set of photo-reactions, rendering the disk surface a photodissociation region (PDR; \citealt{HollenbachTielens97}).
\begin{marginnote}
\entry{PDR}{Photodissociation region, or photon- dominated region.} 
\end{marginnote}Many theoretical studies tend to avoid the complexity of modeling the PDR by adopting a simplified prescription: the ionization fraction increases sharply from the disk value to $\gtrsim10^{-5}$ above a threshold column density $\Sigma_{\rm FUV}$. \citet{PerezBeckerChiang11b} estimated $\Sigma_{\rm FUV}$ to be $0.01-0.1$g cm$^{-2}$ for the disk vertical column, a value that sensitively depends on the assumed abundance of ``very small grains" (VSGs).

Finally, thermal ionization of alkali metals becomes dominant in the innermost disk regions when temperature exceeds $\approx10^3$K, substantially enhancing coupling to magnetic field \citep{Jin96,Gammie96}. Particularly relevant are \ce{K} and \ce{Na}, with first ionization potentials of 4.34 eV and 5.14 eV and solar abundances (per H atom) of $1.5\times10^{-7}$ and $2.4\times10^{-6}$, respectively \citep{Lodders03}. Assuming pure thermal ionization of \ce{K} (i.e., $n_e=n_{\ce{K+}}$), the Saha equation yields \citep{BalbusHawley00,Fromang_etal02}
\bgeq
\frac{n_e}{n_n}\approx5.0\times10^{-12}\bigg(\frac{n_K/n_H}{1.5\times10^{-7}}\bigg)^{1/2}
\bigg(\frac{T}{10^3{\rm K}}\bigg)^{3/4}\bigg(\frac{n_n}{10^{14}}\bigg)^{-1/2}
\bigg[\frac{\exp{(-25188{\rm K}/T)}}{1.15\times10^{-11}}\bigg]\ .
\edeq
We will see that this level of ionization is sufficient to initiate coupling between gas and magnetic fields in the dense midplane region of the innermost disk.

\subsubsection{Ionization-recombination chemistry: recent development}

The standard prescriptions for disk ionization processes discussed above are subject to various uncertainties, particularly on CR ionization, which have been investigated over the recent years.

First, it is unclear to what extent and how galactic CRs reach protoplanetary disks. It is well known that the solar wind modulates the CR flux reaching Earth in the heliosphere, and by extrapolating this effect to the more powerful winds of T Tauri stars (``T-Tauriosphere"), \citet{Cleeves_etal13} argued that the CR ionization rate could be reduced to $\zeta_{\rm CR}\ll10^{-18}$s$^{-1}$, rendering it largely negligible even compared with radionuclides.
Alternatively, if magnetized disk winds are present (Section \ref{ssec:magwind}), the ionizing CR particles (with energy $\lesssim$GeV, gyro-radii $\ll H$) largely travel along open magnetic field lines threading the disk. The CR flux can be enhanced by magnetic focusing of converging field lines while reduced by magnetic mirroring effect. These two effects practically cancel, resulting in little net change to the ionization rate \citep{Silsbee_etal18}, unless magnetic field geometry becomes more complex (e.g., forming ``magnetic packets").\begin{marginnote}
\entry{Magnetic mirroring effect}{A charged particle entering a region with increasing magnetic field can be reflected due to conservation of magnetic moment $p_\perp^2/2B$.
}
\end{marginnote}Empirically, disk winds also likely push CRs away from the disk analogous to stellar winds, reducing the CR flux primarily for low-energy particles, though detailed models are lacking, presumably due to major uncertainties in CR diffusion physics.

Second, CR propagation and ionization are more complex than the standard exponential attenuation model. Through detailed calculations, \citet{Padovani_etal18} showed the ionization rate declines with column density in a non-exponential manner, with an attenuation length increasing from $\approx112$ to $\approx285$g cm$^{-2}$ for $\Sigma=100$--$2100$g cm$^{-2}$.
Furthermore, \citet{FujiiKimura22} demonstrated that when ionization is dominated by protons and the associated secondaries, the relevant column in Equation (\ref{eq:zetaCR}) should be that along magnetic field lines. Given that disk fields are typically toroidal-dominated ($B_\phi \gg B_p$; see Section \ref{sssec:magthermal}), this substantially increases the CR path-lengths compared to pure vertical penetration, implying a much reduced CR vertical penetration column.

Third, protostellar systems themselves can be sources of cosmic-rays \cite[see][for a review]{Padovani_etal20}. Empirically, the X-ray luminosity of young stars implies a $10^5$-fold enhancement in the production of energetic protons in solar flares \citep{Feigelson_etal02}, which may potentially become a dominant disk ionization source \citep{TurnerDrake09}.
Physically, the accretion shock at the stellar surface can be an efficient particle accelerator via diffusive shock acceleration \citep{Padovani_etal15,Padovani_etal16,GachesOffner18}, 
capable of accelerating protons to GeV energies. Magnetic reconnection at the star-disk interface has been proposed as another source of energetic particles \citep{Brunn_etal23,Brunn_etal24}.
Regardless of their origin, the transport of these energetic particles into the disk is highly uncertain and model-dependent. By treating the transport process as isotropic diffusion, studies generally find they could dominate ionization in the inner disk surface layers ($\lesssim$1,AU) \citep{Rab_etal17,Rodgers-Lee_etal17,Rodgers-Lee_etal20,Offner_etal19}, though test-particle simulations in turbulent fields suggest a more limited role \citep{Fraschetti_etal18}.
Overall, these studies call for more investigations in the high-energy processes in protostellar systems.

Turning to X-ray ionization, an important characteristic is its strong flaring activity. Flares can boost the X-ray luminosity by a factor of $\sim$100, accompanied by harder spectra, with typical durations of a day and recurrence times of a week \cite[e.g.,][]{Wolk_etal05}. These events can strongly enhance the instantaneous disk ionization (with tentative observational evidence; cf. \citealp{Cleeves_etal17}), though the time-averaged impact is more modest \citep{IlgnerNelson06c,WaggonerCleeves22}. The primary effect likely stems from the harder spectrum, which allows for deeper penetration. Indeed, \citet{ErcolanoGlassgold13} updated the Monte Carlo calculations of \citet{IG99}, recommending an observationally-motivated two-temperature X-ray model that can yield ionization rates differing by up to an order of magnitude from conventional single-temperature models.

In the very inner disk region, \citet{DeschTurner15} recognized that at high temperatures, dust grains themselves can become ionization sources via thermionic and ion emission—the inverse processes of electron and ion recombination onto grains. The rates of these processes depend on material work function, which can be very uncertain. Incorporating these reactions into the \citet{OD74} network, they found these additional channels effectively lower the first ionization potential of potassium to between $3.5-4$ eV, thus lowering the temperature threshold for thermal ionization. In the meantime, grains become highly charged, with the mean charge increasing linearly with grain size \citep{WilliamsMohanty25}.

\subsubsection{Non-ideal MHD effects: physical origin}\label{sssec:nimhd}

In well-ionized gas, it is well-known that magnetic field is frozen into the gas (flux freezing), described by ideal MHD. This can be understood as charged particles, representing the bulk gas, must gyrate around the magnetic field. Mathematically, this is described by a motional electric field ${\mb E}=-({\mb v}/c)\times{\mb B}$. In poorly ionized gas, charged particles still gyrate around magnetic fields, but they also collide with the neutral molecules, which makes them drift to neighboring field lines, leading to non-ideal MHD effects. 

Non-ideal MHD effects are quantified by the Ohm's law. In the frame co-moving with the neutrals, let ${\mb E}'$ be the electric field, and ${\mb v}'_{s}$ be the drift velocity of charged species $s$. These charged particles experience the Lorentz force and the collisional drag with the neutrals, which must balance each other
\begin{equation}
Z_se({\mb E}'+\frac{{\mb v}'_s}{c}\times{\mb B})
=\gamma_s\rho m_s{\mb v}'_s\ ,\label{eq:balance}
\end{equation}
where $Z_s$ represents particle charge, and $\gamma_s\equiv\langle\sigma v\rangle_s/(m_n+m_s)$ with $\langle\sigma v\rangle_s$ being the rate coefficient for momentum transfer with the neutrals, and $m_n=\mu m_H$ is the mean mass of the neutrals. The relative importance between the Lorentz force and the neutral drag is characterized by the ratio between the gyro-frequency and the momentum exchange (collision) rate, known as the Hall parameters \citep{WardleNg99} (not to be confused with the Hall effect)
\begin{equation}
\beta_s\equiv\frac{Z_seB}{m_sc}\frac{1}{\gamma_s\rho}\propto\frac{B}{\rho}\ .\label{eq:betaj}
\end{equation}
A charged species $s$ is strongly coupled with neutrals (magnetic fields) if $|\beta_s|\ll1$ ($\gg1$). 

The Ohm's law expresses the linear relation between total current density ${\mb J}=e\sum_sn_sZ_s{\mb v}'_s$ and ${\mb E}'$. It directly derives from Equation (\ref{eq:balance}): ${\mb J}=\sigma_O{\mb E}'_\parallel + \sigma_H\hat{\mb B}\times{\mb E}'_\perp+\sigma_P{\mb E}'_\perp$, where $\parallel$ and $\perp$ indicate vector components parallel and perpendicular to the magnetic field,  $\hat{\mb B}$ is the unit vector along magnetic field, and $\sigma_{O,H,P}$ are the Ohmic, Hall and Pedersen conductivities \cite[cf.][for their expressions]{WardleNg99,Bai11a}. Essentially, the presence of magnetic field makes the conductivity a tensor. Inverting this relationship gives the non-ideal electric field:
\bgeq\label{eq:nielec}
{\mb E}'=\frac{4\pi}{c^2}[\eta_O{\mb J}+\eta_H({\mb J}\times{\hat{\mb B}})
+\eta_A{\mb J}_\perp]\ ,
\edeq
where $\eta_{O,H,A}$ are the Ohmic, Hall and ambipolar diffusivities. They incorporate the abundances of all charged species and serve as a bridge between the ionization-recombination chemistry and gas dynamics.

To illustrate the underlying physics, we assume electrons and a representative molecular ion species are the main charge carriers, with number densities $n_e=n_i\ll n_n$ (charge neutrality).
For reference, the rate coefficients for momentum transfer can be estimated by $\langle\sigma v\rangle_e\approx8.3\times10^{-9}{\rm max}[1,(T/100{\rm K})^{1/2}]$cm$^{3}$ s$^{-1}$ \citep{Draine_etal83,Bai11a}, $\langle\sigma v\rangle_i\approx2.4\times10^{-9}(m_H/\mu)^{1/2}$cm$^{3}$ s$^{-1}$ \citep{Draine11,Lesur21JPP}. There is also $\langle\sigma v\rangle_{\rm gr}\approx2.6\times10^{-3}(a/\mu m)(T/100{\rm K})^{1/2}$cm$^{3}$ s$^{-1}$ when considering charged grains.
We see that the $\gamma_s$ coefficients are of the same order for electrons and ions, but the large mass ratio leads to $|\beta_e|\sim10^3\beta_i\gg\beta_i$. This hierarchy naturally leads to three non-ideal MHD regimes:
\begin{itemize}
\item Ohmic-dominated regime ($\beta_i\ll|\beta_e|\ll1$; high density, weak field): both electrons and ions are strongly coupled to the neutrals via collisions. Ohm's law takes the familiar form ${\mb J}=\sigma_O{\mb E}'$.
\item Hall-dominated regime ($\beta_i\ll1\ll|\beta_e|$): electrons are coupled to magnetic fields while ions are tied to the neutrals. Ohm's law is set by this electron-ion drift, with ${\mb E}'\approx-({\mb v}'_e-{\mb v}'_i)\times{\mb B}/c={\mb J}\times{\mb B}/cen_e$.
\item Ambipolar-dominated regime ($1\ll\beta_i\ll|\beta_e|$; low density, strong field): both electrons and ions are coupled with magnetic fields. Ohm's law is set by ion-neutral drift, with ${\mb E}'\approx-{\mb v}'_i\times{\mb B}/c$, where ${\mb v}'_i$ is set by balancing the Lorentz force ${\mb J}\times{\mb B}/c$ acting on the ion fluid and the ion-neutral drag $-\gamma_i\rho\rho_i{\mb v}'_i$.
\end{itemize}
In both the Hall- and ambipolar-dominated regimes, flux freezing is maintained, except that magnetic field is tied to the most mobile species—the electrons. The magnetic diffusivities are given by
\bgeq
\begin{split}\label{eq:etas}
\eta_O&=\frac{cB}{4\pi en_e}\frac{1}{\beta_i+|\beta_e|}\approx\frac{c^2m_e\gamma_e\rho}{4\pi e^2n_e}\approx2.3\times10^3x_e^{-1}\max\bigg[1,\bigg(\frac{T}{100{\rm K}}\bigg)^{1/2}\bigg]{\rm cm}^2{\rm s}^{-1}\ ,\\
\eta_H&=\frac{cB}{4\pi en_e}\frac{|\beta_e|-\beta_i}{\beta_i+|\beta_e|}\approx\frac{cB}{4\pi en_e}=\eta_O|\beta_e|\ ,\\
\eta_A&=\frac{cB}{4\pi en_e}\frac{|\beta_e|\beta_i}{\beta_i+|\beta_e|}\approx\frac{B^2}{4\pi\gamma_i\rho\rho_i}=\eta_O|\beta_e|\beta_i\ .
\end{split}
\edeq
Crucially, all diffusivities scale as $\eta \propto x_e^{-1}$. In terms of their relative strengths, $\eta_O$ is independent of $(B/\rho)$, $\eta_H\propto(B/\rho)$, and $\eta_A\propto(B/\rho)^2$. Given $x_e$, the Ohmic-dominated regime prevails in dense regions with weak magnetic field, the ambipolar-dominated regime occurs in tenuous regions with strong magnetic field, and the Hall-dominated regime lies in between.

\begin{figure*}[h]
\centering\label{fig:diffusivity}
\includegraphics[width=\textwidth]{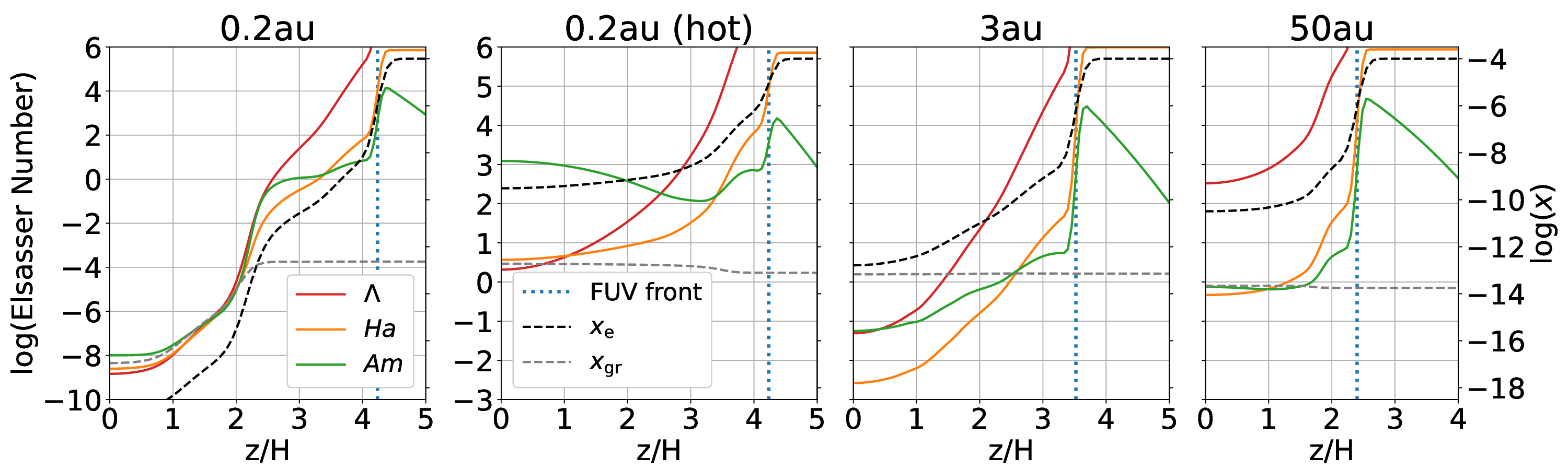}
\caption{Vertical profiles of the ionization fraction and the Elsasser numbers (Eq. \ref{eq:Elsasser}) in our base disk model at 0.2, 3 and 50 AU (1st, 3rd and 4th panels from left), together with a model at 0.2 AU but with temperature boosted to 1200K (2nd panel). Vertical density profile is Gaussian and $z$ is normalized by disk scale height $H$. The calculations are based on a complex chemical reaction network using standard ionization prescriptions for cosmic-rays (Equation \ref{eq:zetaCR}) with $\zeta_0=10^{-17}$s$^{-1}$, X-rays with $L_X=10^{30}$erg s$^{-1}$ and $T_X=3$ keV. FUV ionization is assumed to penetrate over $\Sigma_{\rm FUV}=0.03$g cm$^{-2}$ to boost the ionization fraction to $x_e=10^{-4}$. A population of $0.1\mu$m sized dust with abundance of $10^{-4}$ is assumed, with a work function of $5.0$ eV. A constant field strength corresponding to midplane plasma $\beta=500$ is assumed for Elsasser number calculation. Figure provided by X. Zheng.}
\end{figure*}

Dust grains affect magnetic diffusivities in two ways: (1) they enhance recombination, lowering $x_e$ and increasing all magnetic diffusivities proportionally. We may define $x_{\rm gr}\equiv|\langle Z\rangle|n_{\rm gr}/n_n$ with $\langle Z\rangle$ being the mean grain charge, and grains can become the dominant charge carrier with $x_{\rm gr}\gg x_e$ in the inner disk midplane region (Figure \ref{fig:diffusivity}). (2) More subtly, when $x_{\rm gr}\gtrsim x_e$, the large inertia of grains ($\beta_g \ll \beta_i$) alters the scaling relations \cite[e.g.][]{Bai11b}. While $\eta_O$ remains independent of $B$, the scalings of $\eta_H\propto B$ and $\eta_A\propto B^2$ no longer hold. In particular, 
$\eta_H$ can turn negative above a threshold field strength $B_{\rm th}$ (typically between $\beta_e=1$ and $\beta_i=1$, \citealp{XuBai16}), accompanied by a reduction in $\eta_A$.
This occurs mainly at the Ohmic-dominated disk inner midplane region under typical physical parameters.\footnote{As grains recombine electrons much faster than ions in general, one expects $n_{{\rm gr}^-}>n_{{\rm gr}^+}>n_i>n_e$ when charged grains dominate. With the negative charge carrier being more massive on average, sign change in $\eta_H$ becomes possible (as implied from Equation \ref{eq:etas}).}  Nevertheless, \citet{XuBai16} found that the field strength there rarely reaches the threshold for sign-reversal of $\eta_H$.

Special circumstances can arise when the current density ($\nabla\times{\mb B}$) demands charged particles to drift at super-thermal speeds ($v_s > v_{\rm th}$). One direct consequence is the enhancement of collision rates $\langle\sigma v\rangle_s$ (resulting from both high drift speed and enhanced electron heating, as discussed in \citealp{OkuzumiInutsuka15}),
which in turn increases Ohmic resistivity and reduces ambipolar diffusivity \citep{Hopkins_etal25}, as implied by Equation~(\ref{eq:etas}). Detailed calculations show this effect introduces nonlinearity into Ohm's law and can potentially lead to electric discharge under extreme conditions \citep{InutsukaSano05,OkuzumiInutsuka15,Okuzumi_etal19}. This mechanism could be significant in the inner disk regions ($\lesssim10$,AU) in the presence of moderately strong currents, though its full dynamical implications remain an open area of investigation \cite[see initial exploration by][]{Mori_etal17}.

\subsubsection{Non-ideal MHD effects: physical characters} 

Ohmic resistivity and ambipolar diffusion dissipate magnetic energy, and lead to Joule heating. The dissipation rate is given by
\begin{equation}\label{eq:Joule}
q_{\rm Joule}=\frac{1}{c}{\mb E}'\cdot{\mb J}=\frac{4\pi}{c^2}(\eta_OJ^2+\eta_AJ_\perp^2)\ ,
\end{equation}
Specifically, Ohmic diffusion dissipates the total current, leading to magnetic reconnection and smoothing of magnetic field structures. Ambipolar diffusion damps the perpendicular component of the current via ion-neutral drag. While being dissipative, its anisotropic nature may lead to the formation of sharp magnetic structures \citep{BrandenburgZweibel94}.

The Hall effect has two important characters. First, it is dispersive rather than dissipative, and should not be considered as diffusion. Instead of damping magnetic perturbations, it leads to {\it rotation} of perturbed magnetic vectors (as can be shown from the induction equation), whose consequence will be discussed in Sections \ref{ssec:mri} and \ref{ssec:inner}. Second, the Hall effect has a polarity dependence. The set of MHD equations is unchanged by altering the sign of magnetic field, except for the Hall term as evident from Equation (\ref{eq:nielec}).

In protoplanetary disks, the importance of the magnetic diffusivities can be characterized by the dimensionless Elsasser numbers, defined as
\bgeq\label{eq:Elsasser}
\Lambda\equiv\frac{v_A^2}{\eta_O\Omega}\ ,\quad
Ha\equiv\frac{v_A^2}{\eta_H\Omega}\equiv\frac{v_A}{l_H\Omega}\ ,\quad
Am\equiv\frac{v_A^2}{\eta_A\Omega}\ ,
\edeq
where $v_A\equiv B/\sqrt{4\pi\rho}$ is the Alfv\'en wave speed.\begin{marginnote}
\entry{Alfv\'en wave}{Transverse MHD wave whose restoring force is magnetic tension, which is exactly analogous to string vibration.
}
\end{marginnote}The non-ideal MHD effects are considered strong when their Elsasser numbers are of order unity or less. For the Hall effect, there is also the Hall length $l_H\equiv\eta_H/v_A$ below which the Hall effect dominates \citep{KunzLesur13}. With the standard scaling, it should be noted that $\Lambda\propto B^2(x_e/\rho)$, $Ha\propto Bx_e$, and $Am\propto B^0(x_e\rho)$. Therefore, $Am$ and $l_H$ conveniently have one-to-one correspondence to ionization fraction in disks. For other Elsasser numbers, one needs to assume certain magnetic field strength, often characterized by the plasma $\beta$ parameter. For the bulk protoplanetary disks, one typically expects $\beta>1$, with $\beta=10^{2-3}$ as a common range.\begin{marginnote}
\entry{Plasma $\beta$}{Ratio of gas to magnetic pressure, which can also be written as $\beta=2c_s^2/v_A^2$.} 
\end{marginnote}

\subsubsection{Ionization and magnetic diffusivities in the base disk model}

As an illustration, Figure \ref{fig:diffusivity} shows the vertical profiles of ionization fraction and the associated Elsasser numbers at 0.2, 3 and 50 AU for our base disk model with standard ionization prescriptions. Also shown is a case at 0.2 AU but with temperature boosted to $1200$K to activate thermal ionization. In all cases, we have chosen a uniform field strength that corresponds to $\beta=500$ at the disk midplane. 

With thermal ionization activated in the very inner disk, the disk largely approaches ideal MHD conditions, though not by a wide margin.
In the absence of thermal ionization, the midplane disk ionization fraction $x_e$ sharply drops to below $10^{-15}$, and then increases monotonically with radius to $\gtrsim10^{-10}$. Examining the Elsasser numbers that characterize magnetic coupling, we highlight a few salient features. (1). With decreasing density, the dominant non-ideal MHD effect at the midplane transitions from Ohmic-dominated at 0.2 AU, to Hall-dominated at 3 AU, and eventually to ambipolar diffusion-dominated at $\gtrsim50$ AU. A similar transition occurs in the vertical direction. While the locations where the transitions occur can shift depending on the disk model and field strength, we note that the Hall-dominated midplane occupies a wide radial range.  (2). The gas in the disk midplane is very poorly coupled to magnetic field ($\Lambda\ll1$) in the inner 10 AU. The coupling becomes marginal further out, with $Am\sim1$ in the outer disk. (3). Vertically, while the ionization fraction increases by orders of magnitude, the surface layers are still only marginally coupled due to its low density with $Am\sim1$ until reaching the FUV layer. 

\subsection{Thermodynamics: Heating and Cooling Processes}\label{ssec:thermo}

Protoplanetary disks can be heated passively by external irradiation, mainly from the central star, and actively from energy dissipation in the accretion process itself. They cool from dust thermal emission, as well as gas emission lines (mainly in the optically thin disk atmosphere). The heating-cooling balance sets the disk temperature structure, with dynamical consequences to be discussed in  
\ifbool{supplement}
{Section \ref{ssec:hydro} and Supplemental Text Section 1.}
{Sections \ref{ssec:hydro} and \ref{ssec:pe}.}

\subsubsection[]{Gas and dust opacity}\label{sssec:opacity}

A general understanding of the disk temperature structure can be obtained by solving the equations of radiative transfer. This approach usually assumes LTE conditions where gas and dust emit thermal radiation satisfying Kirchhoff's law $\epsilon_\nu=\kappa^{\rm abs}_\nu B_\nu(T)$, where $\epsilon_\nu$ is emissivity, $B_{\nu}$ is the Planck function, $\kappa^{\rm abs}_\nu$ is the absorption opacity, all at frequency $\nu$.\begin{marginnote}
\entry{LTE}{Local thermodynamic equilibrium.} 
\end{marginnote}The coupling between matter (here the dust-gas mixture assuming they share the same temperature) and radiation is fully characterized by $\kappa^{\rm abs}_\nu$ and the scattering opacity $\kappa^{\rm sca}_\nu$.
However, integrating over frequencies via frequency-dependent (or multigroup) radiative transfer calculations are highly computationally costly. It is common to adopt the grey approximation,\begin{marginnote}
\entry{Gray approximation}{The absorption/extinction opacity is frequency-independent.} 
\end{marginnote}using the frequency-averaged Rosseland mean and Planck mean opacities. 

\begin{textbox}
\section{Frequency-averaged Opacity Tables}
The coupling between gas and radiation is mainly encapsulated into the Rosseland mean and Planck mean opacities,
defined as \cite[e.g.][]{Mihalas284}
\bgeq\label{eq:meanopacity}
\kappa_R(\rho,T)^{-1}=\frac{\int d\nu(\kappa^{\rm abs}_\nu+\kappa^{\rm sca}_\nu)^{-1}\pa B_\nu(T)/\pa T}{\int d\nu\pa B_\nu(T)/\pa T}\ ,\quad
\kappa_P(\rho,T)=\frac{\int d\nu\kappa^{\rm abs}_\nu B_\nu(T)}{\int d\nu B_\nu(T)}\ .
\edeq
In the comoving frame with the fluid, they enter the equations of radiation hydrodynamics describing energy and momentum exchange rate between the fluid (gas-dust mixture) and radiation
\bgeq
G^0\approx\kappa_P\rho c(E_{\rm rad}-a_RT^4)\ ,\quad
{\mb G}\approx\kappa_R\rho \mb F_{\rm rad}/c\ ,
\edeq
where $E_{\rm rad}=a_RT_{\rm rad}^4$ is the radiation energy density, ${\mb F}_{\rm rad}$ is the radiation energy flux, $a_R$ is the radiation constant, and $T_{\rm rad}$ is radiation temperature. 

The fluid is heated by radiation at a rate $G^0$ at the cost of radiation energy. The process only involves absorption and emission, and the Planck mean opacity is simply weighted by the radiation energy spectrum. Written above assumes $T_{\rm rad}=T$, known as single-temperature opacity. More generally, the Planck mean opacity is given as $\kappa_P=\kappa_P(\rho, T, T_{\rm rad})$ by replacing $B_\nu(T)$ by $B_\nu(T_{\rm rad})$ in Equation (\ref{eq:meanopacity}). This is known as two-temperature opacity applicable under broader conditions.

The Rosseland mean opacity primarily sets how efficient radiation gets transported through matter, which is the most accurate in the diffusion (optically thick) regime. Recall that in this regime, radiation is approximately isotropic and the radiative energy flux at each frequency is given by ${\mb F}_{\nu, {\rm rad}}=-c/(3\kappa_\nu\rho)\nabla E_{\nu,{\rm rad}}$. The Rosseland mean is thus a flux-weighted average so that ${\mb F}_{\rm rad}=-c/(3\kappa_R\rho)\nabla E_{\rm rad}$. As $T_{\rm rad}\approx T$ in this regime, opacity can be expressed as $\kappa_R=\kappa_R(\rho, T)$.
The fluid also gains momentum through ${\mb G}$ with $\kappa_R$ as the coupling coefficient; it is largely negligible in protoplanetary disks whose luminosities are far too low. 
\end{textbox}

Frequency-averaged opacities are functions of matter density and temperature (sometimes also radiation temperature), and they are usually displayed as opacity tables or fitting formulae. At low temperature ($\lesssim10^3$K) dust continuum opacity dominates. This depends on dust properties such as composition, size distribution, and structure (e.g., shape, porosity). As dust sublimates at higher temperatures ($\sim1500$K), gas opacity gradually takes over, in the form of molecular and atomic line opacities before the gas becomes fully ionized. Calculating such opacities requires specialized effort. Below we list some popular opacity tables developed in protoplanetary disk and planet formation context,
including the piecewise power law fitting formula of the Rosseland mean opacity by \citet{BellLin94,Zhu_etal09a}, full opacity table by \citet{Semenov_etal03}, dust opacity table by \citet{Pollack_etal94,DAlessio_etal01}, and gas opacity table by \citet{Malygin_etal14}, etc., obtained by fixing dust and/or gas properties. More recently, flexible tools have emerged that allow users to compute opacities with adjustable dust and gas properties. These include gas opacity code {\sf OPTAB} by \citet{Hirose_etal22} and {\sf ÆSOPUS} by \citet{Marigo_etal24}, frequency-dependent dust opacity used by the {\sf DSHARP} project \citet{Birnstiel_etal18}, and a more comprehensive dust opacity tool {\sf OpTool} by \citet{Woitke_etal16,Tazaki_etal18,Dominik_etal21}. By compiling several of these tables, \citet{Zhu_etal21} assembled an opacity table covering a broad range of matter density and temperature.

\subsubsection[]{The radial temperature profile}\label{sssec:diskTemp}

We begin with the radial profile of the midplane temperature, which serves as a proxy for the bulk disk temperature. Let $L_*$ be the stellar luminosity, and for illustration, we assume a constant opacity $\kappa_R = \kappa_P \equiv \kappa$. First consider pure irradiated disk in the optically thin limit. Balancing heating and cooling, we obtain $L_*/4\pi R^2 = a_R c T_{\rm thin}^4$, or $T_{\rm thin} \approx 280{\rm K} (L_*/L_\odot)^{1/4} R_{\rm AU}^{-1/2}$.\footnote{This estimate assumes grey opacity. In reality, small (sub-micron-sized) grains are more efficient at absorbing stellar light ($\kappa_P(T^*)\gg\kappa_P(T_{\rm dust})$) and get super-heated.} This gives the temperature scaling of our base disk model (Equation~\ref{eq:basedisk}).

More realistically, protoplanetary disks are highly optically thick to stellar irradiation. The temperature scaling of our base disk model already suggests that protoplanetary disks are flared, with $H/R$ increasing with $R$. Following \citet{ChiangGoldreich97}, stellar photons incident upon the disk at some grazing angle $\mu_{\rm irr}\approx d(H_p/R)/d\ln R$ and get absorbed, where $H_p$ is the height where the optical depth to stellar irradiation $\tau_V\sim1$. 
Roughly half of the dust re-emission is directed to the disk interior, whose temperature we denote as $T_{\rm irr}$. Balancing heating and cooling, we obtain $(\mu_{\rm irr}/2)(L_*/4\pi R^2)=\sigma_{\rm SB} T_{\rm irr}^4$, where $\sigma_{\rm SB}=a_Rc/4$ is the Stefan-Boltzmann constant. When combined with $H/R\approx c_s/v_K\propto (M_*T_{\rm irr}/R)^{1/2}$, and assuming $H_p\propto H$ (taken to be $H_p=4H$ here), we obtain
\bgeq
T_{\rm irr}\approx120{\rm K}(L_*/L_\odot)^{2/7}(M_*/M_\odot)^{-1/7}R_{\rm AU}^{-3/7}\ . 
\edeq
This temperature is much cooler compared to $T_{\rm thin}$ due to the oblique irradiation geometry, and the temperature slope is flatter which leads to a more flared disk. 

We next consider accretion heating within the pure viscous framework, assuming a constant $\alpha_S$. For a given accretion rate $\dot{M}_{\rm acc}$, the surface density must satisfy Equation (\ref{eq:macc_R}), which simplifies to $\dot{M}_{\rm acc}=2\pi\alpha_S\Sigma c_s^2/\Omega$ in steady state. When accretion heating dominates, the disk temperature is determined by balancing effective viscous dissipation with radiative cooling. The effective viscous dissipation rate is $Q_{\rm vis} = \nu\Sigma (Rd\Omega/dR)^2 \approx (3/2)\alpha_S c_s^2\Sigma\Omega$. For an optically thick, geometrically thin disk, cooling occurs via vertical radiative diffusion. The radiative energy flux from each disk surface is approximately $F_{\rm rad} \approx (4/3)\sigma_{\rm SB}T_{\rm vis}^4/\tau$, where $\tau \approx \kappa_R\Sigma/2$ is the optical depth at the disk midplane and $T_{\rm vis}$ is the midplane temperature. Balancing $Q_{\rm vis} = 2F_{\rm rad}$ and solving for $\Sigma$ and $T_{\rm vis}$ yields:
\bgeq\label{eq:T_vis}
T_{\rm vis}(R)\approx3.5\times10^2{\rm K}\bigg(\frac{\dot{M}}{10^{-8}M_\odot{\rm yr}^{-1}}\bigg)^{2/5}
\bigg(\frac{\alpha_S}{0.01}\bigg)^{-1/5}\bigg(\frac{M_*}{M_\odot}\bigg)^{3/10}
\bigg(\frac{\kappa_R}{5\ {\rm cm}^2{\rm g}^{-1}}\bigg)^{1/5}R_{\rm AU}^{-9/10}\ .
\edeq
\bgeq\label{eq:Sigma}
\Sigma(R)\approx1.6\times10^2{\rm g}\ {\rm cm}^{-2}
\bigg(\frac{\dot{M}}{10^{-8}M_\odot{\rm yr}^{-1}}\bigg)^{3/5}
\bigg(\frac{\alpha_S}{0.01}\bigg)^{-4/5}\bigg(\frac{M_*}{M_\odot}\bigg)^{1/5}
\bigg(\frac{\kappa_R}{5\ {\rm cm}^2{\rm g}^{-1}}\bigg)^{-1/5}R_{\rm AU}^{-3/5}\ .
\edeq
As we fix $\dot{M}_{\rm acc}$, larger $\alpha_S$ leads to smaller $\Sigma$, making the disk more optically thin which reduces $T_{\rm vis}$. While these scalings do not necessarily hold if disk wind dominates angular momentum transport, the thermal balance alone (assuming constant $\kappa$) yields $T_{\rm vis}\propto\Sigma^{2/3}R^{-1/2}$. This profile is steeper than pure irradiated disks as long as $\Sigma$ decreases with $R$. Therefore, effective viscous heating dominates only in the disk very inner region, while the bulk disk beyond $\sim$AU scale is primarily passively heated by stellar irradiation, where disk temperature is largely independent of accretion rate.
This justifies the common simplification of treating disk thermodynamics as vertically isothermal and using prescribed radial temperature profiles.

As we will discuss shortly, strong effective viscous heating (large $\alpha_S$) is expected only when thermal ionization is activated, at $T\gtrsim10^3$K. For a typical accretion rate of $\dot{M}_{\rm acc} \sim 10^{-8} M_\odot$ yr$^{-1}$, this condition is met at $R \lesssim 0.3$ AU. As stellar irradiation hardly heats the disk to such temperature beyond this radius, an abrupt radial temperature transition is expected (mediated by radiative diffusion). 
Inside this transition radius is the disk {\it inner rim}, characterized by the sublimation of silicate dust. The rim's thermodynamics are complex: in addition to effective viscous heating, it receives direct stellar illumination, and dust sublimation/condensation is tightly coupled with opacity which feedback into radiation transport (see review by \citealp{DullemondMonnier10} and a semi-analytic model by \citealp{Ueda_etal17}). 
Overall, this region is the most complex and least understood in protoplanetary disks, as we shall discuss further in Section \ref{ssec:innermost}.

\subsubsection[]{The vertical temperature structure}\label{sssec:diskTemp}

In a purely irradiated disk, the surface is heated to a temperature on the order of $T_{\rm thin}$, below which lies a nearly isothermal region extending to the midplane at a cooler temperature $\sim T_{\rm irr}$. Accretion heating deposits additional energy, raising the temperature not only at the layer where the heat is released, but potentially over the entire disk column through radiation transport.

More precisely, the disk vertical temperature structure can be solved using an approach analogous to classical grey stellar atmosphere models, generalized to include an arbitrary volumetric heating profile $q$ \citep{Hubeny90,Mori_etal19}. Irradiation heating is given by $q_{\rm irr}=\rho\kappa_P(L_*/4\pi r^2)\exp(-\tau_{\rm irr})$, where $\tau_{\rm irr}$ is the optical depth to stellar radiation (depending on disk geometry $\mu_{\rm irr}$) and $\kappa_P$ is the two-temperature Planck opacity. This heating peaks where $\tau_{\rm irr} \sim 1$ and diminishes toward the midplane. The most familiar form of accretion heating is effective viscous heating from turbulent dissipation, with $q_{\rm vis} \approx (3/2)\alpha_S c_s^2\rho\Omega$ under the $\alpha$-prescription. Without strong turbulence, another (weaker) form of accretion heating is Joule heating (\ref{eq:Joule}), resulting from local dissipation of magnetic fields \citep{Mori_etal19,BethuneLatter20}. 

This discussion above primarily holds for setting``dust temperature", as dust dominates the absorption of starlight and thermal emission (cooling). At sufficiently high density, dust and gas temperatures equilibrate thanks to frequent collisions. However, at the low-density and optically thin disk surface layer, gas and dust are subject to different heating and cooling processes, often involving complex thermal chemistry coupled with UV and X-ray irradiation. This generally leads to significantly elevated gas temperatures at high altitudes (\citealp{KampDullemond04}, Section \ref{sssec:thermatm}).

\begin{figure*}[h]
\centering
\includegraphics[width=\textwidth]{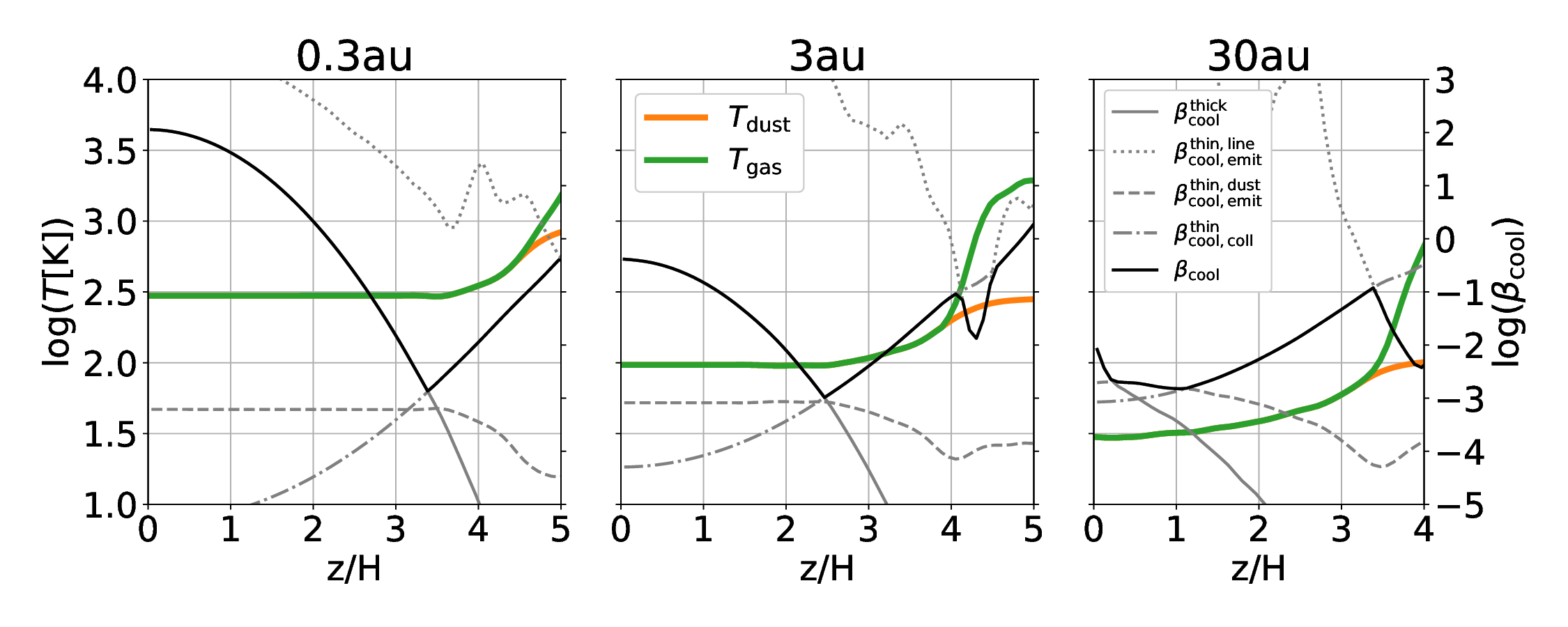}
\caption{Vertical profiles of the dust and gas temperature in our base disk model at 0.3, 3 and 30 AU (from left to right), calculated using the DALI code \cite[][with permission from its developers]{Bruderer_etal12,Bruderer13} under passive stellar heating around a solar-mass star. Stellar parameters include $L_*=2.1L_\odot$, $L_{\rm UV}=0.86L_\odot$, $L_X=5\times10^{29}$erg s$^{-1}$, $T_X\approx3$keV. For illustrative purposes, the density profiles are taken from the base disk model without re-calculating hydrostatic equilibrium. A well mixed population of dust with a size distribution $n(a)\propto a^{-3.5}$ between $a=0.005\mu$m-$1$mm is included with abundance $Z=0.01$.
Also shown are the associated cooling time, expressed as dimensionless $\beta_{\rm cool}$ following the approximate procedure in Section \ref{sssec:trelax}, except that for $\beta_{\rm cool,emit}^{\rm thin}$, we separate the contribution from dust cooling and gas cooling. Figure provided by X. Zheng.
\label{fig:T}}
\end{figure*}

As an illustration, we take the density profile of our base model, and calculate the temperature structure of the disk without recalculating hydrostatic equilibrium. No viscous or Joule heating is included. Figure \ref{fig:T} shows the vertical temperature profiles at representative disk radii. It separately calculates the dust and gas temperature with more self-consistent Monte-Carlo frequency-dependent (instead of grey) radiative transfer, and hence the results do not coincide with simple analytic estimate. The dust temperature is largely vertically isothermal in the optically thick disk interior, and rises toward the super-heated disk surface. The gas and dust temperatures start to decouple when their mutual collision is no longer sufficiently frequent ($\beta_{\rm cool, coll}^{\rm thin}\gtrsim1$, see Equation \ref{eq:betathin}), and gas temperature can become much higher due to UV/X-ray heating processes (Section \ref{sssec:thermatm}).

\subsubsection[]{Thermal relaxation time}\label{sssec:trelax}

Closely linked to the temperature structure is the thermal response of the gas to perturbations, characterized by an effective ``relaxation time" $t_{\rm relax}$,\begin{marginnote}
\entry{Thermal relaxation}{The process by which a system returns to its equilibrium temperature after a perturbation.} 
\end{marginnote}also referred to as ``cooling time". It is often convenient to express this timescale in dimensionless form:
\bgeq
\beta_{\rm cool}\equiv\Omega t_{\rm relax}.\label{eq:betacool}
\edeq
With this prescription, disk thermodynamics may be approximated as $dT/(\Omega dt)=(T-T_{\rm eq})/\beta_{\rm cool}$, where $T_{\rm eq}$ is the equilibrium temperature. This approach, also known as {\it $\beta$ cooling}, provides a convenient (though not always physically consistent) alternative to solving complex equations of radiative transfer,
and the value of $\beta_{\rm cool}$ critically influences several physical processes in protoplanetary disks (see Sections \ref{ssec:hydro}, \ref{ssec:gi}).

Estimating $t_{\rm relax}$ under realistic situations requires approximate treatment of radiative transfer (see detailed discussion in \citealp{Malygin_etal17}), and one must distinguish between optically thick and optically thin regimes. In optically thick regions, thermal relaxation operates through radiative diffusion, with diffusion coefficient $D_{\rm rad}=c/(3\kappa_R\rho)$. Accounting for radiation-gas coupling, with $\eta \equiv E_{\rm rad}/(E_{\rm rad}+E_{\rm int})$ where $E_{\rm int} = P/(\gamma-1)$ is the gas internal energy, the gas thermal diffusivity becomes $\tilde{D} \equiv f D_{\rm rad}$ with $f = 4\eta/(1+3\eta)$. The relaxation time is then scale-dependent: for perturbation with wave number $k$, $t_{\rm relax, thick}=1/(\tilde{D}k^2)$. This highlights the difference between thermal relaxation and radiation transport. For a representative estimate, one may examine the midplane region, taking $k\sim 2\pi/H$ in our base disk model with a constant $\kappa_R\approx 5$ cm$^2$g$^{-1}$, to obtain 
\bgeq
\beta_{\rm cool}^{\rm thick}
\approx24\bigg(\frac{\kappa_R}{5\ {\rm cm}^2{\rm g}^{-1}}\bigg)\bigg(\frac{\rho}{\rho_{\rm mid}}\bigg)^2R_{\rm AU}^{-2}\ . 
\edeq
In optically thin regions, particularly at the disk surface, $t_{\rm relax, thin}$ is determined by the slower channel of two processes. The first is cooling by direct emission, with timescale $t_{\rm emit} \approx C_V/(16\kappa_P\sigma_{\rm SB}T^3)$. 
The second is cooling through collisions with dust particles, with timescale $t_{\rm coll} \approx (n_d\sigma_c v_{\rm th,n})^{-1}$, where $n_d$ is the dust number density, $\sigma_c \sim \pi a^2$ is the geometric collision cross-section for dust of size $a$, and $v_{\rm th,n} \approx \sqrt{3k_BT/\mu m_H}$ is the gas thermal speed. 
For reference, adopting our base disk model with $a_d = 0.1\mu$m, solid density $\rho_s=1$g cm$^{-3}$ and dust-to-gas mass ratio $f_{d2g} \sim 10^{-4}$ yields
\bgeq\label{eq:betathin}
\beta_{\rm cool,emit}^{\rm thin}
\approx1.3\times10^{-4}\bigg(\frac{\kappa_P}{5\ {\rm cm}^2{\rm g}^{-1}}\bigg)^{-1}\ ,\quad
\beta_{\rm cool,coll}^{\rm thin}
\approx3.9\times10^{-4}\bigg(\frac{a_d}{0.1\mu{\rm m}}\bigg)\bigg(\frac{f_{d2g}}{10^{-4}}\bigg)^{-1}\bigg(\frac{\rho}{\rho_{\rm mid}}\bigg)^{-1}R_{\rm AU}\ .
\edeq
While the results can be highly model dependent, we see that cooling in the optically-thick bulk disk becomes nearly instantaneous beyond a few tens of AU. In the optically thin regime, cooling is mainly controlled by dust-gas collision, and can be slow in the disk surface (also see Figure \ref{fig:T}).

\subsubsection{Thermodynamics at the disk atmosphere}\label{sssec:thermatm}

The disk surface is directly exposed to stellar high-energy radiation, including X-ray ($\sim$0.1–10 keV), extreme UV (EUV; $\sim$13.6–100 eV), and far-UV (FUV; $\sim$6–13.6 eV) bands. Such radiation leads to a series of photoionization and photodissociation reactions with significant energy deposition. With complex thermo-chemical physics at play, the thermodynamics is generally non-LTE. This region is also where disk outflows are launched and accelerated, making thermal physics critical for both outflow dynamics and diagnostics
\ifbool{supplement}
{(Section \ref{ssec:magwind} and Supplemental Text).}
{(Sections \ref{ssec:pe}, \ref{ssec:magwind} and \ref{ssec:winddiag}).}

EUV heating is the most well understood, primarily through photoionization of hydrogen atoms. With greatest absorption cross section near the $13.6$eV threshold, penetration is limited to $N_H\sim10^{20}$cm$^{-2}$. The major uncertainty arises from the very poorly constrained EUV luminosity, estimated to be $10^{41-44}$ photons s$^{-1}$ for T Tauri stars \citep{Alexander_etal05}, which might interpolate between FUV and X-ray luminosities \citep{HollenbachGorti09}. The small penetration column makes EUV photons mostly absorbed in the disk upper atmosphere (generally in the outflow), where the gas is heated to $\sim10^4$K analogous to the H\rm{II} regions, as a result of photoionization heating balancing forbidden line cooling \citep{Hollenbach_etal94}. 

Heating by stellar X-rays is closely linked to ionization processes introduced in Section \ref{sssec:ionfid}. Incident photons primarily ionize the K-shell of metal species by the Auger effect, with the ejected photoelectrons triggering secondary ionization and energy deposition. The penetration depth of X-rays is a rapidly increasing function of photon energy, and is on the order of $N_H\sim10^{22}$cm$^{-2}$ for $\sim1$keV photons.

FUV heating is the most complex, similar to PDRs \citep{TielensHollenbach85}. The stellar FUV is expected to arise primarily from accretion hot spots with typical luminosity $L_{\rm FUV}\sim10^{31-32}$erg s$^{-1}$ \citep{Valenti_etal03,Hartmann_etal16}, likely dominated by Ly$\alpha$ emission \citep{Schindhelm_etal12}. FUV drives photoionization (e.g., of \ce{C}, \ce{Fe}, \ce{S}, \ce{Mg}, etc.) and photodissociation (e.g., of \ce{CO}, \ce{OH}, \ce{H2O}), and leads to FUV-pumping of \ce{H2}. All these processes contribute to heating, and drive the transition from molecular to atomic gas in the disk upper layers. 
In the meantime, FUV penetration is primarily regulated by dust attenuation, especially by very small grains (VSGs) and polycyclic aromatic hydrocarbons (PAHs), making it highly sensitive to the uncertain dust properties in the atmosphere (typical $N_{\rm H} \sim 10^{21}$–$10^{23}$ cm$^{-2}$). The VSGs/PAHs are subject to photoelectric effects (with work functions of 4.4eV for graphite and $\sim$8eV for silicate), which typically dominates FUV heating \citep{BakesTielens94,WeingartnerDraine01}. 

Important cooling processes in the disk atmosphere include molecular ro-vibrational line cooling (e.g., \ce{CO}, \ce{OH}, \ce{H2O}, \ce{H2}) and dust-gas collisional cooling in the lower layers and atomic line cooling (e.g., Ly$\alpha$, [O I], [S I] lines) at higher altitudes. This highlights the critical role of chemistry in setting atmospheric temperatures, requiring comprehensive modeling with a chemical network for major coolants, coupled with X-ray/UV radiative transfer and calculating non-LTE atomic/molecular level populations.
A number of numerical tools have been widely adopted for this purpose, such as DALI \citep{Bruderer_etal12,Bruderer13} and ProDiMo \citep{Woitke_etal09,Kamp_etal10}.
Typical temperatures range from $\sim10^3$ K in the molecular layer (at $\sim$1 AU) to several thousand Kelvin in the atomic layer \citep{Walsh_etal12,Woitke15}. Under hydrostatic density profiles, sample surface temperature profiles in the gas can be found in Figure \ref{fig:T}.

The heating-cooling balance under realistic conditions is more complex, as the disk launches outflows, where density profiles deviate significantly from hydrostatic equilibrium. Adiabatic expansion provides additional cooling, while considering magnetic field, ambipolar (Joule) heating often dominates over other mechanisms \citep{Panoglou_etal12,Wang_etal19}, which will be elaborated in Section \ref{sssec:innertherm}.

\begin{figure*}[h]
\centering
\includegraphics[width=\textwidth]{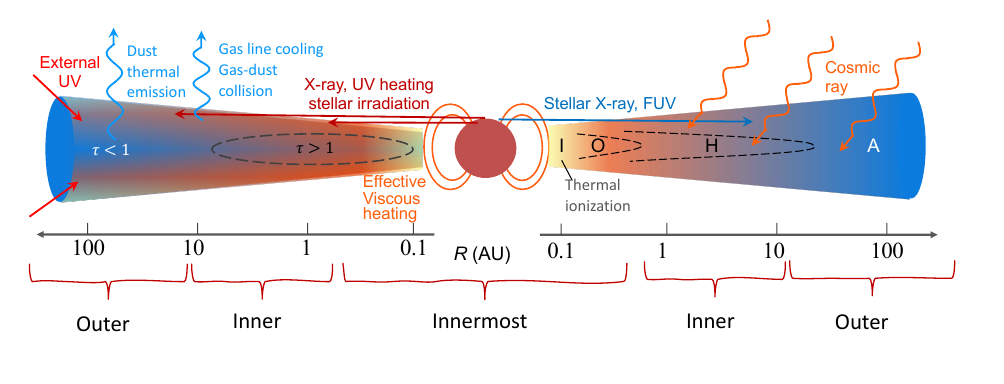}
\caption{Sketch of a protoplanetary disk, which can be divided into three regions governed by distinct microphysics according to different regimes of magnetic coupling (right side, I and O/H/A denote the ideal MHD regime, and Ohmic/Hall/ambipolar diffusion dominated regimes) and heating mechanisms (left side, effective viscous heating and irradiation heating), detailed in Section \ref{ssec:division}. [{\it See published version for a more polished figure.}]
\label{fig:division}}
\end{figure*}

\subsection{Division into Three Regions}\label{ssec:division}

Following the discussions in Sections \ref{ssec:mag} and \ref{ssec:thermo}, protoplanetary disks can be approximately divided into three regions, each characterized by distinct microphysics, as illustrated in Figure \ref{fig:division}.
\begin{itemize}
\item The innermost disk region. This region features high gas temperature ($\gtrsim10^3$ K), enabling strong magnetic coupling through thermal ionization. Disk temperatures are predominantly sustained by effective viscous heating from the MRI turbulence (Sections \ref{ssec:mri}, \ref{ssec:innermost}). This region only extends to $\lesssim1$AU depending on disk parameters, yet the physics is extremely rich as (1) magnetic coupling and thermodynamics are tightly interdependent through thermal ionization, and (2) this region also encompasses dust sublimation and disk truncation within small dynamical range.

\item The inner disk region. Immediately outside the innermost region, the drop in disk temperature quenches thermal ionization, allowing all three non-ideal MHD effects to operate. The midplane becomes largely decoupled with magnetic field due to strong Ohmic resistivity, forming the canonical ``dead zone" \citep{Gammie96}. Meanwhile, ambipolar diffusion prevails in the surface layer, leading to partial magnetic coupling ($Am\sim1$). Lacking vigorous MRI turbulence, gas temperatures are primarily determined by stellar irradiation, with minor contributions from Joule heating.

\item The outer disk region. This region is located further out ($\gtrsim10-30$AU) and the disk is marginally coupled with magnetic field (mainly dominated by ambipolar diffusion), with $Am$ typically of order unity. Gas temperatures are primarily governed by stellar irradiation, while the ambient radiation field of the star-formation environment may also contribute. This region is generally accessible to spatially resolved observations for sources in nearby star-forming regions.
\end{itemize}

This division sets the stage for detailed discussions of protoplanetary disk physics in Sections \ref{sec:process} and \ref{sec:fulldisk}, which exhibit distinct characteristics across different regions. Notably, the transition between the innermost and inner regions is expected to be sharp—commonly known as the ``dead zone inner boundary"—marked by the quenching of MRI turbulence and the consequent loss of effective viscous heating (Section \ref{sssec:DZIB}). In contrast, the transition between the inner and outer regions is more gradual, without a clear dividing line.

\section{GAS DYNAMICAL PROCESSES}\label{sec:process}

In this section, we outline the important gas dynamical processes in protoplanetary disks, which are consequences of various microphysics described in Section \ref{sec:micro}. These include several fluid instabilities (Sections \ref{ssec:mri}-\ref{ssec:gi}, summarized in Table \ref{tab:instability}), and
\ifbool{supplement}
{magnetized disk wind/outflows (Section \ref{ssec:magwind}). We also separately discuss photoevaporation in Supplemental Text Section 1.}
{disk wind/outflows (Sections \ref{ssec:pe}-\ref{ssec:magwind}).}

\begin{table}[]
\caption{Summary of important fluid instabilities in protoplanetary disks.}
\label{tab:instability}
\begin{center}
\begin{tabularx}{\textwidth}
{|>{\centering\arraybackslash}m{.105\textwidth}
|>{\raggedright\arraybackslash}m{.202\textwidth}
|>{\centering\arraybackslash}m{.120\textwidth}
|>{\raggedright\arraybackslash}m{.258\textwidth}
|>{\centering\arraybackslash}m{.18\textwidth}|}
\hline
{\bf Instability}$^{\rm a}$ & {\bf Onset criterion / linear properties}$^{\rm b}$ &  {\bf Controlling parameters}$^{\rm b}$ & {\bf Nonlinear properties}$^{\rm b}$ &  $\alpha_S$ \\\hline\hline
MRI: ideal & $d\Omega/dR<0$; $\beta_z\gtrsim10$; growth rate $(3/4)\Omega$ & Vertical field $B_z$ ($\beta_z$) & Vigorous turbulence, flux concentration \& zonal flows & $0.01-1$ (increase with net $B_z$) \\\hline

MRI: Ohmic & Reduced growth rate, $\lambda_c\lesssim2H$ (Eq. \ref{eq:lambdac}) & $\eta_O$ & Sustained turbulence for $\Lambda\gtrsim1$ & Weaker or suppressed \\\hline

MRI: ambipolar & Reduced growth rate; weaker field required  & $Am$ & Sustained turbulence for $\beta>\beta_{\rm min}(Am)$ (Eq. \ref{eq:betamin_am}) & Weaker or suppressed \\\hline

MRI: Hall & Strongly altered for $Ha\lesssim1$ & $l_H$ or $Ha$ & Field amplification/reduc- tion for $\pm$ polarity (${\mb B}\cdot{\mb\Omega}$) & Enhanced/reduced for $\pm$ polarity\\\hline\hline

 VSI &   $\beta_{\rm cool}\lesssim\beta_{c,{\rm VSI}}$ for fastest growth & Vertical shear, $\beta_{\rm cool}$ & Vertical oscillation (corru- gation \& breathing modes); vortices \& zonal flows (?) & $\lesssim10^{-3}$ \\\hline
COS/SBI  & $N_R^2<0$, $\beta_{\rm cool}\sim1$ & $N_R^2$, $\beta_{\rm cool}$ & Vortices \& zonal flows & $\lesssim10^{-3}$ \\\hline
VSI/COS (general)  & Always (but growth can be slow) & vertical shear, $\beta_{\rm cool}$ & To be explored & Likely weaker than standard VSI/COS  \\\hline
ZVI   & $\beta_{\rm cool}\gtrsim10$, initial vortex trigger & $\beta_{\rm cool}$ & Vortices \& zonal flows & Likely $\lesssim10^{-3}$ \\\hline\hline
GI  & $Q\lesssim1$ & $Q, \beta_{\rm cool}$ & Gravito-turbulence ($\beta_{\rm cool}$ $\gtrsim3$) or fragmentation & {\rm min}$(1.2\beta_{\rm cool}^{-1}, 1)$\\\hline
\end{tabularx}
\end{center}
\begin{tabnote}
$^{\rm a}$ These instabilities are discussed in Sections \ref{ssec:mri}-\ref{ssec:gi}; $^{\rm b}$ Readers may consult Table \ref{tab:symbols} for the meanings of the symbols.
\end{tabnote}
\end{table}

\subsection{Magnetorotational Instability}\label{ssec:mri}

The magnetorotational instability (MRI; \citealp{BH91,BH98}) is a cornerstone of accretion disk theory. Its growth and nonlinear saturation generate vigorous turbulence that efficiently transports angular momentum radially outward \cite[e.g.,][]{HGB95}, supplying the necessary effective viscosity. The MRI is often considered universal in accretion disks, as its linear instability arises under three general conditions: (1) a rotating disk with radially decreasing angular velocity ($d\Omega/dR < 0$), (2) sufficient ionization to couple magnetic fields to the gas, and (3) a sub-thermal magnetic field ($\beta > 1$). However, non-ideal MHD effects prevalent in protoplanetary disks can significantly modify or entirely suppress the MRI.

A fundamental parameter for the MRI is the strength of net vertical magnetic field threading the disk, characterized by $\beta_z$ (the midplane plasma $\beta$ for the vertical field). Neglecting vertical stratification,\begin{marginnote}
\entry{Stratified vs. unstratified}{Stratified models include vertical gravity of the central star to capture disk vertical structure; unstratified models treat the disk as a uniform slab as a proxy for the midplane region.}
\end{marginnote} the linear MRI operates only above a critical wavelength $\lambda_c$ \cite[e.g.,][]{Wardle99}:
\bgeq\label{eq:lambdac}
\lambda_c/H\approx5.13\beta_z^{-1/2}(1+\Lambda_z^{-2})^{1/2}\ ,
\edeq
where the factor $5.13=2\sqrt(2/3)\pi$ arises from the MRI dispersion relation, and $\Lambda_z=v_{Az}^2/\eta_O\Omega$ is the Ohmic Elsasser number (\ref{eq:Elsasser}) evaluated using the vertical field. The most unstable wavelength is approximately twice the critical wavelength, with growth rate $s\sim(3/4)\Omega$ under ideal MHD.
When $\lambda_c\gtrsim 2H$ ($\beta\lesssim10$ in ideal MHD regime), unstable wavelengths can hardly fit within the disk due to the stabilizing effect of magnetic tension. This testifies that the MRI is a weak-field instability [condition (3) above]. The direct growth of the MRI under a net vertical field leads to the development of ``channel flows" \citep{GoodmanXu94},\begin{marginnote}
\entry{MRI channel flow}{The linear MRI eigenmode with a net vertical field, which exhibits vertically alternating layers flowing radially inward and outward and grows exponentially even in non-linear regime.} 
\end{marginnote}which saturate into turbulence through parasitic instabilities \cite[e.g.][]{Pessah10}. 

The MRI turbulence generates both Maxwell and Reynolds stress (Equation \ref{eq:Trphi}) that mediate angular momentum transport, with Maxwell stress generally dominating. Local vertically stratified shearing-box simulations of the MRI in ideal MHD indicate that the resulting $\alpha_S$ value increases nearly linearly with net vertical field strength for $\beta_z\lesssim10^{5}$, reaching $\alpha_S\sim1$ when $\beta_z\sim10^2$ \citep{BaiStone13a,Salvesen_etal16}.\footnote{Without net vertical field, the outcome of the MRI depends on background field configuration and microscopic dissipation. The latter is set by the Prandtl number $\textrm{Pm}$, ratio of (microscopic) viscosity to resistivity (stratified simulations fail to converge without explicit dissipation \citep{Bodo_etal14}). Protoplanetary disks always have $\textrm{Pm}\ll1$ even in the ideal MHD regime, and unstratified simulations suggest that the MRI cannot be self-sustained \citep{Mamatsashvili_etal20} unless there is additional mean toroidal field \citep{Meheut_etal15}. With strong initial toroidal field $\beta\sim1$, on the other hand, stratified simulations by \citet{Squire_etal25} indicate sustained dynamo action leading to a highly magnetically-dominated state with $\alpha_S\sim1$.}
\begin{marginnote}
\entry{Shearing-box simulation}{A computational method modeling a local, co-rotating disk patch in Cartesian coordinates with shearing-periodic radial boundary conditions.} 
\end{marginnote}
This is accompanied by increasing disk magnetization, reaching midplane $\beta \sim 1$ for $\beta_z \sim 10^2$ (here $\beta$ represents average over a local volume), where the magnetic field is always toroidally dominated due to radial shear. 
Global simulations also show consistent findings \cite[e.g.,][]{SuzukiInutsuka14,ZhuStone18}. This establishes a fundamental link between angular momentum transport and the magnetic flux threading the disk.
Additionally, a useful empirical relation of the MRI turbulence is:
\bgeq\label{eq:alphbeta}
\alpha_S\beta \approx 1/2\ ,
\edeq
which holds in both ideal MHD \citep{HGB95} and non-ideal MHD \citep{BaiStone11} regimes. It connects the rate of angular momentum transport (dominated by Maxwell stress) to the magnetic energy density, reflecting the geometric properties of MRI-generated magnetic fields \cite[e.g.,][]{Hawley_etal11}. 

\subsubsection{MRI with non-ideal MHD effects}\label{sssec:nimri}

Incorporating Ohmic resistivity introduces a correction factor $(1+\Lambda_z^{-2})^{1/2}$ that increases the critical wavelength in Equation (\ref{eq:lambdac}), where
note that $\Lambda_z\propto(\eta_O\beta_z)^{-1}$.
Resistivity strongly influences the MRI when $\Lambda_z\lesssim1$, where unstable wavelengths scale as $\propto\eta_O$, and the fastest growth rate is reduced as $\propto\eta_O^{-1}$. The Elsasser number criterion also extends to the non-linear regime. Vertically stratified simulations showed that the MRI turbulence operates when $\Lambda\gtrsim1$ \citep{Turner_etal07,IlgnerNelson08}, where $\Lambda$ is evaluated using the total field strength amplified by the MRI itself.

With ambipolar diffusion, the linear MRI properties are identical to the Ohmic case by simply replacing $\Lambda_z$ by $Am$ in Equation (\ref{eq:lambdac}) for pure vertical field. There are additional growing modes in the presence of net toroidal field \citep{Desch04,KunzBalbus04}, associated with the ambipolar shear instability \citep{Kunz08}. The fact that $Am$ is generally independent of magnetic field strength is particularly convenient: Equation (\ref{eq:lambdac}) indicates that with ambipolar diffusion, the MRI requires weaker field to operate than in ideal MHD regime. Similarly, in the non-linear regime, \citet{BaiStone11} identified the maximum sustainable field strength through a set of unstratified simulations, expressed as a fitting formula for minimum $\beta$: 
\bgeq
\beta_{\rm min}(Am)=[(50/Am^{1.2})^2+(8/Am^{0.3}+1)^2]^{1/2}\ .\label{eq:betamin_am}
\edeq 
This can be turned into a maximum achievable value of $\alpha_S$ following Equation (\ref{eq:alphbeta}), depending on field configurations.

The non-dissipative Hall effect modifies the MRI linear properties depending on net vertical field polarity, and becomes significant when $Ha \lesssim 1$ \cite[e.g.,][]{Wardle99,BalbusTerquem01}. Early unstratified simulations with modest Hall diffusivity ($Ha\gtrsim1$) suggested that the Hall effect enhances/reduces MRI turbulence for net vertical field aligned/anti-aligned with disk rotation axis \citep{SanoStone02b}. Strong Hall effects in the aligned case can lead to magnetic field reorganization without strong turbulence \citep{KunzLesur13,Bethune_etal16}, forming zonal flows (see below)—though this phenomenon disappears with vertical stratification. Additionally, \citet{Kunz08} found that the Hall effect combined with shear (without rotation) leads to the {\it Hall-shear instability} (HSI). This instability can be interpreted as the rotation of perturbed magnetic vector $\delta{\mb B}$ by the Hall effect being amplified by shear, and it contributes to the Hall-MRI when rotation is included. 
We will discuss in Section \ref{ssec:inner} that the HSI plays a crucial role in protoplanetary disk gas dynamics.

\subsubsection{Zonal flows in the MRI turbulence}\label{sssec:zonal}

An interesting phenomenon in MRI turbulence is the spontaneous formation of zonal flows, first reported by \citet{Johansen_etal09a} in local ideal MHD simulations with zero or very weak net vertical field. Zonal flows—a concept borrowed from geophysical fluid dynamics—here refer to radial variations in the azimuthal velocity on top of the smooth Keplerian background. These flows are associated with density and pressure variations arising from large-scale stochastic fluctuations in $\alpha_S$. This leads to a balance between the Coriolis force and the radial pressure gradient:
\bgeq
2\rho(v_\phi - v_K)\Omega = \frac{\partial P}{\partial r}.
\edeq
In essence, zonal flows are synonymous with pressure bumps.

In local ideal MHD simulations, zonal flows become stronger with increasing net vertical field, reaching density variations of $\sim30\%$ for $\beta_z = 10^2$ \citep{BaiStone14}. This enhancement results from a redistribution of vertical magnetic flux: flux becomes concentrated within certain radial ranges, leaving adjacent regions nearly devoid of flux. Regions with concentrated flux exhibit higher local $\alpha_S$, expelling gas to neighboring zones. However, the amplitude and radial scale of these zonal flows do not converge with simulation box size in local models \citep{Simon_etal12}. In contrast, global simulations of ideal MRI turbulence in thin disks (with or without net vertical flux) show only modest stochastic variations in the radial surface density profile \cite[e.g.,][]{Flock_etal11,ZhuStone18}, with no clear evidence of magnetic flux concentration. This discrepancy reflects the limitation of local simulations. On the other hand, magnetic flux concentration becomes both robust and significant when ambipolar diffusion is included, as we will discuss in Section \ref{sssec:fluxcon}.

\subsection{Hydrodynamic Instabilities}\label{ssec:hydro}

We begin by extending our base disk model to more general power-law profiles: surface density $\Sigma \propto R^{-p_{\Sigma}}$ and temperature $T \propto R^{-p_T}$. The midplane density then scales as $\rho_{\rm mid} \propto R^{-p_{\rho}}$ with $p_{\rho} = 3/2 + p_\Sigma - p_T/2$. For an ideal gas, the specific entropy is $s \equiv c_V \ln(P/\rho^\gamma)$, where $\gamma$ is the adiabatic index. This allows us to define the radial and vertical Brunt-V\"ais\"al\"a (buoyancy) frequencies:
\bgeq\label{eq:buoancy}
N_R^2 \equiv -\frac{1}{\gamma\rho c_V} \frac{\partial P}{\partial R} \frac{\partial s}{\partial R}, \quad
N_z^2 \equiv -\frac{1}{\gamma\rho c_V} \frac{\partial P}{\partial z} \frac{\partial s}{\partial z}.
\edeq
At midplane, $N_R^2>0$ requires $(\gamma-1)p_\rho>p_T$ (as long as $dP/dR<0$). 
Vertically isothermal disks are stably stratified with $N_z^2>0$ at $z\neq0$.
Additionally, the epicyclic frequency is given by $\kappa^2=R^{-3}\pa_R(j^2)\approx\Omega^2$.

The fact that the MRI can be strongly suppressed or damped has motivated the investigations of pure hydrodynamic instabilities as alternative sources of turbulence and angular momentum transport.
However, the classic Solberg-Høiland criteria for stability \cite[e.g.][]{Tassoul78} 
\bgeq
\kappa^2+N_R^2+N_z^2>0\ ,\quad-\frac{\pa P}{\pa z}\bigg(\frac{\pa j^2}{\pa R}\frac{\pa s}{\pa z}-\frac{\pa j^2}{\pa z}\frac{\pa s}{\pa R}\bigg)>0\ ,
\edeq
indicate that protoplanetary disks are hydrodynamically stable against linear axisymmetric perturbations under adiabatic conditions. Despite this, three major pure hydrodynamic instabilities have been identified over the past two decades: the Vertical Shear Instability (VSI), the Convective Overstability (COS), and the Zombie Vortex Instability (ZVI). Among them, VSI and COS require finite thermal relaxation time, while ZVI requires finite-amplitude, non-axisymmetric perturbations, thus do not conflict with the Solberg-Høiland criteria. The onset of these instabilities crucially depends on disk structure and gas thermal timescales. Collectively, they likely operate across broad regions in disks, generating (weak-to-moderate) turbulence and vortices, and transport angular momentum outward with $\alpha_S\lesssim10^{-3}$. Several recent reviews have described these instabilities in great detail \citep{LyraUmurhan19,FromangLesur19,Lesur_etal23}; here we summarize their essential properties. 

\subsubsection{Vertical shear instability}\label{sssec:VSI}

The VSI \citep{UrpinBrandenburg98,Urpin03} is the counterpart of the Goldreich-Schubert-Fricke (GSF) instability \citep{GoldreichSchubert67,Fricke68} in differentially-rotating stars manifested in disks. It was made widely known thanks to the independent discovery by \citet{Nelson_etal13}. The VSI is an axisymmetric instability that draws its free energy from vertical differential rotation ($\pa v_\phi/\pa z\neq0$). 
Disks with realistic thermodynamics are baroclinic, which naturally leads to vertical shear.\begin{marginnote}
\entry{Baroclinic fluid}{A fluid where the surface of constant pressure and constant density are misaligned.}
\end{marginnote}For vertically isothermal disks (as in our base disk model), $\pa (v_\phi/c_s)/\pa (z/H)\approx (p_T/2)(c_s/v_K)(z/H)$: vertical shear is weak but increases toward surface layer. The VSI operates by amplifying inertial waves that are vertically elongated and radially narrow, which are destabilized against vertical buoyancy by rapid cooling. Using the $\beta$-cooling prescription (Equation \ref{eq:betacool}), \cite{LinYoudin15} showed that the VSI requires
\bgeq
\beta_{\rm cool}\lesssim\beta_{\rm c, VSI}\equiv[p_T/(\gamma-1)](H/R)\ .\label{eq:betacvsi}
\edeq
The linear growth rate goes as $(|p_T|/2)(|z|/R)\Omega$. Notably, \cite{LatterPapaloizou18} found that the linear modes can carry over into the nonlinear regime.

The nonlinear properties of VSI have been primarily studied by hydrodynamic simulations with $\beta$-cooling, often setting $\beta_{\rm cool} = 0$. \citet{Nelson_etal13} first unambiguously identified VSI in such simulations. The VSI is initially characterized by ``surface modes", which grow the fastest near the disk surface. It eventually becomes dominated by ``body modes" exhibiting narrow bands of large-scale vertical motions throughout the disk column, with typical radial wavelengths of $\sim$ few $H^2/R$. Specifically, turbulent power concentrates in the ``fundamental corrugation" modes (vertical oscillation of the entire disk columns) and ``breathing" modes (reflection-symmetric oscillations about the midplane). These modes behave as traveling inertial wave trains \citep{Svanberg_etal22,Ogilvie_etal25}.
The dominance of vertical motion makes the VSI turbulence highly anisotropic \citep{Stoll_etal17}, with $(\delta v_z/c_s)^2\sim10^{-2}$. Meanwhile, the VSI transports angular momentum radially outward at some moderate $\alpha_S\sim10^{-4}$ to $10^{-3}$, which increases with disk aspect ratio \citep{Manger_etal20}. In 3D, early simulations found that the VSI turbulence continuously generates vortices \citep{Richard_etal16,MangerKlahr18}, likely through the Rossby-Wave Instability \citep{Lovelace_etal99} seeded by weak pressure bumps resulting from radial variations in the rate of angular momentum transport.
However, at very high resolution with $\gtrsim100$ cells per $H$, \citet{Lesur_etal25} no longer observe long-lived pressure bumps and vortices despite vigorous turbulence (see also \citealp{Flores-Rivera_etal20}).
\begin{marginnote}
\entry{Rossby-wave Instability}{A non-axisymmetric instability in thin disks occurring at local maxima of inverse vortensity ($\Sigma s^{2/\gamma}/(\nabla\times{\mb v})_z$), which leads to vortex formation.}
\end{marginnote}

\subsubsection{Convective overstability}

The COS is a linear, axisymmetric instability that derives its free energy from a radial entropy gradient, akin to convection \citep{KlahrHubbard14,Lyra14,Latter16}. In unstratified models, COS operates when $N_R^2<0$ and requires cooling on dynamical times ($\beta_{\rm cool}\sim1$). While radial convection is stabilized by rotation to become epicyclic oscillations, cooling near orbital period allows perturbed gas parcels to exchange heat with surroundings and experience buoyant acceleration. As a result, the COS manifests as growing epicyclic oscillations.
When applied to the disk midplane, $N_R^2<0$ corresponds to unusual surface density profiles ($-7/4<p_{\Sigma}\lesssim0$ for $p_T=0.5$ and $\gamma=7/5$) that may only apply in special locations. However, this condition is more easily satisfied toward the surface layer.

Most numerical studies of COS have employed unstratified models. Under axisymmetry and with sufficiently large Reynolds numbers (low viscosity), the nonlinear evolution of the COS eventually leads to the development of zonal flows over hundreds of local orbits \citep{TeedLatter21}. This likely further triggers vortex formation, as observed in 3D simulations \citep{Lyra14,LehmannLin24}. 
The vortices are subject to amplification by the ``Subcritical Baroclinic Instability" (SBI), a process that was actually discovered earlier \citep{KlahrBodenheimer03,Petersen_etal07a}. The term ``subcritical" emphasizes that SBI requires finite-amplitude vortices to start with \citep{LesurPapaloizou10}. The SBI operates under the same physical conditions as COS, but is fundamentally a 2D instability in the horizontal plane, whereas the COS is axisymmetric and requires the vertical dimension. The resulting vortices launch density waves \citep{JohnsonGammie05b}, which lead to moderate outward angular momentum transport. The $\alpha_S$ value can reach up to a few times $10^{-3}$ \citep{LesurPapaloizou10,LyraKlahr11,LehmannLin24}, though it is much reduced for less favorable entropy gradient or cooling time \citep{Raettig_etal13}. In the meantime, the vortices are subject to destruction via the elliptical instability \citep{LesurPapaloizou09,Barge_etal16}.
Depending on thermal diffusivity, \citet{TeedLatter25} found that the saturated state can exhibit either cycles of zonal-flow/vortex creation and destruction or long-lived persistent vortices.

Recently, \citet{Klahr26} pointed out that in vertically stratified disks (which inherently exhibit vertical shear), the GSF/VSI instability and COS represent two unstable branches of the same dispersion relation for any cooling time, a result already present in earlier analysis \citep{GoldreichSchubert67,Shibahashi80,Urpin03}. 
In particular, the GSF/VSI growth rate does not vanish for longer cooling time $\beta_{\rm cool}>\beta_{c, {\rm VSI}}$, but decrease as $1/\tau^*$, where $\tau^*\equiv\gamma\beta_{\rm cool}$. The condition for COS is generalized to $N_-^2<0$, where $N_-^2$ is the smallest squared Brunt-V\"ais\"al\"a (buoyancy) frequency among all meridional directions. As long as the gas is baroclinic, the surfaces of constant pressure and specific entropy are not aligned, there always exist directions with $N^2<0$, i.e., the system is always unstable to the COS. Its growth rate is approximately
\bgeq
\Gamma_{\rm COS}\approx\frac{p_T^2}{8}\bigg(\frac{H}{R}\bigg)^2\frac{\gamma}{\gamma-1}\frac{\tau^*}{1+\tau^{*2}}\Omega\ .
\edeq
It peaks at $\tau^*\sim1$, same as the unstratified case, but the rate is much smaller (by $\sim qH/4R$) than the VSI growth rate under instant cooling. For $\tau^*\gtrsim1$, the GSF/VSI growth rate is always twice that of the COS, underscoring the close connection between the two instabilities. 

This finding opens a new avenue to explore hydrodynamic turbulence in protoplanetary disks, especially in cooling regimes with $\beta_{\rm cool}\gtrsim\beta_{c,{\rm VSI}}$, where VSI/COS growth rates are typically very small $\lesssim10^{-3}\Omega$. Capturing the dominant growing modes generally requires low-dissipation numerical schemes with very high resolution. \citet{Klahr_etal26} carried out numerical experiments confirming these modes in 2D axisymmetric setting with $256$ cells per $H$ resolution. Future very-high-resolution 3D simulations are essential to characterize the saturated state of these instabilities and quantify the resulting $\alpha_S$ for angular momentum transport.

\subsubsection{Zombie vortex instability}

The ZVI is a non-axisymmetric, nonlinear instability operating under near adiabatic conditions ($\beta_{\rm cool}\gg1$). It was first identified in \citet{BarrancoMarcus05}, with the physics clarified later by \citet{Marcus_etal13,Marcus_etal15,Marcus_etal16}, showing that vortices in stably stratified disks can excite ``baroclinic critical layers" at some distance away, which subsequently spawn new vortices. These critical layers are narrow structures resulting from a resonance between the Doppler-shifted (by radial shear) frequency of a Rossby-wave associated with the initial vortex, and the Brunt-V\"ais\"al\"a frequency. The nonlinear dynamics within these layers generate jet-like vertical vorticity fields subject to secondary instabilities \citep{Umurhan_etal16}, completing the process of vortex self-replication. The ZVI is highly sensitive to dissipative processes at small scales which suppress the dynamics in the critical layers, requiring Reynolds number $\gtrsim10^7$ and cooling time $\beta_{\rm cool}\gtrsim10$ \citep{LesurLatter16}. 
Capturing the ZVI numerically thus requires low-dissipative schemes with very high resolution ($\gtrsim256$ cells per $H$, \citealp{Marcus_etal15}). 

The nonlinear development in the critical layer leads to volume-filling vortices exhibiting a Kolmogorov spectrum \citep{Marcus_etal16}, where the injection scale is approximately the critical layer separation (typically $\sim1-2H$). With more realistic stratification ($g\approx-\Omega^2z$),  \citet{Barranco_etal18} found that the ZVI first develops off the midplane ($z\gtrsim1.5H$) where the critical layer is formed with buoyancy frequency $N_z^2\gtrsim\Omega^2$. The ZVI turbulence later penetrates into the midplane region but at reduced turbulent strength. Long-term simulations reveal the emergence of radially-periodic  zonal flows with a radial wavelength $\gtrsim H$, and the system eventually enters an intermittent state that cycles between a quasi-laminar state of zonal flows, and chaotic outbursts of newly generated zombie vortices. Finally, we note that while the existence of the ZVI has been confirmed by compressible, finite-volume codes \citep{Marcus_etal15}, almost all numerical studies of the ZVI use incompressible models, precluding reliable measurement of the Reynolds stress and hence $\alpha_S$. A hypothesized optimistic estimate of $\alpha_S\sim10^{-3}$ was reported by \citet{Marcus_etal15}.

\subsection{Gravitational Instability}\label{ssec:gi}

The physics of gravitational instability (GI) has been discussed in the comprehensive review by \citet{KratterLodato16}. Here, we briefly summarize the essentials, and highlight recent developments. 

Under the Wentzel Kramers-Brillouin (WKB) approximation (assuming tightly-wound spirals), GI in disks is triggered when the Toomre $Q$ parameter,
\bgeq\label{eq:TmrQ}
Q\equiv\frac{c_s\kappa}{\pi G\Sigma}\approx\frac{M^*}{M_d}\frac{H}{R}\ ,
\edeq
falls below unity, where approximately $M_d\approx\pi\Sigma R^2$ for disk size $R$. This occurs when the disk is massive, or cold. Growing GI modes have a typical wavenumber $k\sim1/H$ and manifest as spiral density waves.
GI simulations often adopt $\beta$-cooling to control thermodynamics, and \citet{Gammie01} showed that the non-linear outcome of GI can be either gravito-turbulence or fragmentation when $\beta_{\rm cool}$ is above or below a threshold $\beta_{c, {\rm GI}}$. Most GI simulations adopt adiabatic index $\gamma=5/3$, where it was found $\beta_{c, {\rm GI}}\approx3$ \cite[e.g.][]{Deng_etal17}. The $\beta_{c, {\rm GI}}$ value is likely higher for $\gamma=7/5$ \citep{Rice_etal05}, which is more appropriate for protoplanetary disks.
On the other hand, an emerging view based on theoretical arguments \citep{Paardekooper12,HopkinsChristiansen13} and recent simulations \citep{BrucyHennebelle21,Xu_etal25a} suggest that fragmentation can be stochastic in nature. It can even occur for $\beta_{\rm cool}>\beta_{c, {\rm GI}}$ but at reduced probability. Specifically,
spirals in gravito-turbulence are generically clumpy, and clumps more likely evolve into bound fragments through clump-clump collisions when cooling is fast.

GI transports angular momentum via gravitational and Reynolds stresses (Equation \ref{eq:Trphi}) mainly through the spirals, accompanied by shock dissipation. If turbulent stress and dissipation can be treated as local effective viscosity using the $\alpha$-prescription, then
$\alpha_{S, {\rm GI}}$ can be written as a steep function of $Q$ which caps at $\sim1$, e.g., $\alpha_{S, {\rm GI}}\approx\exp(-Q^4)$ or  $\alpha_{S, {\rm GI}}\approx{\cal H}(1.2-Q)$ \citep{Zhu_etal10a,Xu_etal25b}, where ${\cal H}$ is the Heaviside step function. In the meantime,
balancing effective viscous heating and $\beta$-cooling,
one can derive $\alpha_{S, {\rm GI}}\approx(2/3)[\gamma(\gamma-1)]^{-1}\beta_{\rm cool}^{-1}\approx1.2\beta_{\rm cool}^{-1}$ for $\gamma=7/5$. In this sense, GI is self-regulated: GI dissipation heats up the disk, raising $Q$ towards marginal stability to offset cooling. While largely unexplored, this relation also appears to hold in the fragmentation regime, reaching $\alpha_{S,{\rm GI}}\sim1$ (see \citealp{Xu_etal25b}).

However, it has been recognized that GI transport can be non-local, which questions the use of $\alpha$-disk models, unless the pattern speeds of the spirals match local rotation speeds \citep{BalbusPapaloizou99}. Fortunately, it has been shown that in the nonlinear stage, the spirals undergo shock dissipation typically within a distance of $\sim H$ from corotation radius \citep{Cossins_etal09,Bethune_etal21}. Consequently, contribution from non-local transport is of the order $\sim M_d/M^*$, which is insignificant for $M_d/M^*\lesssim0.1$. This is corroborated by \citet{Xu_etal25b}, who showed quantitatively that over a time-averaged sense, GI can be modeled as a local effective viscosity to zeroth order in $(H/R)$ and $\alpha_S^{1/2}$, and provided prescriptions for first order corrections.

\ifbool{supplement}{}{

\subsection{Photoevaporation}\label{ssec:pe}

Photoevaporation is a hydrodynamic mass loss process driven by heating from energetic radiation \cite[see][for recent reviews]{Gorti_etal16,ErcolanoPascucci17,Pascucci_etal23}. While it does not contribute to angular momentum transport, it has been considered as a major mass loss channel that drives disk dispersal. It can be driven internally by UV/X-ray photons from the central star, or externally by UV radiation from the cluster environment. Here we outline the basic physics together with recent progress.

\subsubsection{Internal photoevaporation}

When gas in the disk atmosphere is heated to a temperature with sound speed $c_{s, {\rm atm}}$, gas beyond radius $r_g=GM_*/c_{s, {\rm atm}}^2$ becomes unbound. A thermal Parker wind can be launched beyond a critical radius $r_{\rm crit}\sim0.1-0.2r_g$ \citep{Adams_etal04}. For pure EUV, X-ray and FUV heating, the resulting $r_{\rm crit}$ is of the order $\sim1$AU, $2-4$AU and $\sim10$AU, respectively 
\ifbool{supplement}
{(see Section 3.5)}
{(see Section \ref{sssec:thermatm})}.
The mass loss rate may be written as $\dot\Sigma_w\approx \rho_bc_{s, {\rm atm}}$, where $\rho_b$ is the gas density at the sonic point, considered as the ``wind base", but accurate estimates require coupling thermochemistry with hydrodynamics.

Early studies generally focus on individual driving mechanisms (EUV, FUV or X-rays). The cases for pure EUV and X-ray photoevaporations are simpler, where gas temperature can be prescribed based on photoionization models to enable full hydrodynamic simulations \citep{Owen_etal12,Picogna_etal19}. Calculations of FUV photoevaporation are considerably more complex and uncertain, being highly sensitive to molecular chemistry and dust properties \cite[e.g.][]{Gorti_etal15}.
Importantly, the heating and cooling processes are coupled with chemistry, radiative transfer and hydrodynamics, therefore, contribution from individual driving sources are not simply additive, but mutually affect each other.

Major advances in photoevaporation theory over the past decade are enabled by coupling thermochemistry with hydrodynamics \citep{WangGoodman17,Nakatani_etal18a,Nakatani_etal18b,Nakatani_etal21,Komaki_etal21,Sellek_etal24}. These hydrodynamic simulations employ ray tracing to track the energetic photons from the central star, coupled with a reduced thermo-chemical reaction network on the fly. \citet{WangGoodman17} found that EUV and Lyman-Werner photons which interact most strongly with molecular/atomic hydrogen play a crucial role in driving mass loss, in spite of small EUV penetration depth. They also found that the static disk atmosphere models commonly employed in the literature tend to overestimate mass loss rate by a factor of several. Both \citet{WangGoodman17} and \citet{Nakatani_etal18b} showed that previous X-ray photoevaporation models tend to overestimate mass loss due to the lack of several (mainly molecular) cooling channels, a result later confirmed by \citet{Sellek_etal24}. 
These results highlight the complex interplay among all different factors, with an emergent picture that FUV largely sets the wind base location, while EUV and X-rays aid flow acceleration. The typical integrated mass loss rates are $10^{-9}$ to $10^{-8}M_\odot$ yr$^{-1}$, and approximately scale with FUV/X-ray luminosity as $\dot{M}_{\rm wind}\propto L_{\rm FUV}^{1/2}$ and $\dot{M}_{\rm wind}\propto L_{\rm X}^{1/4}$. These together lead to $\dot{M}_{\rm wind}\propto M_*^2$ \citep{Komaki_etal21}.
These complex simulations are also starting to motivate improved semi-analytic theory for photoevaporation \citep{Nakatani_etal24}.

\subsubsection{External photoevaporation} \label{sssec:extphoto}

In a cluster environment, disks can experience mass loss due to intense UV radiation from nearby massive stars, with well-known examples being the `proplyds' in the Orion Nebula \cite[e.g.][]{Odell_etal93}.
The external FUV flux is usually measured in Habing unit $G_0$ (low: $<10^2G_0$, intermediate: $10^2$--$10^4G_0$, high: $>10^4G_0$). For reference, with stellar $L_{\rm FUV}=10^{31}$erg s$^{-1}$, the FUV flux at $R=100$AU is about $220G_0$. Therefore, intermediate-to-high level external FUV flux is expected to play a dominant role driving photoevaporation in outer disks. External photoevaporation has received renewed interest over the past decade, and we refer to \citet{WinterHaworth22} and \citet{Allen_etal25} for in-depth coverage on this subject. Here, we outline recent theoretical advances.
\begin{marginnote}
\entry{Habing unit}{Flux integral over $912-2400$\AA  at the solar neighborhood, being $1.6\times10^{-3}$erg cm$^{-2}$ s$^{-1}$.} 
\end{marginnote}

External photoevaporation shares similar physics with FUV photoevaporation. The general physical picture was outlined in \citet{Johnstone_etal98} assuming quasi-spherical flow geometry. Conventional hydrodynamic models of external photoevaporation in 1D  assume steady state under chemical and thermal equilibrium based on PDR calculations \citep{Adams_etal04,Facchini_etal16}, showing mass loss rates sensitive to the disk outer truncation radius, the intensity of external radiation field and dust properties.
Recent studies employ full hydrodynamic simulations in 2D (again assuming equilibrium conditions) \citep{Haworth_etal16b}, reporting mass loss rates consistent with 1D calculations within a factor of $\sim4$ \citep{HaworthClarke19}. This further enabled an extensive survey (in 1D) over a wide range of parameters \citep{Haworth_etal18,Haworth_etal23}, showing mass loss rates from down to well below $10^{-10}M_{\odot}$ yr$^{-1}$ up to $\sim10^{-5}M_{\odot}$ yr$^{-1}$ under extreme conditions (disk size $\sim$a few $10^2$AU with FUV flux $\gtrsim10^4G_0$).
As the physics is closely tied to PDRs, major uncertainties are associated with dust physics in the outflow: grain growth (deeper UV penetration) and higher PAH abundances (more photoelectric heating) can strongly enhance mass loss by up to two orders of magnitude.

}

\subsection{Magnetically-Driven Wind}\label{ssec:magwind}

Disks threaded by a large-scale poloidal field are subject to magnetic launching of disk outflows, which represents a fundamental mechanism for angular momentum transport. Initially developed in the 1980s \citep{BlandfordPayne82}, the theory of magnetically-driven disk winds has subsequently been applied to outflows from young stellar objects and protoplanetary disks \cite[e.g.][]{PudritzNorman83,UchidaShibata85,WardleKoenigl93}. It was later overshadowed by the MRI, but became revived thanks to recent theoretical advances (Section \ref{ssec:inner}).
Here, we outline the basic physics of magnetically-driven disk winds, and additionally highlight the role of thermodynamics (and hence the name magnetothermal wind).

\subsubsection{Basic physics of magnetically-driven wind}\label{sssec:windbasics}

The physical picture of MHD wind launching and angular momentum extraction can be understood as follows. First, starting from an hour-glass-shaped poloidal field anchored to the disk, its radial component can be sheared to build up a toroidal field $\pa B_\phi/\pa t\approx(-3/2)\Omega B_r$.
Second, the vertical gradient in $B_\phi$ leads to a Lorentz force $F_z=(1/c)({\mb J}\times{\mb B})_z\propto-(\pa B_\phi/\pa z)B_\phi$ which pushes the gas away to launch the wind. Third, magnetic fields always tend to straighten thanks to magnetic tension, which resists being wound by exerting a force against rotation/shear $F_\phi=(1/c)({\mb J}\times{\mb B})_\phi\propto-(\pa B_\phi/\pa z)B_z$, thus extracting disk angular momentum and driving accretion \citep{BaiStone13b}
\bgeq
-\frac{1}{2}\rho\Omega v_R\approx-\frac{B_z}{4\pi}\frac{dB_\phi}{dz}\ .\label{eq:windaccflow}
\edeq
With $B_z\approx$ constant across the disk, we can recover the wind-driven accretion rate (Equation \ref{eq:Mdotz}) by integrating this equation over the disk column.

\begin{textbox}
\section{Conservation Laws in a Steady Axisymmetric Flow in Ideal MHD}
A steady and axisymmetric flow in ideal MHD obeys four conservation relations. To begin with, we decompose the velocity and magnetic fields into poloidal (with subscript `$_p$') and toroidal components in cylindrical coordinates ($R, \phi, z$):
${\mb B}={\mb B}_p+B_\phi{\mb e}_\phi$, ${\mb v}={\mb v}_p+R\Omega(R){\mb e}_\phi$,
where ${\mb e}_\phi$ is a unit vector along the toroidal direction, and $\Omega$ is angular velocity of the flow.
Under these conditions, it can be shown that the poloidal flow direction ${\mb v}_p$ must be everywhere parallel to ${\mb B}_p$, and the conservation relations are constants along the poloidal field/stream lines.

The first conserved quantity along poloidal field is expressed as
\begin{equation}
k\equiv\frac{4\pi\rho v_p}{B_p}\ ,\label{eq:k}
\end{equation}
which states that the ratio of mass flux to magnetic flux is conserved.

The second conserved quantity is 
\begin{equation}
\omega\equiv\Omega-\frac{k B_{\phi}}{4\pi\rho R}\ ,\label{eq:omega}
\end{equation}
which states that the angular velocity of magnetic flux surface is constant.
Note that when $k=0$ (no mass loading), equation (\ref{eq:omega}) reduces to Ferraro's law of isorotation $\Omega=\omega$.
With these results, the flow velocity along a field line can be conveniently written as
${\mb v}=k{\mb B}/(4\pi\rho)+\omega R{\mb e}_\phi$.
In the frame corotating with the field line, the flow is everywhere parallel to the magnetic field.

The third conserved quantity is 
\begin{equation}
l\equiv\Omega R^2-\frac{RB_\phi}{k}
=\Omega R^2-\frac{RB_\phi B_p}{4\pi\rho v_p}\ ,\label{eq:l}
\end{equation}
which expresses the specific angular momentum in the wind flow, which consists of a classic kinetic part and a magnetic part stored in the magnetic stress.

Assuming adiabaticity, the last conserved quantity is
\begin{equation}
e\equiv\frac{v^2}{2}+h+\Phi-\frac{\omega RB_{\phi}}{k}\ ,\label{eq:e}
\end{equation}
being the specific energy along a field line. In the above, $h\equiv\int dP/\rho$ is specific
enthalpy, $\Phi$ is gravitational potential. This equation (\ref{eq:e}) can be combined with Equation (\ref{eq:l}) to yield the Bernoulli constant
\begin{equation}
\mathcal{B}\equiv e-\omega l=\frac{v^2}{2}-\omega Rv_\phi+h+\Phi
=\frac{v_p^2+(v_\phi-\omega R)^2}{2}+h+\Phi_{\rm eff}\ ,\label{eq:bernoulli}
\end{equation}
where $\Phi_{\rm eff}\equiv\Phi-\omega^2R^2/2$ is the effective potential. Note that
$\mathcal{B}$ is independent of magnetic field, and the flow becomes unbound when $\mathcal{B}>0$.
\end{textbox}

Given the geometry and strength of poloidal field, general wind properties can be understood from four conservation relations under steady-state, axisymmetric, ideal MHD assumptions \cite[e.g.,][see text box below]{Spruit96}. These relations are considered to apply in the disk atmosphere, where the gas generally approaches (though not necessarily exactly satisfies) ideal MHD conditions thanks to XUV irradiation. Among them, $\omega\approx\Omega_K$, set by the rotation of field line footpoints.
The other three conserved quantities are determined by regularity conditions at ``critical points", where the flow velocity matches characteristic MHD wave speeds in the wind flow.
In particular, consider a poloidal field line anchored at radius $R_0$. By combining equations (\ref{eq:l}) and (\ref{eq:k}), we obtain
$(\Omega-\omega)R^2=(l-\omega R^2)/(1-4\pi\rho/k^2)$.
Singularity arises when $4\pi\rho=k^2$, i.e., $v_p^2=B_p^2/4\pi\rho=v_{Ap}^2$, which corresponds to the Alfv\'en point. Its cylindrical radius,$R_A$ is known as the Alfv\'en radius. Regularity condition for the numerator then demands $l=\omega R_A^2$. The lever arm parameter introduced in Section \ref{sec:diskevol} is simply given by $\lambda=(R_A/R_0)^2$.

What determines the poloidal field geometry and strength? Poloidal field strength is primarily determined by the distribution of magnetic flux across the disk. This flux is likely inherited from disk formation (Section \ref{ssec:early}),
and subsequently evolved via poorly understood processes of magnetic flux transport (Section \ref{ssec:Bfluxtrans}).
The field geometry is set by force balance across neighboring field lines. Its steady state solution is described by the {\it Grad-Shafranov} equation, a second-order non-linear partial differential equation which is very challenging to solve.
Special solutions can be obtained by imposing the self-similarity {\it ansatz}\begin{marginnote}
\entry{Self-similar MHD wind solution}{Wind solution based on an infinitely-extended disk where all physical quantities follow appropriate power-law scalings with radial distance.} 
\end{marginnote}\cite[e.g.][]{BlandfordPayne82},
but solution properties, such as mass loss rates and wind collimation, strongly depend on the imposed magnetic flux profile and boundary conditions \citep{Ostriker97}. Alternatively, general wind solutions can be obtained via time-dependent simulations with imposed magnetic flux profile towards (quasi-)steady state \cite[e.g.][]{Krasnopolsky_etal99}, which has become routine since the 2000s.

\begin{figure*}[h]
\centering
\begin{minipage}[l]{0.496\textwidth}
\includegraphics[width=\textwidth]{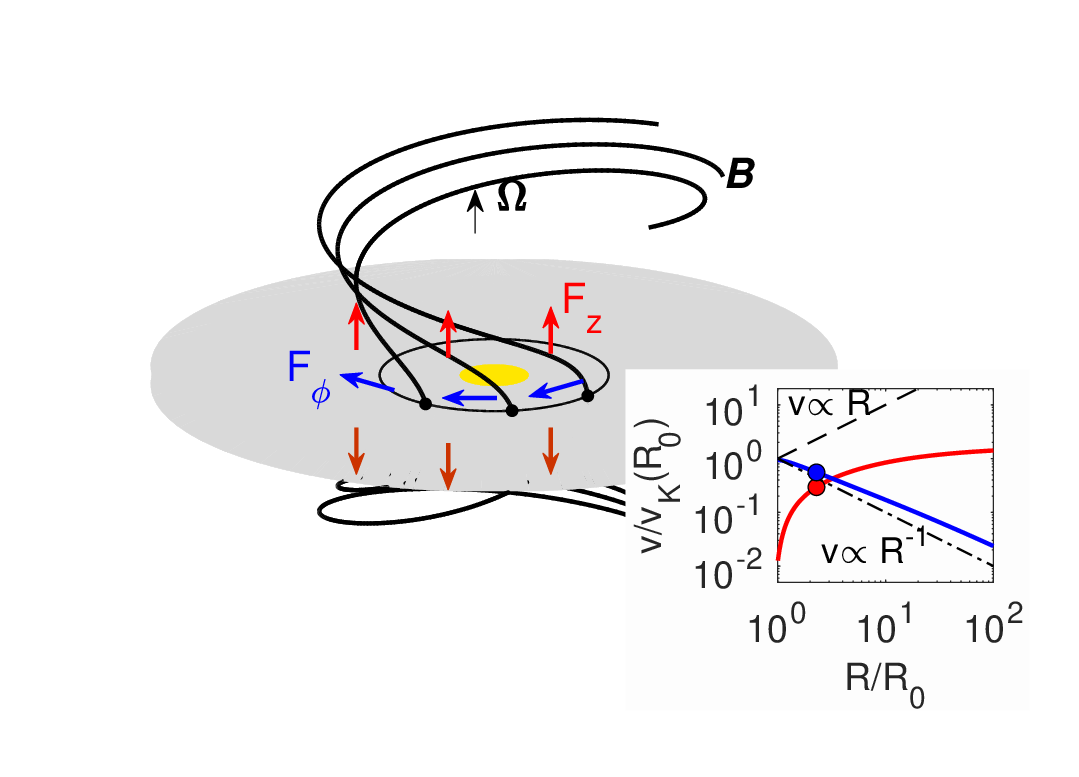}
\end{minipage}
\begin{minipage}[l]{0.496\textwidth}
\includegraphics[width=\textwidth]{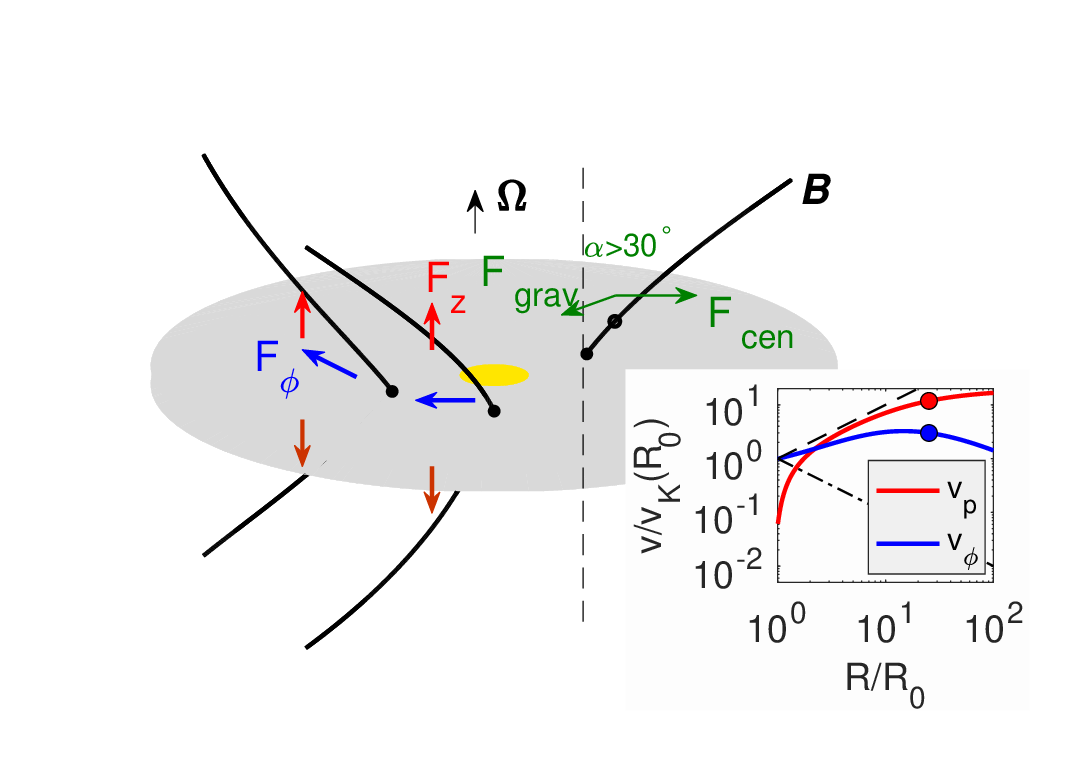}
\end{minipage}
\caption{Sketch of magneto-thermal wind (left) and magneto-centrifugal wind (right) launched from radius $R_0$. In the former case, the field is weak and field lines get tightly-wound from the disk surface. Wind launching is driven by vertical magnetic and thermal pressure gradient ($F_z$), and the flow is characterized by small Alfv\'en radius (dots in the inset). A braking torque ($F_\phi$) is exerted as a result of bent field lines. In the latter case, the field is sufficiently strong to enforce corotation (blue lines in the insets). The field lines remain largely straight before reaching the Alfv\'en radius, which is much larger, leading to centrifugal acceleration. The Lorentz force still assists wind launching and drives accretion. The left panel is adapted from \citet{Weiss_etal21} (CC BY-NC 4.0).
\label{fig:windsketch}}
\end{figure*}

\subsubsection{Conventional cold MHD wind}

Classic magnetically-driven wind models \cite[e.g.][]{BlandfordPayne82} assume the wind to be cold, with thermal energy being negligible in the energy budget. Early models treated the disk as razor-thin, i.e., a boundary condition. Subsequent models/simulations incorporated disk vertical structure \cite[e.g.][]{WardleKoenigl93,CasseKeppens02,Zanni_etal07}, imposing poloidal magnetic field that reaches equipartition ($\beta_z\sim1$) at the disk midplane, and hence the wind flow is always magnetically dominated. One useful constraint for such cold MHD wind can be obtained by diminishing $h$ while requiring $\mathcal{B}>0$ at the wind base, leading to a lower limit of $\lambda>3/2$ for Keplerian disks.

Such cold wind models are known as ``magneto-centrifugal" wind (right panel of Figure \ref{fig:windsketch}). Being magnetically dominated, the field configuration becomes essentially ``force-free".\begin{marginnote}
\entry{Force-free}{When highly magnetically dominated, force balance essentially reduces to Lorentz force being zero.} 
\end{marginnote}As a result, poloidal fields anchored to the disk are very stiff and become nearly straight (instead of being tightly wound). Gas parcels loaded to the field are enforced to co-rotate at the angular velocity at the field footpoints, and can be accelerated centrifugally (in the corotating frame) like ``beads-on-a-wire" when poloidal field is inclined by more than $30^\circ$ relative to disk rotation axis \citep{BlandfordPayne82}.
Corotation cannot be sustained indefinitely, and must break down before reaching the Alfv\'en point, where by definition, the flow's ram pressure matches magnetic pressure, and the field must be wound up afterwards. 

A major drawback of classic wind models is the accretion rate. Following
\ifbool{supplement}
{Supplemental Text Section 2.3.1}
{Section \ref{sssec:Bexp}},
we adopt geometric parameters $m\sim1$ and $f'\sim1$ in 
\ifbool{supplement}
{Supplemental Text Equation (6)}
{Equation (\ref{eq:Bzphi})} 
for cold disk wind, and expressing field strength using the midplane plasma $\beta$ with our base disk model, we obtain $\dot{M}^z_{\rm acc}\approx8\sqrt{2\pi}\beta_{z,{\rm mid}}^{-1}\Sigma c_sR\approx2\times10^{-4}\beta_{z,{\rm mid}}^{-1}R_{\rm AU}^{-1/4}M_\odot$ yr$^{-1}$. This value far exceeds accretion rates of typical T-Tauri disks if $\beta_{z,{\rm mid}}$ is of order unity. Cold wind models can still be applicable in two cases: disks in early stages where accretion rate can be much higher (Section \ref{ssec:early}), and in highly depleted disk regions such as transition disk cavities (Section \ref{sssec:truncation}).

\subsubsection{Recent development and magneto-thermal wind}\label{sssec:magthermal}

More recent development in magnetically-driven wind models/simulations take into account the following three key ingredients.
First, 
disks are threaded by weak vertical field with midplane $\beta_z\sim10^{4-5}$, which helps achieve reasonable wind-driven accretion rate.
Second, the proper prescription of non-ideal MHD effects in disk-wind transition: the disk region is subject to strong non-ideal MHD effects
while the disk surface can be closer to ideal MHD conditions thanks to XUV ionization.
Third, the same stellar/external XUV also leads to significant heating, which would already drive photoevaporation in the absence of magnetic field. 
Therefore, launching of magnetized wind from protoplanetary disks inevitably involves 1). a transition from the non-ideal disk region which does not obey the wind conservation relations, and 2). an ingredient from thermal driving.

Motivated from these considerations, \citet{Bai_etal16} constructed a simple ``magneto-thermal" wind model, where launching occurs in an ideal MHD atmosphere above a transition zone ($|z| \sim 3-5H$). The main parameters are $v_{Ap}\equiv B_p^2/4\pi\rho$ and $c_s$, representing poloidal field strength and temperature at the wind base, with wind solution obtained by solving conservation relations along prescribed poloidal field lines (assumed to be straight and inclined) in the wind zone. 
The classic magneto-centrifugal wind is recovered when $v_{Ap}\gg c_s$. Under more realistic conditions, $v_{Ap}$ and $c_s$ are expected to be comparable. In this regime (left panel of Figure \ref{fig:windsketch}), the poloidal field is insufficient to enforce corotation. Instead, the flow approaches the $v_\phi\propto R^{-1}$ profile, and the field becomes tightly wound ($B_\phi\sim1-10B_p$) already in the wind base. Wind launching is primarily driven by the vertical gradient of $B_\phi^2$, with minor assistance from thermal pressure gradient. Most importantly, the lever arm is small (typically $R_A/R_0\lesssim2$), indicating the wind is heavily loaded with significant mass loss (see Section \ref{ssec:diskevolsol}). In other words, warm disk winds enhance mass loss, pointing to a smooth transition from pure photoevaporative wind to fully magnetically-driven wind.

An example of such transition was demonstrated in \citet{Rodenkirch_etal20}, who studied the interplay between magnetically-driven wind and X-ray-driven photoevaporation. The mass loss rate reaches  a floor value of $\dot{M}_{\rm wind}\sim10^{-8}M_\odot$ yr$^{-1}$ for very weak vertical field ($\beta_{z,\rm mid}\gtrsim10^7$), corresponding to pure X-ray photoevaporation.
Further increasing poloidal magnetic flux leads to rapid increase in disk mass loss rate. On the other hand, \citet{Lesur21} designed a 1D simulation approach to systematically obtain global self-similar magnetized wind solutions.
Using prescribed non-ideal MHD diffusivity profiles but without surface heating, he demonstrated a similar trend of how wind properties vary with net poloidal field strength. 
Most solutions exhibit small lever arm $\lambda\lesssim2$, thus surface heating is not essential to obtain low-$\lambda$ wind solutions.

Finally, it should be noted that while wind-driven accretion rate is generally set by robust physics, other properties (e.g., mass loss rate) are likely sensitive to the physics in the disk-wind transition zone, which different authors treat differently. Therefore, caution must be exercised when quoting the absolute values. 

\subsection{Other Non-Turbulent Processes}\label{ssec:nonturb}

There are several other processes that lead to angular momentum transport without sustained turbulence. Hydrodynamically, these processes generally require external forcing to drive large-scale spiral density waves, which nonlinearly steepen into spiral shocks as they propagate \cite[e.g.][]{Larson90,GoodmanRafikov01}. Shock dissipation leads to irreversible heating, together with a jump in angular momentum across the shock, which induces accretion. Averaging over azimuth and time, the heating rate per unit radius and the accretion rate are given by
\bgeq
\frac{d\dot{Q}}{dR}=[\Omega_p-\Omega(R)]\frac{dF_J}{dR}, \quad
\dot{M}_{\rm acc}=-\bigg(\frac{dl}{dR}\bigg)^{-1}\frac{dF_J}{dR}\approx-\frac{2}{v_K}\frac{dF_J}{dR},
\edeq
where $\Omega_p$ is the pattern speed of the spiral (normally the angular frequency of the perturber which drives the spiral), $dF_J/dR$ is the rate of angular momentum deposition per unit radius, given by \citep{Rafikov16}
\bgeq
\frac{dF_J}{dR}={\rm sgn}[\Omega_p-\Omega(R)]m_\phi R\Sigma c_s^2\psi_Q(\Pi),
\edeq
where $m_\phi$ represents the number of spiral arms, $c_s$ is the pre-shock isothermal sound speed, $\Pi$ is ratio of post-shock to pre-shock pressure $p/p_0$, and $\psi_Q(\Pi)=(\Pi\Xi^\gamma-1)/(\gamma-1)$, with $\Xi(\Pi)\equiv[(\gamma+1)+(\gamma-1)\Pi]/[\gamma-1+(\gamma+1)\Pi]$, and $\gamma$ is the adiabatic index. In the weak shock limit, it reduces to $\psi_Q\approx\gamma(\gamma+1)(\Delta\Sigma/\Sigma_0)^3/12$ \cite[e.g.][]{Savonije_etal94}, where $\Delta\Sigma$ is the surface density jump across the shock. Compared to effective viscously driven accretion in steady state (from Equation \ref{eq:macc_R}), approximately $\alpha_S\approx m_\phi\psi_Q/\pi$. Therefore, a spiral shock with $\Delta\Sigma/\Sigma_0$ of order unity effectively yields $\alpha_S\sim0.1$. These results have been verified by recent hydrodynamic simulations \citep{RyanMacFadyen17,ArzamasskiyRafikov18}.

There are several categories of external forcing that drive large-scale spirals:
\begin{itemize}
\item[--] Tidally-induced spirals. This is the most well-studied case particularly in the context of planet-disk interaction: a perturber drives density waves inside and outside of its orbit from Lindblad resonances \cite[see reviews by][]{KleyNelson12,Paardekooper_etal23}. In particular, Jovian-mass or stellar-mass companions can induce secondary or even tertiary spiral arms inside its orbit by constructive interference \cite[e.g.][]{Dong_etal15,BaeZhu18}. The same physics applies to stellar flybys, which induce transient but strong spirals and can temporarily raise the accretion rate by up to an order of magnitude \citep{Cuello_etal19}.

\item[--]Infall-induced spirals. During disk early stages, infall forms an accretion shock with strong azimuthal shear. \citet{Lesur_etal16} and \citet{Hennebelle_etal17} found that the accretion shock becomes unstable to KHI/RWI-like instabilities, launching spiral density waves that propagate inward. Such spirals lead to significant angular momentum transport with an effective $\alpha_S$ up to $10^{-2}$ (at high infall rate of $10^{-6}M_\odot$ yr$^{-1}$) near the accretion shock, though it rapidly declines inward.

\item[--]Shadow-driven spirals. Some protoplanetary disks exhibit asymmetric shadow features in scattered light observations \citep{Benisty_etal23}, likely resulting from a misaligned inner disk. Because the outer disk is primarily heated by stellar irradiation (Section \ref{sssec:diskTemp}), shadowing thus creates periodic thermal forcing. \citet{Montesinos_etal16} first demonstrated spiral formation from thermal forcing in the context of transition disk HD 142527. In fact, shadows can generate a variety of substructures including spirals, rings and vortices depending on shadow parameters (cooling time, effective viscosity, etc.), with spirals being the initial linear response, which share the same pattern speed of the shadow \citep{SuBai24,Ziampras_etal25}. \citet{Zhu_etal25} showed that spirals are launched in a way similar to a dynamical perturber from Lindblad resonances, but often lead to anomalous (oscillatory) wave transport with a tendency to form rings.
The waves may dissipate effectively in transition disk cavities, leading to strong $\alpha_S\sim10^{-3}-10^{-2}$ \citep{QianWu24,ZhangZhu24}. Considering the vertical dimension, thermal forcing can further drive strong vertical oscillation and dissipation, leading to $\alpha_S$ up to order unity \citep{Zhang_etal25}.
\end{itemize}

How far the externally induced spirals can propagate into the disk is uncertain. \citet{Hennebelle_etal16} constructed self-similar solutions of locally-isothermal spiral shocks, suggesting that they can propagate deep into the disks. On the other hand, \citet{Bae_etal16} identified that the spirals are subject to the ``spiral-wave instability"
which couples a pair of inertial waves and the spiral wave. The growth of inertial waves at the expense of the spiral wave leads to small-scale turbulence, potentially destroying the spirals. 

Finally, radial angular momentum transport can also be mediated by ``magnetic spirals", i.e., a laminar Maxwell stress $-B_RB_\phi$. In a laminar disk, a net $B_R$ will be constantly sheared to produce a large-scale $-B_\phi$ (Section \ref{sssec:windbasics}). The generation of $B_\phi$ can be balanced by non-ideal MHD effects which dissipate $B_\phi$, and/or disk outflow which advects $B_\phi$ away. Therefore, the efficiency of angular momentum transport by laminar Maxwell stress can be sensitive to local microphysics and wind launching (Section \ref{sssec:innerdyn}).

\section{FULL DISK APPLICATIONS}\label{sec:fulldisk}

In this section, we move toward the ``reality" characterized by the interplay among multiple physical processes. We focus on the Class II stage and separately discuss three distinct regions outlined in Section \ref{ssec:division}.
While a complete picture remains elusive, tremendous progress has been made, primarily through numerical simulations that incorporate complex physics to address nonlinear phenomena. A major establishment is the emerging consensus that angular momentum transport and disk evolution are likely dominated by magnetically-driven disk winds, on top of which additional physics is being built.

\subsection{The Inner Disk Region}\label{ssec:inner}

Recent studies of the disk inner region feature incorporation of multiple non-ideal MHD effects with proper magnetic diffusivity prescriptions (based on ionization chemistry calculations) in stratified disk models.  Also included are heating/cooling processes with varying level of complexity. 

\subsubsection{Non-ideal MHD physics and disk-wind connection}\label{sssec:innerdyn}

Early studies of non-ideal MHD physics in protoplanetary disks focused on Ohmic resistivity under the framework of MRI-driven accretion, leading to the physical picture of layered accretion \citep{Gammie96,Armitage11}:
a midplane ``dead zone" (MRI-suppressed by Ohmic resistivity) and surface ``active layers" with vigorous MRI turbulence.
However, when including ambipolar diffusion, even allowing the MRI to operate at maximum efficiency (see Section \ref{sssec:nimri}), semi-analytic models show that the resulting $\alpha_S$ values are too small for the disk to achieve nominal accretion rate of $10^{-8}M_\odot$ yr$^{-1}$ \citep{Bai11a,Mohanty_etal13,Delage_etal22}.

Through local shearing-box simulations that incorporate both Ohmic resistivity and ambipolar diffusion based on a pre-computed diffusivity table, \citet{BaiStone13b} showed that in the absence of net vertical magnetic field threading the disk, there are only minimum MRI activities in the disk upper layer with $\alpha_S\sim10^{-6}$. In contrast, including a weak net vertical magnetic field with midplane $\beta_z=10^5$ (Section \ref{sssec:magthermal}) in a minimum-mass solar nebula disk model, the MRI turbulence is completely suppressed. The disk enters a laminar state, launching a magnetically-driven disk wind from the surface that drives accretion at the desired rate of $\sim10^{-8}M_\odot$ yr$^{-1}$. They found that wind launching occurred exactly at the FUV front (though not always necessary), prescribed as a sharp ionization transition from $Am\sim1$ to the ideal MHD regime ($Am\gtrsim100$). The general wind properties are mostly set by the net vertical field threading the disk and are insensitive to dust abundance, but deeper FUV penetration enhances wind mass loss and transport.

The simulations also exposed intrinsic limitations of the shearing-box framework. First, as the vertical domain is truncated, mass loss rate depends artificially on the vertical domain size. Second, as the shearing box model ignores disk curvature, there is no distinction between radially inward/outward, the outflows tend to exhibit symmetry issues. Under the ``correct" symmetry, where winds launched from both upper and lower surfaces bend in the same radial direction, the horizontal field components (dominated by $B_\phi$ due to shear) must flip across the disk. However, strong Ohmic resistivity near the midplane demands the field to stay straight, forcing the flip to occur in a thin surface layer dominated by ambipolar diffusion, which carries strong current. From Equation (\ref{eq:windaccflow}), this layer is where most magnetic stress is exerted to drive bulk disk accretion. The accretion flow velocity can approach sonic speed due to the low density at the disk surface \citep{BaiStone13b}. However, shearing-box model has a tendency to develop the ``wrong" symmetry solution, where the horizontal field does not flip, leading to zero net accretion.

These findings are corroborated by semi-global (radially global, vertically truncated) simulations of \citet{Gressel_etal15}. While the mass loss rate remains uncertain, they firmly established that magnetically-driven disk winds with the ``correct" symmetry are launched. They highlighted that the initial configuration is in fact unstable to the MRI in the ambipolar-diffusion-dominated upper layer, but it does not saturate into turbulence. Instead, magnetically-driven wind can serve as an alternative channel for MRI saturation in the low-$\beta_z$ regime  \citep{Lesur_etal13}.

The Hall effect substantially alters the magnetic field structure within the disk, depending on the polarity of the net vertical magnetic field, as first explored by local simulations of \citet{Lesur_etal14} and \citet{Bai14}. As the Hall effect diminishes towards the disk surface, wind properties remain similar to the Hall-free case discussed earlier. When net vertical field is aligned with disk rotation axis (i.e., aligned case), the Hall-shear instability strongly amplifies horizontal magnetic field in the disk interior, reaching up to $100$ times the net vertical field. This leads to substantially enhanced radial Maxwell stress $-B_RB_\phi$, making radial angular momentum transport comparable to wind transport.
Nevertheless, strong field amplification in shearing-box simulations always leads to ``wrong" wind symmetry. For anti-aligned vertical field, the opposite happens: horizontal field is reduced towards zero, leading to nearly pure vertical field configuration in the disk interior. In shearing box simulations, this was found to be unstable and tend to yield oscillations in wind directions \citep{Bai15} or bursty behaviors \citep{Simon_etal15}.

\begin{figure*}[h]
\centering
\includegraphics[width=\textwidth]{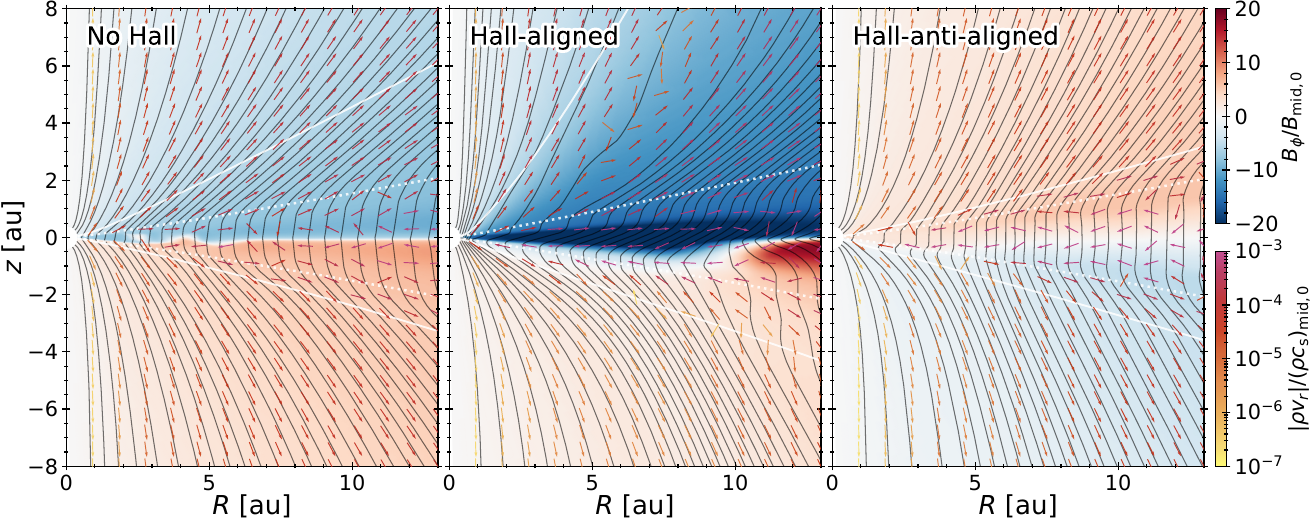}
\caption{Magnetic field configuration (color for $B_\phi$ and solid line contours for $B_p$) and gas flow structures (colored arrows) in inner disk simulations. The simulations incorporate all non-ideal MHD effect and approximated radiation transport, except for the left panel where the Hall effect is excluded. Middle and right panels correspond to simulations with aligned and anti-aligned field polarity, which exhibit strong field amplification and reduction relative to the Hall-free case, respectively. White dotted and solid lines represent irradiation and FUV fronts. Note the significant asymmetry in the aligned case. Figure provided by S. Mori, adapted from \citet{Mori_etal25} (CC BY 4.0).\label{fig:diskinner}}
\end{figure*}

Recent works incorporated all three non-ideal MHD effects in fully global simulations. Under aligned poloidal field geometry, \citet{Bethune_etal17} found diverse behaviors with the operation of HSI, including both accreting and non-accreting regions with the latter possessing the ``wrong" field symmetry, and highly turbulent, one-sided winds.
In contrast, \citet{Bai17} obtained a unified set of results\footnote{The discrepancy is likely due to two reasons. First, the simulation domain in \citet{Bethune_etal17} only extends to $\sim\pm60^\circ$ above/below the midplane, leading to artificial truncation of the wind flow. Second, the Hall shear instability makes system evolution dependent on initial condition, and \citet{Bai17} imposes the Hall effect gradually in an inside-out manner to mimic disk formation.} (see Figure \ref{fig:diskinner}). For the aligned case, the horizontal field strength maximizes at the midplane thanks to field amplification by the Hall shear instability, leading to the emergence of accretion/decretion flow just above/below the midplane according to Equation (\ref{eq:windaccflow}). The net accretion flow, on the other hand, is still carried by the one-sided strong current layer in the disk surface. Toward outer radii, the strong current layer gradually moves to the midplane as resistivity weakens. For the anti-aligned case, the reduction of horizontal field makes the field configuration less stable, and can also lead to the formation of a one-sided strong current layer at the disk surface. In both aligned and anti-aligned cases, the wind is largely laminar and magnetothermal in nature with strong mass loss ($\lambda<2$), but wind kinematics and mass loss rate can still be sensitive to the prescriptions of magnetic diffusivities and thermodynamics near the wind launching region (Section \ref{sssec:innertherm}).

One common feature revealed by global simulations is that the disk interior flow/field structure, and the outflows themselves, can be highly asymmetric between the two disk hemispheres (Figure \ref{fig:diskinner}). The level of asymmetry again can depend on the disk diffusivity profiles especially at the wind base, but generally tends to become less prominent when the disk is more strongly magnetized.

\subsubsection{Role of thermodynamics}\label{sssec:innertherm}

The aforementioned studies established the dominance of magnetized wind in driving disk evolution, though they typically adopt simplified treatment of thermodynamics with prescribed temperatures (locally isothermal or $\beta$-cooling).
Recent studies have begun incorporating more realistic radiation transport in the disk (Section \ref{sssec:diskTemp}), and/or more comprehensive treatment of heating/cooling processes in the wind zone (Section \ref{sssec:thermatm}), though the results are less matured.

The fact that the inner disk interior is largely magnetically inactive leaves room for hydrodynamic instabilities to develop (Section \ref{ssec:hydro}). \citet{Malygin_etal17} and \citet{PfeilKlahr19} mapped instability zones for given disk models based on local growth rates and thermal relaxation times (Section \ref{sssec:trelax}). While the details highly depend on the disk model and dust properties, it is expected that the ZVI likely operates in the inner $\sim1$AU, the COS and its nonlinear counterpart SBI could operate in the $\lesssim1$AU - $\gtrsim10$ AU region, provided there is favorable radial entropy gradient, while the outer radii $\gtrsim10$AU is more prone to the VSI. However, the close relationship between COS and VSI highlighted by \citet{Klahr26} necessitates updates to these mappings. It is likely that the VSI and COS co-exist in the disk inner region, although the level of turbulence and the associated $\alpha_S$ remain to be quantified.

Under inner disk conditions, it is yet to demonstrate how such hydrodynamic instabilities interplay with wind-driven accretion. 
Recently, \citet{Mori_etal25} extended the simulations of \citet{Bai17} with all three non-ideal MHD effects and approximated radiation transport for irradiation and XUV heating (Figure \ref{fig:diskinner}). General results are consistent with \citet{Bai17}. While they did not identify hydrodynamic instabilities in the disk interior (which can be subject to COS/VSI but would require much higher resolution), they found that surface layer accretion can become episodic. This is conceptually similar to the irradiation instability \cite[][though refuted by radiation hydrodynamic simulations by \citealp{MelonFuksmanKlahr22}]{Ueda_etal21,WuLithwick21}, but results from the interplay between surface layer accretion and irradiation heating. They also show that Joule heating is insignificant in the disk interior, whereas irradiation heating can be enhanced by the wind due to elevated irradiation front that intercepts more stellar luminosity.

Incorporating more realistic heating/cooling processes in the disk atmosphere is essential to properly quantify wind kinematics, though with major complications. Only a few studies have implemented different combinations of thermal and non-ideal MHD prescriptions, yielding results that are not easily comparable.

\citet{Wang_etal19} combined the non-ideal MHD (with Ohmic and ambipolar diffusion) simulation setup similar to \citet{Bai17} and time-dependent thermochemistry with four-band (from FUV to X-ray) ray tracing similar to \citet{WangGoodman17} in the disk atmosphere. They showed that wind launching is primarily driven by magnetic stresses. While radiative heating is significant near the wind base, Joule (ambipolar) heating dominates the wind region with $Am\sim10-100$. The mass loss rate, though smaller than in \citet{Bai17} by a factor of a few, remains significant with $\lambda\sim2$, and is much higher than pure photoevaporative mass loss rate given similar parameters. They also showed that EUV heating produces a fast, ionized wind component at high latitudes, but reduces the overall mass loss rate.

\citet{Gressel_etal20} and \citet{Sarafidou_etal24} separately incorporated PDR (FUV) physics and X-ray photoevaporation physics \citep{Picogna_etal19}, though thermal and ionization physics are decoupled. \citet{Gressel_etal20} generally found rather high location of the wind base ($\sim6.5H$), leading to a wind that is in the centrifugal regime. Mass loss rate is only a fraction of $\dot{M}_{\rm acc}$, and increases with FUV luminosity. They also reported that Joule heating is subdominant compared to thermochemical heating. \citet{Sarafidou_etal24} included the Hall effect, although it was found to have little direct impact on wind properties. Their results are consistent with \citet{Rodenkirch_etal20}, showing a transition from pure photoevaporative to magnetically-enhanced mass loss rates.

\subsection{The Outer Disk Region}\label{ssec:outer}

The outer disk physics is simpler as ambipolar diffusion is the main dominant non-ideal MHD effect. Its modest strength ($Am$ of order unity) allows the MRI to operate, and potentially interact with disk winds and hydrodynamic instabilities, most notably the VSI. Their interplay makes thermodynamics another important player in this region. Furthermore, in massive disks, GI develops more readily in the outer disk, which contains the bulk of disk mass, especially during the early Class 0/I stages.

\subsubsection[]{Interplay between wind and disk instabilities}\label{sssec:outerwind}

The outer disk was once thought to be fully MRI turbulent. By incorporating ambipolar diffusion (with $Am\sim1$) with an ideal MHD FUV layer, \citet{Simon_etal13b} showed using shearing-box simulations that the MRI turbulence is damped in the bulk disk except for the upper FUV layer. With net vertical magnetic field, the disk also naturally launches outflows, and the MRI turbulence is enhanced with increasing net vertical field.
The coexistence of the weak MRI turbulence and magnetized wind was firmly established by global simulations of \citet{CuiBai21}, showing that angular momentum transport is dominated by MHD winds, together with significant contribution from the MRI turbulence (with $\alpha_S\sim10^{-3}$ for $Am\sim1$ in the bulk disk and midplane $\beta_z\sim10^4$). In contrast, in ideal MHD, a highly turbulent disk tends to develop fast ``surface accretion". This is owing to the inward dragging of surface poloidal field by the accretion flow, which in turn enhances the magnetic stress $T_{r\phi}$ to further drive accretion. The wind is launched from much higher location and becomes ineffective in driving accretion \citep{ZhuStone18,Mishra_etal20,JacqueminIde_etal21}.

\begin{figure*}[h]
\centering
\begin{minipage}[l]{0.495\textwidth}
\includegraphics[width=0.99\textwidth]{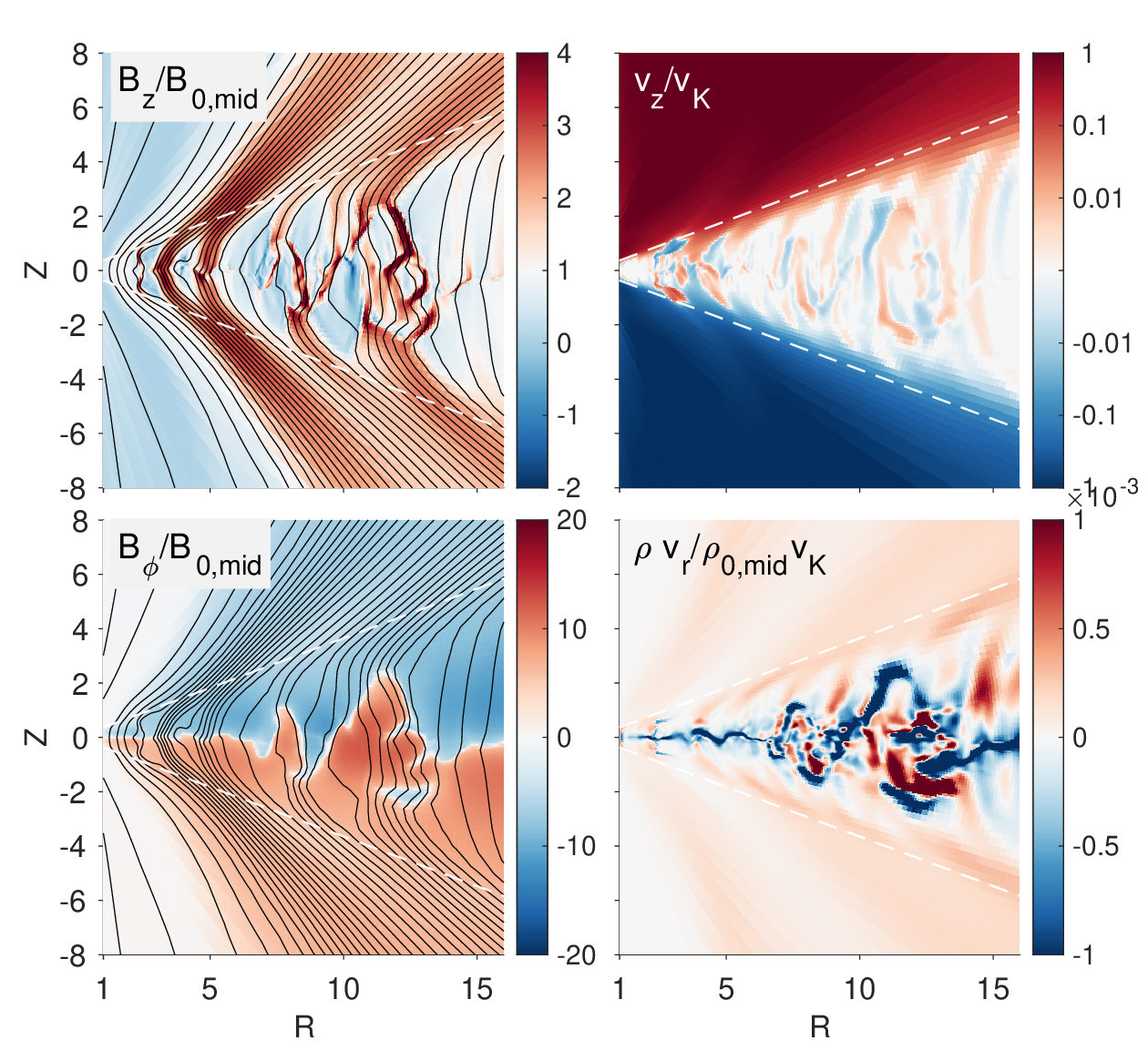}
\end{minipage}
\begin{minipage}[l]{0.261\textwidth}
\includegraphics[width=0.99\textwidth]{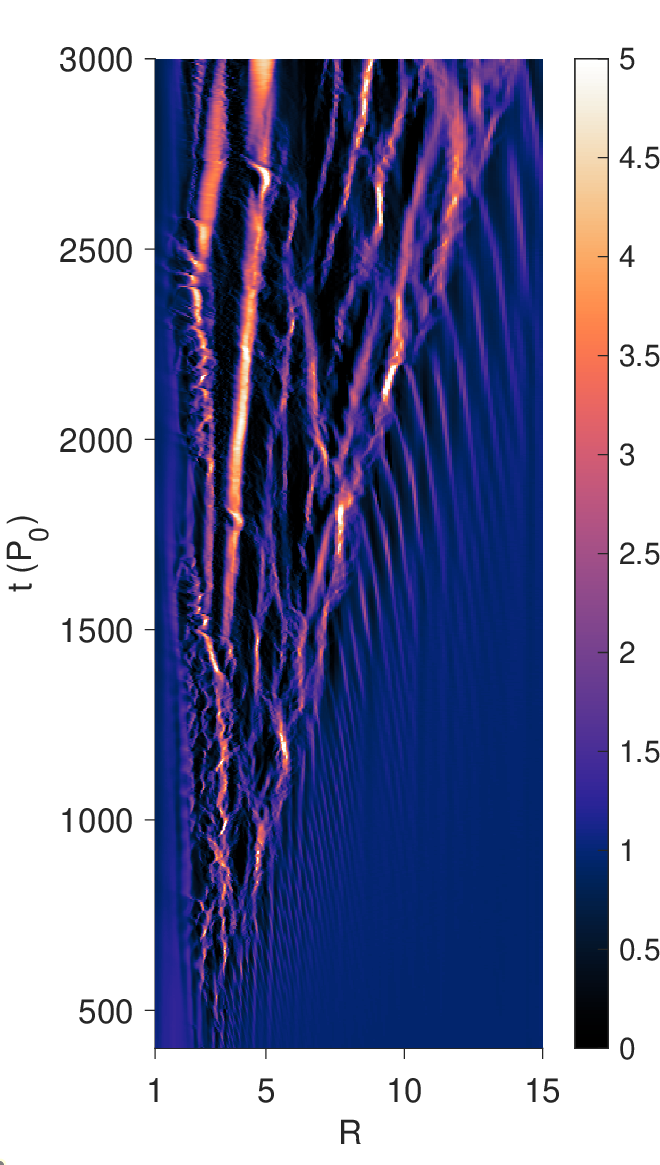}
\end{minipage}
\begin{minipage}[l]{0.23\textwidth}
\includegraphics[width=0.99\textwidth]{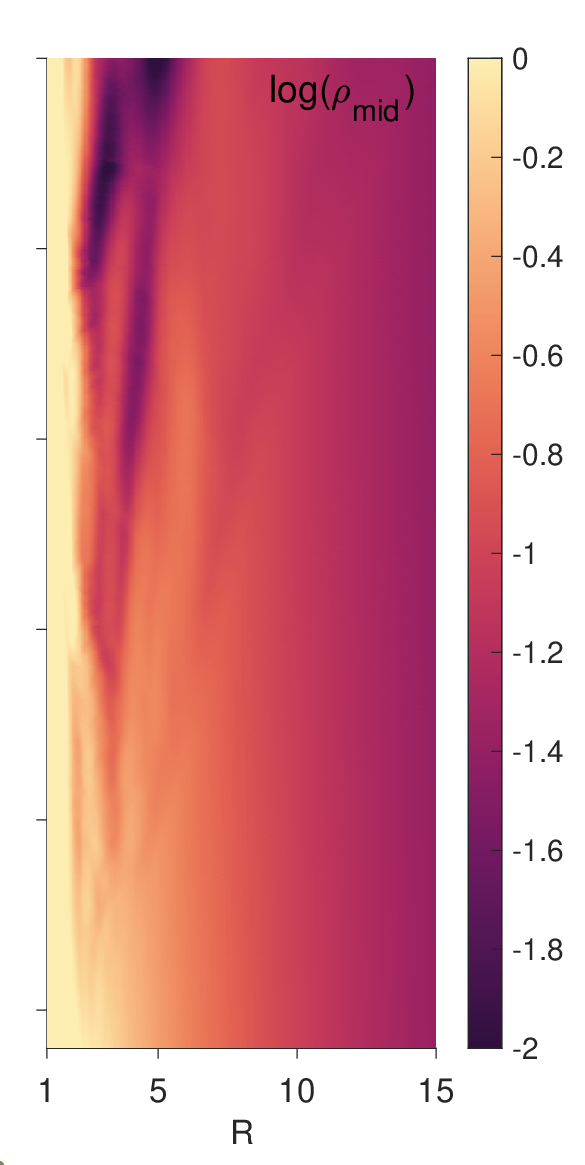}
\end{minipage}
\caption{A 3D simulation of gas dynamics in the outer protoplanetary disk. The four panels on the left show azimuthally- averaged vertical magnetic field, toroidal magnetic field, vertical velocity and radial mass flux, and the right two panel shows the time evolution of the radial profiles of mean vertical magnetic field and density in the midplane up to 3000 orbits at $R=1$. The disk is weakly MRI-turbulent with ambipolar diffusion, and in the meantime is subject to the VSI (with signatures of corrugation mode identifiable in vertical velocity structure). Accretion is primarily driven by magnetized disk winds, with net accretion flow mainly residing in regions where toroidal field flips. Magnetic flux concentration develops stochastically and leads to ring/gap formation, and the flux sheet can migrate, merge and split over time. Figure adapted from \citet{CuiBai22} (CC BY 4.0) with data provided by C. Cui.
}\label{fig:diskouter}
\end{figure*}

The presence of magnetic fields raises questions on whether hydrodynamic instabilities can survive. Early studies showed that the SBI can no longer survive in ideal MHD \citep{LyraKlahr11}. For the VSI, linear analysis in the ideal MHD regime shows the VSI can be suppressed by even mild magnetic tension ($\beta\lesssim(R/H)^2$), or give way to the development of the MRI \citep{LatterPapaloizou18}, with the surface mode being more susceptible to suppression \citep{CuiLin21}. Non-ideal MHD effects help the VSI survive by suppressing magnetic tension and damping the MRI, though the details depend on diffusivities and field configuration \citep{CuiLin21,LatterKunz22}. Non-ideal MHD simulations by \citet{CuiBai20} and \citet{CuiBai22} showed that the VSI can coexist with both magnetized disk winds and the MRI, where in the latter case, VSI body modes can be identified in the weakly turbulent MRI background for $Am\lesssim1$ (see Figure \ref{fig:diskouter}), but get overwhelmed by the MRI turbulence for $Am\gtrsim10$.

Another major line of research for hydrodynamic instabilities is to incorporate more realistic thermodynamics. Most numerical studies of the VSI assume a locally-isothermal equation of state, i.e., instant cooling, which is most favorable for the VSI to operate. More recent simulations use more realistic estimates of the cooling time \citep{PfeilKlahr21,Fukuhara_etal23}, or directly incorporate radiation transport \citep{StollKley14,Flock_etal17b,MelonFuksman_etal24a,MelonFuksman_etal24b,Zhang_etal24}. The detailed results are sensitive to the vertical profile and size distribution of dust \cite[e.g.][]{Fukuhara_etal21}, but generally fall into two categories characterized by the vertical profile of $\beta_{\rm cool}$ \citep{Fukuhara_etal23,Zhang_etal24}. First, $\beta_{\rm cool}<\beta_{c,{\rm VSI}}$ over the bulk disk. This usually occurs when the disk is highly optically thin (e.g., cooling dominated by gas-dust collision), and the behavior is similar to locally isothermal simulations, with highly anisotropic turbulence dominated by the corrugation mode. Second, the $\beta_{\rm cool}$ profile has a bump in the midplane region that exceeds $\beta_{c,{\rm VSI}}$, which can be due to cooler midplane temperature and/or the disk being optically thick. The result is that the VSI operates only in the surface layers, driving surface layer accretion.
These subtleties highlight the need for simulations with self-consistent dust dynamics and opacity \citep{Fukuhara_etal25}.

\subsubsection[]{Magnetic flux concentration and substructure formation}\label{sssec:fluxcon}

Magnetic flux concentration and zonal flows were first identified in local MRI simulations (Section \ref{sssec:zonal}). Under outer disk conditions dominated by ambipolar diffusion, local stratified simulations also reported strong flux concentration within a thickness of $\sim H$ \citep{Bai15}, though the results could be affected by limitations of the shearing-box.

Subsequent global non-ideal MHD simulations of protoplanetary disk winds have consistently demonstrated robust magnetic flux concentration into thin, quasi-axisymmetric flux sheets that create gas gaps \citep{Bethune_etal17,Suriano_etal18,Suriano_etal19,Riols_etal20,CuiBai21,CuiBai22,Hsu_etal24}. These simulations share similar setups: smooth initial conditions and smooth diffusivity profiles characterized by ambipolar-diffusion-dominated midplane region transitioning to near-ideal MHD conditions in the disk surface (Figure \ref{fig:diskouter}). Flux sheets typically form with thickness $\sim H$ at random locations separated by a few $H$. \citet{Hu_etal19,Hu_etal21} further found that considering dust sintering near snow lines \citep{Okuzumi_etal16}, the resulting non-smooth magnetic diffusivities can make the snow line regions preferred sites for flux concentration and ring/gap formation.

Flux concentration naturally forms multiple gas rings and gaps. When angular momentum transport is dominated by the MRI turbulence, higher $\alpha_S$ in flux-concentrated zones leads to lower surface density in steady state (Section \ref{sssec:zonal}). When magnetic wind dominates transport, flux concentration similarly depletes density but for a different reason: it enhances both angular momentum extraction and mass loss locally. This effect can form relatively deep gaps, but not indefinitely: when flux sheet locations stay stationary, meridional flows tend to develop, which significantly impacts dust dynamics \citep{Hu_etal22}. In 3D, \citet{Hsu_etal24} found that the ring-gap contrasts can be sufficiently large to trigger the RWI, making the rings non-axisymmetric. On the other hand, when the MRI turbulence is resolved, \citet{CuiBai21} showed that flux sheets can migrate, merge, and split on secular timescales, and the disk does not form very deep gaps (Figure \ref{fig:diskouter}). This is arguably consistent with super-resolution imaging of ALMA disks showing many additional shallow ring/gap substructures \citep{Jennings_etal22}.

Currently, a solid theoretical understanding of magnetic flux concentration is lacking. Different mechanisms have been proposed, though none is necessarily rigorous and universal.
In the case of the MRI, \citet{BaiStone14} speculated that recurrent channel flows followed by magnetic reconnection in MRI turbulence keeps pumping gas away from flux-rich regions. In the wind-dominated case, \citet{Suriano_etal18} found that the midplane accretion flow (where $B_\phi$ reverses) advects poloidal field inward to form a pinched configuration subject to reconnection. This similarly leads to segregation between gas and magnetic flux. \citet{RiolsLesur19} found a secular linear instability to concentrate magnetic flux from the joint action of wind-driven accretion, mass loss, turbulent diffusion and magnetic flux advection.
Overall, this phenomenon is robust in wind-dominated disks with ambipolar diffusion, but a comprehensive theoretical framework requires further development.

\subsubsection{Disk truncation}\label{sssec:truncation}

Transition disks are a special class of protoplanetary disks featuring large inner cavities. They are conventionally considered as in the late phase of disk evolution with inside-out disk clearing resulting from effective viscous evolution and photoevaporation \cite[see review by][]{ErcolanoPascucci17}. However, many transition disks are actively accreting with accretion rate as high as full disks \citep{Manara_etal14}, challenging conventional understandings.

The highly depleted inner cavity brings the gas into the ambipolar-diffusion-dominated regime, sharing similar physics to outer disks. \citet{WangGoodman17a} calculated the ionization structure in transition disk cavities, showing that $Am$ in the cavity is of order unity for typical disk parameters. They suggest that wind-driven accretion can sustain observed rates with accretion velocity reaching sonic speed,  akin to the ``jet-emitting disk" model of \citet{CombetFerreira08}. This scenario was investigated numerically by \citet{MartelLesur22}, where under a prescribed $Am$ profile of order unity, they confirmed that a steady state wind can be sustained: the system smoothly transitions to a highly magnetized inner cavity characterized by wind-driven accretion with sonic accretion speed, sub-Keplerian rotation and long lever-arm. Recent work with more realistic treatment of ionization and thermodynamics confirms these findings \citep{Sarafidou_etal25}. On the other hand, it is still yet to address how transition disks form at first place.

Disk outer truncation is conventionally attributed to initial conditions at disk formation, followed by viscous spreading (Section \ref{sec:diskevol}) and external photoevaporation 
\ifbool{supplement}
{(Supplemental Text Section 1.2)}
{(Section \ref{sssec:extphoto})}. 
Under the wind-driven accretion framework, however, most studies assume the disk to be infinitely extended (e.g., self-similar). \citet{YangBai21} carried out 2D non-ideal MHD simulations of magnetically-driven disk wind with outer disk truncation, using a prescribed profile of $Am$. They found that near and beyond the truncation region, poloidal magnetic fields collapse toward the midplane, and reconnect to form closed field loops. Most interestingly, as vertical field changes sign beyond loop center, the direction of wind-driven flow reverses (Equation \ref{eq:windaccflow}), leading to decretion flow with considerable mass loss. This is phenomenologically analogous to, but physically distinct from viscous spreading and external photoevaporative mass loss. Nevertheless, 
longer-term simulations with more realistic ionization and thermodynamics are needed to definitively understand outer disk evolution.

\subsubsection{Gravitational instability in massive disks} 

The disk outer region generally contains most of the disk mass, and can be prone to the GI. 
This is particularly the case toward younger Class 0/I disks, which are known to be more massive \citep{Tychoniec_etal20},
while some Class II disks also show GI signatures \cite[e.g.][]{Perez_etal16,Speedie_etal24,Yoshida_etal25NA}. Under more realistic conditions, the outcome of GI is subject to additional complications through its interactions with magnetic fields and radiation.

Incorporating magnetic field, one might expect to observe the interplay between GI and MRI turbulence.
However, recent local studies revealed that the GI spiral density waves (especially vertical motion associated with the spiral wake) drive a mean-field dynamo \citep{RiolsLatter18,RiolsLatter19} which is distinct from the MRI turbulence (unless for very long cooling time $\beta_{\rm cool}\gtrsim100$). Global simulations confirm the presence of the dynamo, highlighting its large-scale nature \citep{Deng_etal20,BethuneLatter22}. This dynamo has been studied mostly under the zero-net-vertical magnetic flux configuration in both ideal and resistive MHD conditions. In the optimal case (magnetic Reynolds number $Rm\equiv c_sH/\eta_O$ around $5-20$), magnetic field amplification saturates at near equipartition field strength, where the large-scale toroidal field dominates over small-scale turbulence. It leads to very efficient radial angular momentum transport with $\alpha_S\gtrsim0.1$, much stronger than the MRI turbulence itself under the same field configuration. 

Current studies of the GI dynamo remain idealized compared to the expected physical conditions in the outer disk. Toward reality, \citet{Riols_etal21} incorporated ambipolar diffusion in local simulations of the GI dynamo under zero-net-vertical magnetic flux, showing that it operates in a similar manner and can yield equipartition field for $Am\sim30-100$. The resulting field amplification decreases for smaller $Am$, but remains stronger than what can be achieved by the MRI itself. It remains to examine what happens in the presence of net vertical magnetic flux \cite[e.g.][]{Tsung_etal25}, eventually in a global simulation.

In addition, while GI has been primarily studied under the $\beta$-cooling approximation for simplicity, \citet{Xu_etal25a,Xu_etal25b} recently carried out hydrodynamic simulations of the GI including full radiation transport. Focusing on the gravito-turbulence regime (modest-to-long cooling time), the disk dynamics can be captured using an effective cooling time with first-order corrections.
More realistic radiation transport primarily reduces the amplitude of temperature perturbations, especially in the optically thick regime. On the other hand, radiation transport remains crucial in tracking the formation and evolution of clumps in the fragmentation regime \citep{Ni_etal25}, which is beyond the scope of this review.

\subsection{The Innermost Disk Region}\label{ssec:innermost}

\begin{figure*}[h]
\centering
\includegraphics[width=0.99\textwidth]{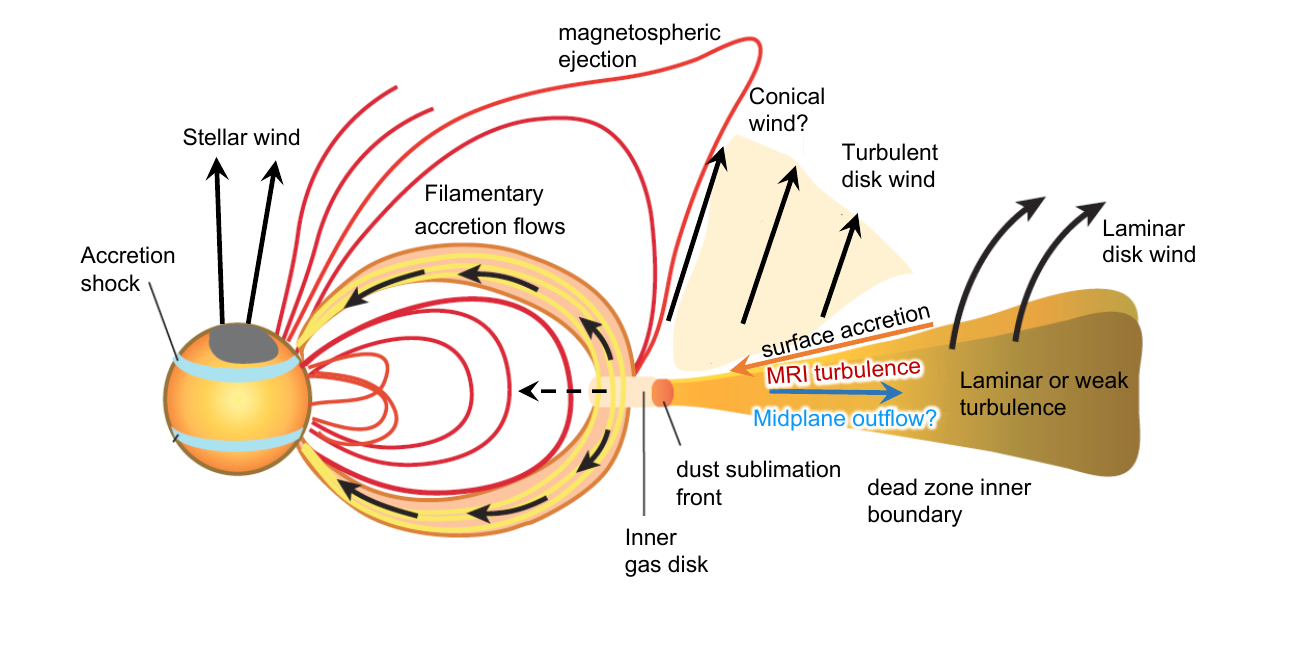}
\caption{Schematic view of the disk innermost region. Inside the dead zone inner boundary, the disk temperature reaches $\gtrsim10^3$K to become fully MRI turbulent, and likely undergoes surface accretion potentially accompanied by midplane outflow. Slightly inward, the silicate dust sublimates at the disk inner rim. Unless for very high accretion rate, the disk is truncated by the stellar magnetosphere, leading to magnetospheric accretion. For slow rotators, the disk-magnetosphere interface is subject to the magnetic interchange instability, causing the formation of penetrating accreting filaments. A good fraction (depending on stellar spin) of the accreting material can be ejected in the form of stellar wind and (episodic) magnetospheric ejection, which likely make up the high-velocity jet. It is surrounded by lower-velocity wind, launched from the turbulent innermost disk, surrounded by more laminar wind further out. Figure adapted from \citet{Hartmann_etal16}, with permission from Annual Reviews. [{\it See published version for a more polished figure.}]
}\label{fig:diskinnermost}
\end{figure*}

The innermost disk region presents the greatest complexity with substantial scale separations, making comprehensive studies incorporating all necessary physics currently unfeasible. We outline major physical processes in this region in Figure \ref{fig:diskinnermost}. Over recent years, research in this area has revived thanks to advances in numerical simulations, typically focusing on one problem at a time.

\subsubsection{The dead zone inner boundary and dust inner rim}\label{sssec:DZIB}

The dead zone inner boundary is characterized by an abrupt increase in $\alpha_S$ due to thermal ionization activating the MRI. Under the conventional scenario of pure MRI-driven accretion considering Ohmic resistivity, this transition would lead to an abrupt change in surface density, which is further subject to the Rossby wave instability to generate vortices \citep{LyraMacLow12,Faure_etal15,Flock_etal17}. The coupling between effective viscous heating and thermal ionization makes the location of this boundary bi-stable, which can migrate inward/outward depending on the physical conditions \citep{LatterBalbus12}. Moreover, density waves launched from this boundary could lead to heating in the dead zone, pushing this boundary outward \citep{Faure_etal14}.

The physical picture of this region gets more much more complex
when considering additional non-ideal MHD effects with wind-driven accretion in the dead zone.
Recently, \citet{Iwasaki_etal24} and \citet{Roberts_etal26} carried out 3D simulations across the dead zone inner boundary with net poloidal field incorporating both Ohmic resistivity and ambipolar diffusion, and using prescribed disk temperatures to mitigate further complications. 
Both works found a redistribution of magnetic flux that leads to abrupt surface density jump across the dead zone inner boundary. However, the dynamics are highly complex.
\citet{Iwasaki_etal24} identified a ``transition zone" devoid of magnetic flux (and hence magnetic wind), whereas in \citet{Roberts_etal26}, with minor differences in simulation setup such as the diffusivity profiles, it is the MRI active zone where magnetic flux gradually depletes. This further leads to major differences in the innermost disk structure and evolution.
The reason behind the differences is not yet clear, but these works are starting to unveil the physical richness of this region.

With more realistic thermodynamics, the disk in this region becomes thermally unstable: the system should relax to either a hot branch where the disk is thermally ionized, MRI active, with effective viscous heating sustaining thermal ionization, or a cold branch where the disk is MRI inactive and cannot sustain thermal ionization.
This effect is known in the context of FU Orionis outbursts (e.g., \citealp{Zhu_etal10a}), and is different from the classic thermal instability resulting from opacity transitions at higher temperatures
\cite[e.g.][]{Hirose15}. Under the effective viscously-driven accretion framework, global radiation hydrodynamic simulations of \citet{CecilFlock24} demonstrated that thermal instability in this region leads to cyclic transitions between active and quiescent phases. Recently, \citet{Wang_etal26} conducted local shearing-box simulations that self-consistently incorporated radiation transport, and Ohmic and ambipolar diffusion with thermal and non-thermal ionization. They found that MRI-driven accretion dominates over wind-driven accretion in the hot branch with $\alpha_S\gtrsim0.03$,
whereas the cold branch resembles the inner disk described in Section \ref{ssec:inner}.
They also mapped out an ``S-curve",\begin{marginnote}
\entry{Thermal instability}{The system can undergo runaway heating or cooling when thermally perturbed.}
\entry{``S-curve"}{Equilibrium curve of temperature versus surface density with effective viscous heating balancing radiative cooling, which is S-shaped when thermally unstable.}
\end{marginnote}showing that reducing surface density can bring the system from the hot to cold branch, but increasing surface density does not raise disk temperature due to inefficient Joule heating (which concentrates in the disk surface). It is thus unclear whether the dead zone inner boundary should exhibit the classic limit-cycle behavior.

Finally, the characteristic temperature for thermal ionization ($\sim10^3$K) is close to the sublimation temperature of silicate dust ($\sim1500$K), linking the dead zone inner boundary and the dust inner rim (Section \ref{sssec:diskTemp}). Viscous radiation hydrodynamic simulations of \citet{Flock_etal16} incorporated a prescription\footnote{The actual sublimation process can be highly complex, yielding a broad sublimation front \citep{XuWang_etal26}.} for dust sublimation coupled with dust opacity and a transition in $\alpha_S$, focusing on Herbig AeBe disks. They showed that the sublimation front is triangular-shaped extending from the inner midplane up to several tenths of an AU in the surface (approaching the dead zone inner boundary), which significantly enhances irradiation heating in this region. Subsequent radiation resistive-MHD simulations of \citet{Flock_etal17} (with MRI turbulence but no wind) generally confirmed these findings, with the disk puffing-up further due to a magnetically supported disk surface. Additionally, dust can be lifted in a magnetized disk wind, especially in the inner disk \citep{RodenkirchDullemond22}, which could similarly enhance the interception of stellar light and heating \citep{BansKonigl12,Mori_etal25}. Clearly, more self-consistent calculations are needed to decipher the complex dynamics in this region.

\subsubsection{Magnetospheric accretion and jet launching}\label{sssec:magnetosphere}

The final journey of the accretion flow is the process it channels toward the proto- or pre-main-sequence star. For typical accretion rate of $\sim10^{-8}M_\odot$ yr$^{-1}$, the disk is truncated by the stellar magnetosphere, funneling the gas through magnetospheric accretion. 
The stellar magnetic field is often treated as a dipole with $B=\mu_*/R^3$, where $\mu_*=B_*R_*^3$ is the magnetic dipole moment.
The truncation radius, a.k.a. magnetospheric radius $R_m$, 
can be estimated by balancing ram pressure $\rho v_K^2$ and stellar magnetic pressure $B^2/8\pi$ under quasi-spherical accretion \citep{Lamb_etal73},
or through torque balance \citep{DAngeloSpruit10,Takasao_etal22}. The result is
\bgeq
R_m\approx\bigg[\frac{\mu_m^4}{2GM_*\dot{M}_{\rm acc}^2}\bigg]^{1/7}
\approx6.5\bigg(\frac{B_*}{1{\rm kG}}\bigg)^{4/7}\bigg(\frac{R_*}{2R_\odot}\bigg)^{5/7}\bigg(\frac{M_*}{0.5M_\odot}\bigg)^{-1/7}\bigg(\frac{\dot{M}_{\rm acc}}{10^{-8}M_\odot {\rm yr}^{-1}}\bigg)^{-2/7}R_*\ .\label{eq:Rm}
\edeq
Moreover, stellar rotation introduces the corotation radius
\bgeq
R_{\rm co}\equiv(GM/\Omega_*^2)^{1/3}, \label{eq:Rco}
\edeq
where disk Keplerian frequency matches stellar rotation frequency $\Omega_*$, and the fastness parameter $\omega_s\equiv\Omega_*/\Omega_K(R_m)$. Stellar rotation is considered fast when $\omega_s\gtrsim1$, entering the ``propeller" regime.

Magnetospheric accretion is a major topic in astrophysics, and has recently been reviewed by \citet{RomanovaOwocki15} and \citet{Hartmann_etal16}. 
The key questions include, first, how does the disk gas, which is in ideal MHD conditions attached to disk magnetic field, get loaded onto the stellar magnetosphere and accrete onto the central star? Second, what drives mass ejection (jets and winds)? The ejection process extracts angular momentum from the accretion flow, which sets the evolution of stellar spin.
Addressing these questions requires advanced simulations, which is challenging mainly for two reasons. First, the magnetosphere is so strongly magnetically-dominated and becomes nearly force-free, which is difficult for MHD simulations (very low-$\beta$, high Alfv\'en speed). Second, the disk-magnetosphere connection is key to understanding these questions, which requires high resolution to properly resolve the MRI turbulence in the innermost disk. In addition, the process involves significant dissipation of kinetic and magnetic energy, necessitating further effort to incorporate thermodynamics and potentially plasma kinetic physics.

Here, we highlight recent progress in theory/simulations without offering a comprehensive review on this topic. In particular, recent high-resolution 3D simulations of magnetospheric accretion for aligned rotators can properly resolve the MRI turbulence in the disk \citep{Takasao_etal22,Takasao_etal25a,Zhu_etal24,Zhu25,Tu_etal26a}. While a wide range of parameters remain to be explored, these works have revealed major new physical insights. 

For slow rotators, these simulations reveal that the disk develops filaments at the truncation radius due to the magnetic interchange instability, which is a Rayleigh-Taylor-type instability in magnetically-supported plasma \cite[e.g.][]{Spruit_etal95,Romanova_etal08}. Such filaments can penetrate deep into the magnetosphere, forming multiple accretion columns. Depending on the persistence of individual filaments, such accretion can be further categorized into  ``ordered unstable" and ``chaotic unstable" regimes as spin increases \citep{Blinova_etal16}, which exhibit different observational signatures in time variability. Outside $R_m$, the gas is highly dynamic in the ordered unstable regime (where $\omega_s\lesssim0.5$), forming ``magnetic bubbles" \citep{Zhu_etal24} resembling the ``magnetically-arrested disk" (MAD, e.g., \citealp{Tchekhovskoy_etal11}). As stellar spin increases, the region near $R_m$ becomes more rotationally supported, and the interchange instability is suppressed for $\omega_s$ reaching order unity (stable regime). The accretion flow gets loaded to the stellar magnetosphere simply by turbulent dissipation \citep{Takasao_etal25a}, leaving a clean magnetospheric cavity. 
\citet{Zhu_etal24} and \citet{Zhu25} found that accretion proceeds primarily from the disk surface layer, similar to the surface accretion scenario outlined in Section \ref{sssec:outerwind}. The midplane gas exhibits outflowing motion as poloidal field bends inward rather than outward, and this midplane outflow becomes stronger for faster rotators. They also found that $R_m$ becomes smaller than predicted by (\ref{eq:Rm}) but gets close to $R_{\rm co}$ for fast rotators. On the other hand, the parameters adopted in \citet{Takasao_etal22} lead to a smaller $R_m$, and they found that the magnetic field tends to restructure itself so that accretion becomes asymmetric, proceeding primarily through one side of the hemisphere. In this case, the counterpart of the midplane outflow likely becomes a ``failed magnetospheric wind" \citep{Takasao_etal25a}.

Outflows from magnetospheric accretion manifest in three forms:

--Stellar wind, a thermal wind from open field line regions near the magnetic poles. 
If receiving sufficient power from accretion energy, \citet{MattPudritz05} suggested that
it can extract stellar spin to offset the spin-up torque from accretion, although
\citet{ZanniFerreira11} argued that it is energetically challenging.
In simulations, the stellar wind depends on heating prescription and is highly uncertain
\citep{Takasao_etal22}.

--Magnetospheric ejection, which originates from the interface between the disk and the magnetosphere at high altitude \cite[e.g.][]{ZanniFerreira13}. This ejection process results from a cycle of energy storage and release, and is generally episodic in nature. The detailed processes vary in different simulations \citep{Takasao_etal22,Zhu25,Tu_etal26a}, but generally involve the build up of toroidal magnetic field of opposite signs (field ``inflation", \citealp{Lovelace_etal95}) around the interface due to opposite velocity shear, followed by magnetic reconnection.
We emphasize that this process generally involves magnetic fields that connect both to the star and to the disk material, thus plays a crucial role in extracting angular momentum and energy from the star and/or accretion flow. For instance, \citet{Zhu25} reported this wind carries from $1\%$ to $40\%$ of the accretion rate as the stellar spin increases from zero to reaching the propeller regime, establishing an equilibrium stellar spin with $\omega_s\sim0.7$. Toward larger scale, this ejection process likely makes up the YSO jets. In particular, incorporating net poloidal field in the disk, \citet{Tu_etal26a} demonstrated the formation of bipolar jets in a continuous (despite turbulent) manner with velocity exceeding $200$km/s. We also note that the dynamics and mass loading of magnetospheric ejection can be sensitive to thermodynamics, especially given strong energy dissipation in the truncation region through magnetic reconnection \citep{Takasao_etal22}, which are not yet well constrained in these simulations.

--Disk wind. In the absence of net poloidal field in the disks, the disk near the truncation radius is fully MRI-turbulent and lacks strong coherent poloidal field. The wind launched from the disk is insufficient to escape and falls back at larger distance \citep{Takasao_etal22}. On the other hand, with net poloidal field threading the disk, \citet{Tu_etal25b,Tu_etal26a} found that the disk wind can be enhanced as additional magnetic flux opened up from the stellar magnetosphere.

Overall, the physics at the disk-magnetosphere interface is extremely rich, and a wide range of parameters, such as stellar obliquity, remain to be explored \cite[e.g.][]{Romanova_etal21}. 
Concurrently, more observational constraints are in place toward large YSO samples, including interferometric constraints of gas kinematics \cite[e.g.][]{GRAVITY_YSO23}, combined diagnostics of accretion and stellar properties \cite[e.g.][]{HerczegHillenbrand14,Pittman_etal25a,Pittman_etal25b}, and morphological and kinematic information of YSO jets
\ifbool{supplement}
{(Supplemental Text Section 2.2.2)}
{(Section \ref{sssec:windobs})}.
 Together, these data set the benchmark for future theoretical advances.

\subsubsection{Boundary layer accretion}\label{sssec:BL}

For very high accretion rate ($\dot{M}_{\rm acc}\gtrsim10^{-6}M_\odot$ yr$^{-1}$), the accretion flow crushes the magnetosphere ($R_m\lesssim R_*$). Without truncation, a Keplerian disk extends to the protostellar surface, forming a boundary layer. This layer is expected to emit up to half of the total accretion luminosity \citep{LyndenBellPringle74}.

The central question in boundary layer accretion concerns the mechanism mediating angular momentum transport between the star and the disk. This layer is linearly stable to the MRI because $d\Omega/dR>0$. Earlier idealized hydrodynamic/MHD simulations have identified that the boundary layer is subject to shear-acoustic instabilities \citep{BelyaevRafikov12}, akin to the Papaloizou-Pringle instability \citep{PapaloizouPringle84}. These instabilities lead to the excitation of acoustic waves, which transmit angular momentum both to the star and to the disk in a non-local manner \cite[e.g.][]{Belyaev_etal13,Coleman_etal22}. 
However, \citet{BelyaevQuataert18} found that these waves only carry a small fraction of the angular momentum needed for steady state accretion, leading to accumulation of angular momentum and mass at the boundary layer. This conclusion also holds in the MHD case, where they found that an initially imposed vertical magnetic field simply gets amplified through mass pileup, but does not undergo shear amplification to generate significant Maxwell stress.

Recently, \citet{Takasao_etal25} carried out fully global 3D MHD simulations of boundary layer accretion incorporating a convective protostar and an MRI-active disk threaded by net poloidal magnetic field.
They showed that multiple mechanisms operate in concert that drive efficient angular momentum transport in the boundary layer. One main finding is the discovery of magnetically-excited spiral shocks: stellar convection generates magnetic concentrations in the surface analogous to star spots, which drive large-amplitude spiral shocks upon colliding with rotating disk gas. Such spiral shocks have much higher amplitude than the aforementioned acoustic waves, and play a significant role in driving boundary layer accretion \citep{Rafikov16}. They have also identified major contributions from the Maxwell stress and vertical transport by jets. While there is modest angular momentum accumulation at the boundary layer in steady state, it is effectively diffused by the turbulent stellar convection. Overall, a comprehensive understanding of the boundary layer requires global models that capture the interplay of disk magnetism, vertical transport, and stellar surface dynamics.

\ifbool{supplement}{}{

\section{OBSERVATIONAL CONSTRAINTS}\label{sec:obs}

Multiple aspects of disk gas dynamics and angular momentum transport can be put to observational tests, with the observed stellar accretion rates serving as a baseline.
While subject to systematics and uncertainties, multiple lines of evidence are converging toward a consistent general physical picture as described in this review, which also receives supporting evidence from solar system studies.

\subsection{Measurement of Turbulence}

\subsubsection{Direct measurement}
The most direct constraint on disk turbulence is from kinematic information of gas motion. Treating turbulence as isotropic, small-scale random motion, most studies measure turbulent (non-thermal) broadening of molecular emission lines, typically \ce{CO} and its isotopologues. This technique requires high spectral resolution and moderate spatial resolution to reliably subtract the background rotation profile and thermal broadening, usually by fitting parameterized disk models. As different molecules have different emission surfaces, it allows turbulence level to be constrained at various locations, which is expected to increase with height \citep{Simon_etal18}. Thanks to ALMA, this technique has been applied to a handful of disks, yielding a stringent upper limit of $\delta v/c_s\lesssim10\%$ in HD 163296 \citep{Flaherty_etal17}, TW Hydra \citep{Flaherty_etal18,Teague_etal18b}, V4046 Sgr and MWC 480 \citep{Flaherty_etal20}. Denoting a turbulent $\alpha_t\sim(\delta v/c_s)^2$, these correspond to $\alpha_t\lesssim10^{-3}$ to $10^{-2}$ in disk outer regions ($\gtrsim30$AU). On the other hand, evidence for significant turbulence ($\delta v/c_s\gtrsim0.2$) was reported for two very extended disks DM Tau \citep{Flaherty_etal20} and IM Lup \citep{Flaherty_etal24,PanequeCarreno_etal24}, where the emission mainly probes the disk surface layer. However, the recent exoALMA large program \citep{Teague_etal25} has revealed rich set of kinematic substructures among disks deviating from smooth Keplerian rotation. Caution must be exercised that
unresolved kinematic substructures could be misinterpreted as turbulent line broadening.

Turbulence generated by hydrodynamic and MHD instabilities can also exhibit coherent kinematic features at large scales ($\sim H$), which can be probed by high spatial resolution ALMA observations. For instance, highly anisotropic VSI body modes can produce quasi-axisymmetric, radially alternating kinematic ``rings" with negligible non-thermal line broadening \citep{Barraza-Alfaro_etal21}, while GI spirals exhibit as ``wiggles" in velocity channel maps \citep{Longarini_etal21}.
Velocity perturbations in the MRI turbulence can exhibit spiral-like substructures, but they are local and height-dependent \citep{Barraza-Alfaro_etal25}. Comparison to data from the exoALMA program have revealed no compelling VSI evidence, with some disks being laminar or having spiral substructures, while others show intricate velocity patterns \citep{Barraza-Alfaro_etal25}. Such complexities may not be simply attributed to a single mechanism to generate turbulence, but imply multiple sources of turbulence simultaneously at play 
\ifbool{supplement}
{(see Section 5.2.1)}
{(Section \ref{sssec:outerwind})}.

\subsubsection{Indirect measurement}
There have been a variety of other works that infer the strength of disk turbulence indirectly, often relying on information from dust, especially dust substructures. Here, we focus on a few representative approaches, and recommend \citet{Rosotti23} for a comprehensive review.

In the vertical direction, dust vertical scale height $H_d$ is set by the gas vertical diffusion coefficient $D_z$, which may be taken as a proxy for $\alpha_S$ (not necessarily true), given by \citep{YoudinLithwick07}:
\bgeq
H_d/H\approx\sqrt{\alpha_{z}/{\rm St}}\ ,
\edeq
where we define $\alpha_z$ through $D_z\equiv\alpha_z c_sH$, and ${\rm St}\equiv\Omega t_{\rm stop}$ is the dust Stokes number, with $t_{\rm stop}$ being the dust stopping time. Dust settling is best measured in edge-on disks, and strong settling for mm-sized grains has been observed for a number of cases \cite[e.g.][]{Villenave_etal20,Sturm_etal23}, though such systems are uncommon. For inclined disks with ring-like substructures, dust layer thickness impacts the continuum intensity variations or gap contrast between the projected major and minor axes of the disk image. The method has been pioneered by \citet{Pinte_etal16} applied to HL Tau, with some variants derived/adopted by \citet{DoiKataoka21,Liu_etal22} and \citet{Pizzati_etal23} mainly applied to disks from the DSHARP program \citep{Andrews_etal18}. The results overall indicate low level of turbulence with $\alpha_z\lesssim10^{-3}$, some even less than $10^{-4}$, except for the inner ring in HD 163296. There is also evidence that dust in the Class 0/I phase are not well settled \cite[e.g.][]{Lin_etal21,Sheehan_etal22}, indicating higher level of turbulence at earlier evolutionary stages.

In the radial direction, ring-like dust substructures imply dust concentration in gas pressure bumps. The radial dust ring width $w_d$ is set by the balance between dust drift toward the bump against radial turbulent diffusion. Let $D_r\equiv\alpha_rc_sH$ be the gas radial diffusion coefficient (as a proxy for turbulence), the dust ring width  is approximately given by \citep{Dullemond_etal18}
\bgeq
w_d/w\approx(1+{\rm St}/\alpha_r)^{1/2} ,
\edeq
where $w$ is the width of the gas bump. While measuring dust ring width is relatively straightforward, this method requires high-quality data to constrain dust ${\rm St}$ and gas gap width. This method has been applied to disks in the DSHARP program \citep{Dullemond_etal18} and a few individual systems \cite[e.g.][]{Facchini_etal20} under broad constraints on gas gap width. Incorporation of gas kinematic information significantly improves the estimates \citep{Rosotti_etal20,Yoshida_etal25}, and the results typically show $\alpha_r$ on the order of $10^{-3}$ or higher but with large scatter, and typically $\alpha_r\gtrsim \alpha_z$.

When further assuming 1). dust size is limited by the fragmentation barrier, and 2). collision velocities between dust are dominated by turbulence (generally true in dust rings), then the dust Stokes number is directly connected to turbulent velocity $\delta v$ and fragmentation velocity $v_{\rm frag}$ by \citep{OrmelCuzzi07}
\bgeq
{\rm St}\approx\frac{1}{3}\frac{v_{\rm frag}^2}{\delta v^2}\ .
\edeq
Denoting $\delta v^2\equiv\alpha_t c_s^2$, this provides an additional relation between turbulence level $\alpha_t$ and ${\rm St}$, thus breaks the degeneracy from the aforementioned indirect methods. Applying to HD 163296 and AS 209, \citet{Rosotti_etal20} reported high level of turbulence approaching $\alpha_t\sim10^{-2}$ for $v_{\rm frag}=10$m/s. In contrast, based on dust and gas properties inferred from high-resolution multi-wavelengths data, \citet{Jiang_etal24} found that five out of seven disks (which also include HD 163296 and AS 209) show very low level of turbulence with $\alpha_t\lesssim10^{-4}$ for fragmentation velocity of $\lesssim1$m/s. One exception, IM Lup, has stronger turbulence, which is in line with direct line broadening measurement. Another emerging trend from larger samples of structured disks is that most dust rings in outer disks are settled, with medium $\alpha_t\lesssim10^{-3}$ \citep{Villenave_etal25}, while in multi-ring systems, the inner rings are often more puffed up \citep{Jiang_etal25}.

Finally, dust continuum polarization provides additional constraints on dust properties. For inclined disks, azimuthal variation of dust polarization can constrain dust layer thickness, and this technique has been applied to multi-ring systems of HD 163296 \citep{OhashiKataoka19} and HL Tau \citep{Yang_etal25}, providing further evidence for radial variations in turbulent strength.

Overall, observational constraints converge toward relatively low level of turbulence in the outer region of most protoplanetary disks, in line with theoretical expectations discussed in this review.
However, exceptions and complexities remain, and the source of weak turbulence remains uncertain.
As individual systems show great diversity in disk properties, forward-modeling with customized disk microphysics may represent a promising avenue for fully revealing the nature of turbulence in the outer protoplanetary disks.

\subsection{Kinematics of Disk Winds}\label{ssec:winddiag}

\subsubsection{Overview of main observational diagnostics}
Young stellar objects drive outflows across all stages. These are often classified into ``jets" and ``winds" for fast ($\gtrsim50$km s$^{-1}$) and slow ($\lesssim50$km s$^{-1}$, often $\lesssim10$km s$^{-1}$) components, also known as high-velocity and low-velocity components (HVC/LVC) \citep{Hartigan_etal95}. Identification and diagnostics of such outflows mainly rely on atomic and molecular line tracers, most notably the [O I]$\lambda$6300 line, which is nearly ubiquitously detected in accreting T Tauri stars.
Individual lines require specific excitation conditions (many are non-LTE), thus probe specific components/regions of the outflow.
The complexity in modeling line excitation and environmental contamination (e.g., absorption), complicates the interpretation and direct comparison with standard wind models.

Line observations of disk outflows generally fall into three categories. (1). Spatially unresolved observations from high-resolution spectroscopy provide rough but the most common diagnostics by modeling line profiles. (2). Spatially resolved observations, especially by ALMA and JWST, provide rich information down to $\lesssim10$AU scale. (3). The spectro-astrometry technique \citep{WhelanGarcia08,Pontoppidan_etal11}, which measures the spatial centroid offset of the spectrum across a line or other spectral features, can reveal spatial structure on scales well below the diffraction limit along selected spatial directions. 

With these techniques, the primary observables include outflow velocity (line blueshift), mass loss rates (requiring density estimate, with large uncertainties), and launching radius. We elaborate on the latter. If outflow rotation can be inferred from spatial velocity gradient, the asymptotic flow velocity $v_{\phi}$ and $v_{p}$ can be constructed at some large cylindrical radius $R$. As the Bernouli constant 
\ifbool{supplement}
{(Equation 46)}
{(Equation \ref{eq:bernoulli})}
is independent of magnetic field, the wind launching radius $R_0$ can be inferred by equating the Bernoulli constant at $R_0$ and $R$. For a cold wind ($c_s\ll v_K$), $h$ is negligible, this yields \citep{Anderson_etal03}
\bgeq
-\frac{3}{2}\frac{GM_*}{R_0}\approx\frac{v^2}{2}-\sqrt{\frac{GM_*}{R_0^3}}Rv_\phi\ .\label{eq:windR0}
\edeq
Given $R_0$, the lever arm can then be estimated by definition: $\lambda=Rv_\phi/\sqrt{GM_*R_0}$.

\subsubsection{Summary of bulk observational constraints}\label{sssec:windobs}
Concerning global disk evolution, key questions include: (1). What is the radial profile of mass loss rate? (2). What is the physical origin of the disk outflows? Here, we summarize our current understandings from outflow observations, but refer to \citet{Pascucci_etal23} for a recent review. In particular, their Table 1 provides an executive summary on the kinematics of YSO outflows at all stages.

The HVC/jets are highly collimated, with semi-opening angles of only a few degrees \cite[e.g.][]{Agra-Amboage_etal11,Erkal_etal21b}.
There is a strong correlation between the inferred mass loss rate and accretion rate $\dot{M}_{\rm jet}/\dot{M}_{\rm acc}\sim0.1$\cite[e.g.][]{Ellerbroek_etal13,Nisini_etal18,Podio_etal21},
which appears consistent across evolutionary stages \citep{Lee20,Sperling_etal21}. 
When rotation is detected, the jet launching radius is found to be down to $\lesssim0.1$AU scale \cite[e.g.][]{Lee_etal17,Lee_etal25}.
At larger distances, the jets often exhibit time variable knotty structures, which likely arise from episodic variations in the velocity of mass ejection that cause internal shocks, or interaction with the ambient medium. It is clear that the HVC/jet must be launched magnetically in the innermost disk,
and recent 3D simulations outlined in 
\ifbool{supplement}
{Section 5.3.2}
{Section \ref{sssec:magnetosphere}} 
offer a promising path forward for detailed comparison with observations.

Disk outflows exhibit nested, ``onion-like" velocity structures, with the HVC surrounded by layers of slower, wide-angle components identified from different tracers \cite[e.g.][]{Agra-Amboage_etal14,Lee_etal21a,Delabrosse_etal24,Liu_etal25,Pascucci_etal25}, often in
 \ce{SO}, \ce{H2}, \ce{CO}, etc., but also in more complex molecules such as \ce{CH3OH} \citep{Nazari_etal24}. From spatially resolved observations, wide-angle outflows are inferred to be launched from extended regions up to $10-100$AU, with mass loss rates comparable to stellar accretion rates \cite[e.g.][]{Bjerkeli_etal16,Lee_etal18,Louvet_etal18,deValon_etal20}. These observations also reveal that the wind carries excess angular momentum to drive disk accretion, supporting magnetically-driven wind as opposed to alternative scenarios such as entrainment of ambient materials by the central jet/X-wind \citep{Tabone_etal20,deValon_etal22}. The inferred wind lever arm is typically $1.5\lesssim\lambda\lesssim2$, pointing to magneto-thermal-type winds 
\ifbool{supplement}
{(Section 4.5.3)}
{(Section \ref{sssec:magthermal})} 
with significant mass loss. On the other hand, photoevaporation wind models can also reproduce some morphological features of large-scale outflow, such as the ``X-shaped" \ce{H2} emission \citep{Arulanantham_etal24,Nakatani_etal25}.

From spatially unresolved line data (available over large samples), the LVC can be decomposed into a broad component (BC) and a narrow component (NC) \citep{Rigliaco_etal13}, which are stable over decades \citep{Simon_etal16}. The kinematics of BC (FWHM$\sim$100 km/s, centroid at $\sim$10 km/s) and NC (FWHM$\sim$30 km/s, centroid at $\sim$3 km/s) are strongly correlated, with largest blueshift at intermediate inclinations of $\sim35^\circ$, interpreted as the wind opening angle \citep{Banzatti_etal19}. Their line widths (from Keplerian broadening) indicate launching radii of $\lesssim0.5$ AU for the BC and $0.5–5$ AU for the NC \citep{Simon_etal16,McGinnis_etal18}. The LVC line luminosities for both components (e.g., [O I]$\lambda$6300) are correlated with the accretion luminosity \citep{Simon_etal16,Fang_etal18}, and the estimated mass loss rates, particularly from the BC, are generally comparable to $\dot{M}_{\rm acc}$ \citep{Natta_etal14,Fang_etal18}. The data also show an evolutionary trend: as accretion rate decreases, the HVC disappears first, followed by the BC, while the NC generally persists into the transition disk phase \citep{Simon_etal16, Banzatti_etal19,Fang_etal23}.

The BC's small launching radius strongly suggests an MHD wind origin, as photoevaporation struggles to overcome the deep gravitational potential well from the very inner disk \citep{Simon_etal16,Picogna_etal19}. The origin of the NC remains debated: its strong kinematic correlation with the BC implies a common MHD origin \citep{Banzatti_etal19}, but it can also be reproduced by X-ray photoevaporation models \citep{ErcolanoOwen16,Weber_etal20}. Additional spatial information may help break the degeneracy in individual sources. For instance, \citet{Whelan_etal21} found that photoevaporation models can hardly reconcile with the high wind velocity ($\sim30$km s$^{-1}$) close to the midplane ($\Delta z\sim2$AU) in the [O I] line from spectro-astrometry observations of RU Lup. More recently, \citet{Fang_etal23b} reported VLT/MUSE observations of TW Hya, showing that $80\%$ of its [O I] emission (all being NC) arises from within $1$AU radius, which can be well reproduced by magneto-thermal wind models of \citet{Wang_etal19} and rules out photoevaporation models (but see further debate from \citet{Rab_etal23} and \citet{Lin_etal25}). Currently, coupling MHD models with realistic thermal-chemical physics remains challenging for definitive comparison with observations, but encouraging progress is being made \citep{NemerGoodman24,Hu_etal25,Weber_etal25}. 

\subsubsection{Finer diagnostics}

Recent observations have also revealed flow properties in the disk atmosphere that deviate from being smooth and symmetric, which we summarize below:

\begin{itemize}
\item[--] Top-bottom asymmetry. Jet asymmetry in intensity, morphology, velocity, etc. has been well known \cite[e.g.][]{Hirth_etal94}. Recent observations reveal that many of the asymmetries originate from the launching mechanism itself rather than asymmetric ambient environment \citep{Liu_etal12,Erkal_etal21,Bajaj_etal25}. Intrinsic asymmetries have also been identified in the low-velocity wind from observations of edge-on disks \cite[e.g.][]{Louvet_etal18,Pascucci_etal25}. These findings corroborate results from recent simulations of magnetospheric accretion \cite[e.g.][]{Takasao_etal22} and inner disk simulations with the Hall effect \cite[e.g.][]{Mori_etal25}.

\item[--] Surface-layer accretion. As a by-product of wind-driven accretion, it involves a small column of surface gas carrying the bulk accretion flow at trans-sonic speed. It is one-sided in the inner disk especially thanks to the Hall effect \cite[e.g.][]{Bai17}, or on both sides via surface accretion in the fully MRI-turbulent innermost disk \cite[e.g.][]{ZhuStone18}. \citet{Najita_etal21} reported the first detection of such surface accretion flow in the disk GV Tau N. While highly desirable, capturing such flows likely requires specific disk geometry and can be demanding even for edge-on systems.

\item[--] Radial substructures. Magnetic flux concentration 
\ifbool{supplement}
{(Section 5.2.2)}
{(Section \ref{sssec:fluxcon})} 
leads to ring-like disk substructures due to radial variations in wind efficiency, which should also leave an imprint in wind kinematics. Recently, \citet{Bacciotti_etal25} discovered that the CO outflows from the HL Tau disk exhibit nested shell-like structures. Tentatively, the derived magnetic footpoints of these shells (from 
\ifbool{supplement}
{Supplemental Text}{} 
Equation \ref{eq:windR0}) coincide with the location of three dust rings.
A similar ringed wide-angle outflow in DO Tau was also reported by \citet{FernandezLopez_etal20}.
These results offer a promising avenue for investigating the connection between disk and wind substructures.

\item[--] The wind launching region. In recent years, ALMA has enabled detailed kinematic studies of gas emission surfaces in disks. Assuming azimuthal symmetry, \citet{Teague_etal19} and \citet{Galloway-Sprietsma_etal23} reported evidence of disk winds of two disks HD 163296 and AS 209 from their \ce{CO} emission surfaces, though only in localized regions. 
With the exoALMA program, vertical motions are detected in most disks, exhibiting primarily as oscillatory up/down flows or transition from downward to upward motions, with the latter considered as the bases of disk winds \citep{Benisty_etal26}.
The overall kinematic structure of such emission surfaces appears rather complex, calling for more detailed modeling effort and theoretical advances.
\end{itemize}

\subsection{Measurement of Disk Magnetic Field}\label{ssec:magobs}

\subsubsection{Theoretical expectations}\label{sssec:Bexp}

If angular momentum transport is dominated by magnetic stresses, even without knowing detailed microphysics, disk field strength can be estimated through the equation of angular momentum transport 
\ifbool{supplement}
{(7)}
{(\ref{eq:mdotacc})}.
By considering the magnetic components of $T_{R\phi}$ and $T_{z\phi}$, a minimum total field strength (under optimistic field geometry to maximize $B_RB_\phi$ and $B_zB_\phi$) can be estimated for MRI-driven and wind-driven accretion scenarios, respectively \citep{Wardle07,BaiGoodman09}. As wind transport is more efficient than radial transport by a factor of $\sim R/H$ 
\ifbool{supplement}
{(Section 2)}
{(Section \ref{sec:diskevol})},
the minimum field strength required for wind-driven accretion is smaller by a factor of $\sim\sqrt{R/H}$, which can be nearly $10$.

We can refine such estimates by incorporating geometric parameters of disk magnetic field, as in \citet{Weiss_etal21}. Let $B_{\rm mid}$ be the midplane field strength, with subscripts $_{R\phi}$ and $_{z\phi}$ denoting estimates based on purely radial or vertical transport of angular momentum. For radial transport, we assume $|B_\phi|$ is a factor $f$ of $|B_R|$, and magnetic stress is exerted over a vertical extent of $L_z$. For vertical transport, we assume $B_\phi$ at the midplane is a factor $m$ stronger than that in the wind base, and that the midplane $B_\phi$ is a factor $f'$ stronger than $B_z$ (taken to be constant threading a thin disk). This yields:
\bgeq
B_{{\rm mid}, R\phi}\approx0.65{\rm G}\bigg(\frac{M}{M_\odot}\bigg)^{1/4}\bigg(\frac{\dot{M}_{\rm acc}}{10^{-8}M_\odot {\rm yr}^{-1}}\bigg)^{1/2}\bigg(\frac{fH}{L_z}\bigg)^{1/2}R_{\rm AU}^{-11/8}\ .\label{eq:Brphi}
\edeq
\bgeq
B_{{\rm mid}, z\phi}\equiv mB_{{\rm base},z\phi}\approx m0.065{\rm G}\bigg(\frac{M}{M_\odot}\bigg)^{1/4}\bigg(\frac{\dot{M}_{\rm acc}}{10^{-8}M_\odot {\rm yr}^{-1}}\bigg)^{1/2}f'^{1/2}R_{\rm AU}^{-5/4}\ .\label{eq:Bzphi}
\edeq
We highlight that these estimates are independent of disk surface density. The only dependence on disk parameters is through $H$ for radial transport, and we have adopted the scaling using our base disk model.
Moreover, the two equations are not mutually exclusive when both transport mechanisms operate.

Uncertainties in the above are encapsulated in the geometric parameters. Thanks to detailed simulations discussed in 
\ifbool{supplement}
{Section 5}
{Section \ref{sec:fulldisk}}, 
reasonable choices can be made. For inner disks with aligned field geometry (where the HSI operates), we may take $m\sim7$ and $f'\sim10$ in the wind scenario, and $f\sim30$ and $L_z\sim6H$ for radial transport \cite[e.g.][]{Bai17}. They yield similar field strength of $\sim1.4$G at 1 AU, reflecting comparable roles of radial and vertical transport. For anti-aligned field geometry, wind transport dominates, and we may choose $m\sim1$ and $f'\sim10$. This latter choice also carries over to the ambipolar diffusion dominated outer disk. On the other hand, for standard MRI-driven accretion, 
\ifbool{supplement}
{Equation (33)}
{Equation (\ref{eq:alphbeta})}
suggests $f\sim4$. 

\subsubsection{Probing magnetic field morphology}
Magnetic field morphology is traditionally probed via grain alignment: spinning dust tends to spin up around its minor axis by radiative torques, and align its spin axis to background magnetic field \cite[see review by][]{Andersson_etal15}. The dust thermal emission then becomes linearly polarized perpendicular to background field. ALMA observations of protostellar envelopes at sub-millimeter show polarization patterns consistent with magnetic grain alignment, generally exhibiting pinched poloidal field \cite[e.g.][]{Cox_etal18,Maury_etal18,Kwon_etal19}. However, in protoplanetary disks where toroidal fields are expected to dominate, the corresponding polarization patterns are generally absent. Instead, polarization patterns are typically found to be consistent with dust self-scattering \citep{Kataoka_etal15,Yang_etal16}, or azimuthally-aligned grains likely by aerodynamic alignment \citep{Yang_etal19,Lin_etal24}. The lack of magnetically aligned grains is likely due to relatively weak magnetic field and substantially enhanced collisional damping in the dense disk environment \cite[e.g.][]{Tazaki_etal17}.

A notable exception is the transition disk HD142527. \citet{Kataoka_etal16} reported the detection of polarized thermal emission from its southern part consistent with magnetically-aligned grains in a toroidal field. This finding is corroborated by \citet{Ohashi_etal25} with multi-wavelength data. Although the disk itself is highly disturbed with lopsided dust distribution in the ring (peaking in the north), a high polarization fraction up to $15\%$ in the disk southern part is strongly indicative of grain alignment. This is plausible because the southern part is away from dust concentration (thus lower gas density), and the dust size there is found to be relatively small, both conditions being favorable for grain alignment. In the meantime, a few Class 0/I systems show tentative evidence of magnetic alignment in part of the disk \citep{Lee_etal21,Lee_etal24}, though the data are also compatible with self-scattering.

Polarimetry at the infrared wavelength offers an alternative to probe magnetic field geometry. As such wavelengths probe much smaller grains (mainly suspended in the disk surface), their magnetic alignment becomes much easier. \citet{Li_etal16} detected linear polarization at mid-IR ($10\mu$m) in AB Aurigae, suggesting magnetic grain alignment in a ``tilted poloidal" field within $\sim70$AU, though the interpretation is unclear. \citet{YangLi22} showed that near-IR polarimetry of scattered light could probe the 3D field geometry in the disk surface, provided the (sub-)$\mu$m-sized dust contains super-paramagnetic inclusions.
Furthermore, \citet{deLangenTazaki23} found that scattered light by magnetically-aligned grains can also produce circular polarization especially when the field configuration deviates from pure toroidal. These diagnostics call for adequate measurement of total intensity in scattered light as a prerequisite \citep{Ren_etal23}, and could potentially become powerful probes of disk surface magnetic field.

The Goldreich-Kylafis (GK) effect \citep{GoldreichKylafis81} provides another probe of magnetic field morphology. It arises from magnetic sublevels of the excited molecular state deviating from LTE due to anisotropic radiation field, producing linearly polarized molecular line emission (i.e., rotational transitions) parallel or perpendicular to the background magnetic field. However, there is a $90^\circ$ directional ambiguity.
\citet{Lankhaar_etal22} studied the possible signature of GK effect in simple disk models, and concluded that molecules with strong dipole moment and low collision rates, such as \ce{HCN}, are most useful probes, while spectral lines from the most abundant \ce{CO} molecule are easily thermalized and are only significantly polarized in the disk outer regions. \citet{Teague_etal21} firmly detected weak linear polarization of $^{12}$\ce{CO}(3-2) and $^{13}$\ce{CO}(3-2) lines in the line wings toward the TW Hydra, following marginal detections toward two disks \citep{Stephens_etal20}. However, interpreting the polarization morphology remains challenging, which is sensitive to disk geometric properties.

\subsubsection{Probing magnetic field strength}

The Zeeman effect is the classic approach to probe line-of-sight magnetic field strength. In the limit of weak field applicable to many astrophysical environments, it is exhibited as circularly polarized line emission, and is most sensitive to paramagnetic molecules, particularly \ce{CN}.\begin{marginnote}
\entry{Paramagnetic molecules}{Molecules with unpaired electrons which can be ``attracted" by magnetic fields.} 
\end{marginnote}Thermal-chemical models and disk observations suggest FUV-induced CN formation in the disk surface, leading to a peak of CN emission in the $\sim50-150$AU region \citep{Cazzoletti_etal18,Bergner_etal21}.
The Zeeman signal thus probes the area-averaged line-of-sight field of the line-emitting region.

With ALMA, \citet{Vlemmings_etal19} and \citet{Harrison_etal21} reported ($3\sigma$) upper limits of line-of-sight magnetic field of a few mG in TW Hydra and AS 209, corresponding to $\sim30$mG and $\sim10$mG ($1\sigma$) total field upper limits when corrected for disk inclination. 
With their respective stellar masses and accretion rates of [$0.82M_\odot$, $2.5\times10^{-9}M_\odot$ yr$^{-1}$] and [$1.25M_\odot$, $10^{-7}M_\odot$ yr$^{-1}$], we can estimate from 
\ifbool{supplement}
{Supplemental Text}{}
Equation (\ref{eq:Bzphi}) that at $R=50$AU, $B\sim0.7$mG for TW Hydra and $B\sim5$mG for AS 209 (taking $m=1$, $f'=10$ for wind-driven accretion at outer disk).
On the other hand, the Zeeman signal can be subject to cancelations when magnetic field changes sign along the line-of-sight, especially for sources with modest inclination: under wind-driven accretion, the mean radial and toroidal field must flip across the disk. Considering this cancellation, \citet{LankhaarTeague23} found the aforementioned upper limits could be raised to 65mG and 35mG for TW Hya and AS 209, respectively, well above theoretical expectations.

Moreover, \citet{LankhaarTeague23} pointed out that the Zeeman effect also broadens spectral lines and causes their linear polarization.
The major advantage is that both Zeeman-induced broadening and linear polarization depend quadratically on field strength, which largely eliminates the influence of field reversals. The linear polarization signal is sensitive to the plane-of-sky field component, while Zeeman broadening depends on the total field strength.
These effects are detectable in nearby disks after azimuthal averaging given the current ALMA sensitivity, thus it opens the avenue to derive full 3D magnetic field direction and strength in protoplanetary disks.

Applying this technique, \citet{Teague_etal25b} made the first concrete measurement of the Zeeman signal from TW Hydra, reporting a field strength around $10$mG over the radius of $60-120$AU. They also constrained the magnetic field morphology (though with less fidelity) transitioning from poloidal- to horizontal-dominated beyond $\sim80$ AU. This is surprising as the expected field strength assuming wind-driven accretion from 
\ifbool{supplement}
{Supplemental Text}{}
Equation (\ref{eq:Bzphi}) is $\lesssim0.2$mG over this radial range. This discrepancy may be partially reconciled if the wind leads to heavy mass loss with a small lever arm $\lambda\lesssim2$, so that the actual mass accretion rate at $\sim100$AU can be boosted by a factor of $\gtrsim10$ 
\ifbool{supplement}
{(Section 2.1)}
{(Section \ref{ssec:diskevolsol})}.
Moreover, this radial range is near the disk outer edge \citep{Teague_etal22}, and could itself be a peculiar region \citep{YangBai21}. It is thus highly desirable to employ this technique to more representative nearby disk samples to better test the theory of magnetically-driven disk evolution.

\begin{figure*}[h]
\centering
\includegraphics[width=0.75\textwidth]{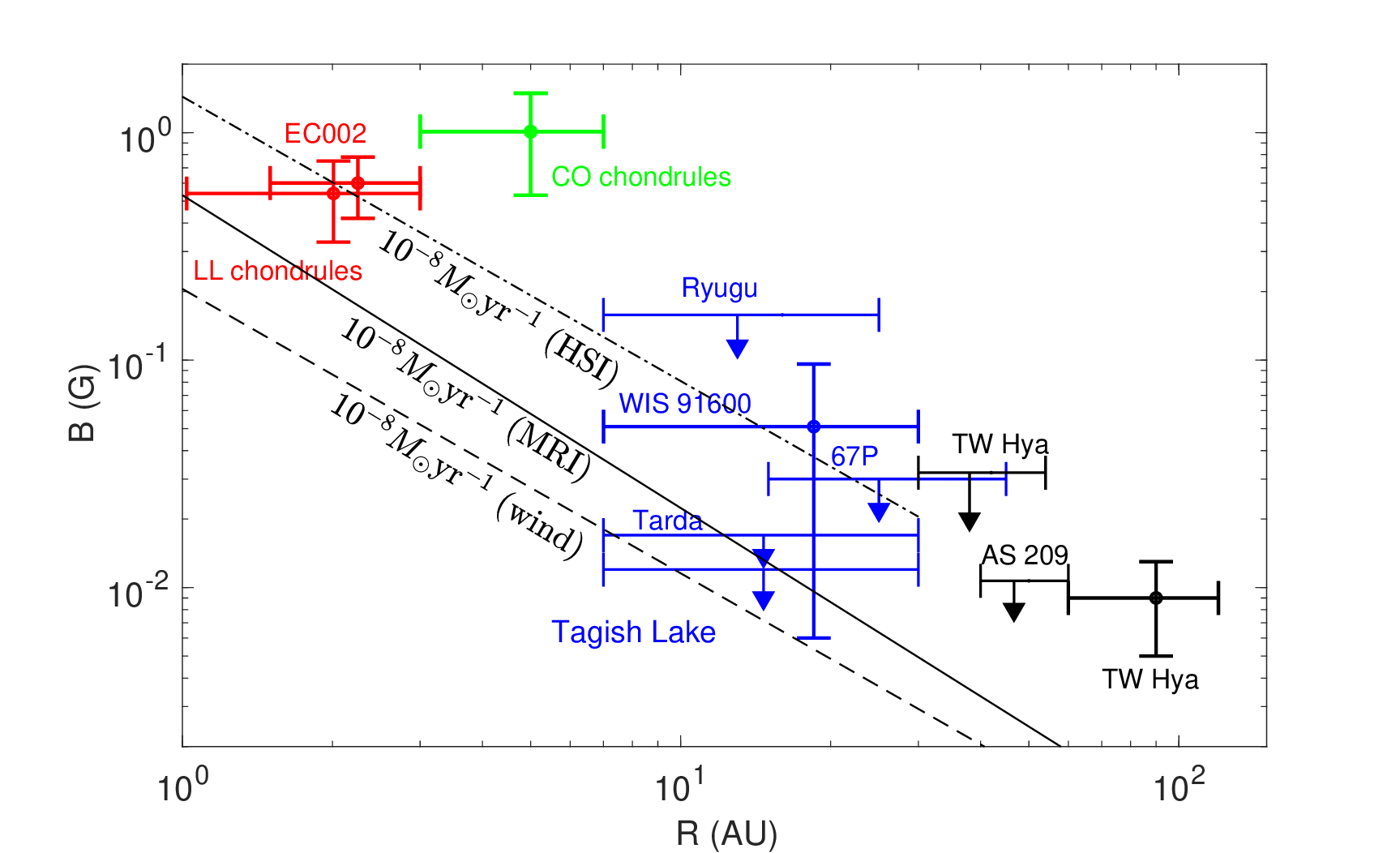}
\caption{A compilation of magnetic field strength measurements as a function of estimated distance to the central star, combining paleomagnetism records of the solar nebula (only including samples older than 3 Myrs after CAI formation) and observations of the Zeeman effect in nearby protoplanetary disks. Paleomagnetic data are color-coded by source: red and green for non-carbonaceous and carbonaceous meteorites, including LL chondrules \citep{Fu_etal14}, Erg Chech 002 \citep{MaurelGattacceca24}, and CO chondrules \citep{Borlina_etal21}; blue for more distant outer solar system bodies, including the Ryugu return sample updated from \citet{Mansbach_etal24}, spacecraft measurements of comet 67P \citep{Biersteker_etal19}, and three ungrouped carbonaceous chondrites Tagish Lake, Tarda and Wisconsin Range 91600 that are expected to experience alteration at larger heliocentric distances \citep{Bryson_etal20,MaurelGattacceca23,Bates_etal24,Mansbach_etal24}. 
Disk measurements from TW Hydra \citep{Vlemmings_etal19,Teague_etal25} and AS 209 \citep{Harrison_etal21} are shown in black.
Theoretical expectations of the field strength for given accretion rate of $10^{-8}M_\odot$ yr$^{-1}$ around a solar-mass star are also shown for comparison 
\ifbool{supplement}
{(based on Supplemental Text Section 2.3.1)}
{(based on Section \ref{sssec:Bexp})}, 
where we consider three standard scenarios with accretion driven by magnetized disk winds, the MRI, together with the Hall-shear instability within $\sim30$ AU.
\label{fig:Bstrength}}
\end{figure*}

\subsection{Long-Term Disk Evolution}\label{ssec:longevol}

The theory of disk angular momentum transport can be tested by comparing samples of stellar and disk properties (including stellar mass, stellar accretion rate, disk mass and size) as a function of age. However, this test is far from being straightforward. Observationally, the measurements of disk mass and size are highly uncertain. It is often convenient to use dust mass and dust disk size as a proxy, but interpreting the results must involve additional modeling effort on dust evolution. Theoretically, while the disk physics is far from homogeneous, models of disk evolution still primarily rely on the self-similar assumption 
\ifbool{supplement}
{(Section 2)}
{(Section \ref{sec:diskevol})}, 
and more realistic prescriptions are still lacking. We refer to the recent review by \citet{Manara_etal23} and results from the AGE-PRO ALMA large program \citep{Zhang_etalAGEPRO} for comprehensive discussions. The primary objective here is to distinguish between effective viscously-driven versus wind-driven disk evolution scenarios, and we highlight key results here.

The most obvious diagnostic is the evolution of gaseous disk size $R_{\rm gas}$. This is typically measured as $R_{\ce{CO},90\%}$, the radius that encloses $90\%$ of the low-$J$ $^{12}$\ce{CO}
emission \citep{Ansdell_etal18}. Note that it does not necessarily trace the theoretical truncation radius $R_t$. Nevertheless, using physical-chemical models with quantitative analysis, \citet{Trapman_etal20,Trapman_etal22} confirm that $R_{\ce{CO},90\%}$ should increase and decrease over time under pure effective viscously-driven and wind-driven evolution models, respectively. 

Observational tests between the two scenarios from gas disk size information alone are currently inconclusive. \citet{NajitaBergin18} analyzed gas disk size from the literature based on various tracers and found that Class II disks are overall larger than younger embedded disks, while \citet{Long_etal22} collected a larger Class II disk sample with a broad range of stellar ages and found no significant evolution in $R_{\rm CO}$. 
\citet{Trapman_etal20,Trapman_etal22} found that the moderate disk sizes in the Lupus star forming region require a small $\alpha_S\lesssim10^{-3}$ for effective viscous evolution models, but the significantly smaller disk sizes in the older Upper Sco star forming region is hard to reconcile even with wind-driven evolution models.
However, caution must be exercised in interpreting these results. Besides the unknown disk initial sizes which likely yield large scatter, even weak levels of external photoevaporation can significantly influence the evolution of disk radii \cite[e.g.][]{Coleman_etal24}.
Furthermore, note that the disk may also expand in the MHD wind scenario \citep{YangBai21}.

Another diagnostic distinguishing between the two scenarios is the evolution of ``apparent disk lifetime" $t_{\rm disk}\equiv M_{\rm disk}/\dot{M}_{\rm acc}$. This metric is observationally more convenient, as $M_{\rm disk}$ can be approximated from dust mass measurement. In effective viscous evolution models with constant $\alpha_S$, $t_{\rm disk}$ represents the viscous time at the truncation radius $R_t$, which increases linearly over time
\ifbool{supplement}
{(Section 2.1)}
{(Section \ref{ssec:diskevolsol})},
though external photoevaporation can reduce it during late-stage disk dispersal \citep{Rosotti_etal17}. In contrast, pure wind-driven evolution generally makes $t_{\rm disk}$ stay constant for constant $\alpha_W$; if $\alpha_W$ increases as disk evolves (e.g., the disk preserves poloidal magnetic flux), $t_{\rm disk}$ can rapidly decrease over time \citep{Tabone_etal22}. 

Observations based on early dust-derived disk masses show no obvious trend of $t_{\rm disk}$ increasing with time, and remain compatible with both effective viscous and wind-driven evolution models \citep{Mulders_etal17,Lodato_etal17}. On the other hand, population synthesis studies show that pure effective viscous models predict smaller spread in $t_{\rm disk}$ over time, but this is not observed \citep{Manara_etal19}, even in the old Upper Sco region \citep{Manara_etal20}. While incorporating a simple model of dust evolution alleviates the situation \citep{Sellek_etal20}, the disk wind scenario is more naturally compatible with data \citep{Tabone_etal22a}.

Recently, the ALMA AGE-PRO program \citep{Zhang_etalAGEPRO} provided high-quality gas disk mass and size measurements for 30 disks across three star-forming regions spanning a wide range of disk ages (0.5–6 Myr). \citet{Tabone_etal25} systematically compared population synthesis models with these data, including the observed disk fraction, disk mass, gas disk size, and accretion rates of the sample systems. They found that while effective viscous disk models with small $\alpha_S\lesssim4\times10^{-4}$ can reproduce disk fractions and gas disk sizes, they tend to severely over-predict $t_{\rm disk}$. On the other hand, wind disk models can successfully reproduce the evolution of bulk disk properties across the three star-forming regions, provided that net poloidal field threading the disk declines with time. This represents the strongest constraint to date from population studies which favors wind-driven disk evolution, though external photoevaporation can alleviate the tension with effective viscous disk models \citep{Anania_etal25}.

\subsection{Evidence from the Solar-System}\label{ssec:paleo}

\subsubsection{The mixing problem}
The solar system provides a rich and complementary dataset for interpreting disk processes. A major piece of evidence is large-scale mixing, demonstrated by the presence of crystalline silicates in comets \cite[e.g.][]{CampinsRyan89,Wooden_etal99}, and refractory materials in the Stardust mission return samples from comet 81P/Wild 2 \citep{Brownlee_etal06}, which must be processed in the hot inner disk. This is corroborated by the widely identified crystalline silicates features in the mid-IR spectra of protoplanetary disks \cite[e.g.][]{vanBoekel_etal04,Watson_etal09}. Moreover, Calcium-Aluminum-rich Inclusions (CAIs), the oldest solar system solids, must be formed in a hot ambient medium ($T\gtrsim1300K$), yet are mostly found in carbonaceous chondrites originating from the outer solar system \cite[see review by][]{Krot19}. 

This problem of large-scale mixing remains a major piece of puzzle in planetary science. It has been commonly interpreted as strong radial turbulent mixing \citep{Bockelee_etal02,HughesArmitage10} or wind ejection \cite[e.g.][]{Shu_etal96}, though each scenario has its own deficiencies \cite[e.g.][]{Desch_etal10}. Recent studies suggest an alternative scenario, where the fully MRI turbulent disk undergoes surface accretion, accompanied by outward meridional flow in the disk midplane \citep{ZhuStone18,JacqueminIde_etal21,Zhu25}. This flow potentially joins the more effective ``pseudo-diffusion" in the Hall-effect-dominated inner disk with alternating radial flows \citep{HuBai21}. Nevertheless, the mixing problem still deserves further investigation with better understanding of the inner/innermost disk gas dynamics.

\subsubsection{Paleomagnetism}\label{sssec:paleo}

Paleomagnetism, a subfield in planetary science, studies the ancient magnetic fields recorded by planetary materials in the form of natural remanent magnetization, with several sub-categories. As a basic principle, planetary materials (containing ferromagnetic inclusions) can acquire remanent magnetization when they cool, crystallize, or accrete under an ambient field $B_{\rm paleo}$, with a certain proportionality
\bgeq
M_{\rm NRM}=\chi B_{\rm paleo},
\edeq
The proportionality coefficient $\chi$ depends on materials and form of magnetization, and can be estimated or calibrated by subsequent laboratory experiments.

Paleomagnetic analysis of meteorites and in-situ magnetometry of small bodies (asteroids and comets) potentially reveal the ambient (i.e., solar nebula) magnetic field strength during their formation or evolution, which provides complementary information to astronomical observations \cite[see][for a comprehensive review]{Weiss_etal21}.
Specifically, paleomagnetic studies can directly measure the total field strength, which can be considered instantaneous or time-averaged depending on the form of magnetization. Moreover, as meteorites can be dated, it potentially allows to trace nebular field evolution. The main downside include the lack of field orientation, and the uncertainty in the location where magnetization was acquired. It is usually assumed that the parent bodies for NC and CC chondrites formed in the inner 1-3 AU and 3-7 AU, respectively.\begin{marginnote}
\entry{NC and CC chondrites}{Non-carbonaceous and carbonaceous chondrites, which are considered to form in the inner and outer solar system, respectively.} 
\end{marginnote}

The first concrete paleomagnetic measurement of nebular field was achieved by \citet{Fu_etal14} from the chondrules in the Semarkona LL chondrite (expected location at $1-3$AU) with an estimated age of $2.0\pm 0.8$ Myr after CAI formation. The inferred nebular field was $0.54\pm0.21$G, likely recorded during rapid cooling following the chondrule formation event. So far, about a dozen meteorites have been identified to contain high-fidelity ferromagnetic minerals to potentially record solar nebular field. We choose those with age older than 3 Myrs after CAI formation and show the inferred nebular field strength or upper limits in Figure \ref{fig:Bstrength}. Also shown are inferred field strengths from in-situ measurement of comet 67P/Churyumov-Gerasimenko \citep{Biersteker_etal19} and the analysis of the Ryugu return sample \citep{Mansbach_etal24}. 

For a typical accretion rate of $10^{-8}M_\odot$ yr$^{-1}$, the inferred field strength in the non-carbonaceous region (1-3AU) is consistent with the Hall-shear instability scenario 
\ifbool{supplement}
{(Supplemental Text Section 2.3.1)}
{(Section \ref{sssec:Bexp})}. 
This requires the presence of net poloidal field with polarity aligned with disk rotation, leading to comparable radial and vertical angular momentum transport \citep{Bai17}. Intriguingly, the inferred field strength from CO chondrules in the CC region (3-7AU) is even stronger, contrary to theoretical expectations. \citet{Borlina_etal21} interpreted this anomaly as the presence of a ``magnetic substructure", although its physical nature remains unknown. Toward the more distal outer solar system, current data confirm the presence of nebular field, but the large uncertainties in both radial distance and field strength preclude distinguishing among different scenarios.

Finally, there are additional paleomagnetic measurements for CC meteorites whose ages are between 3-6 Myrs after CAI formation \cite[e.g.][]{Cournede_etal15,Gattacceca_etal16,Fu_etal21,Borlina_etal22,Bryson_etal24}. They generally record time-averaged fields during alteration, showing mixed results including both upper and lower limits, which might be reconciled by age uncertainties. A stringent upper limit in field strength of about 3 mG is considered to constrain the lifetime of the outer solar nebula (3-7 AU) to about 4.9 Myr \citep{Gattacceca_etal16,Borlina_etal22}. This is joined by an upper limit of about 6 mG at 3.9 Myr for the inner solar nebula \citep{Wang_etal17}.

} 

\section{MISSING PHYSICS IN DISK EVOLUTION}\label{sec:frontier}

In Section \ref{sec:diskevol}, we formulated the master equation (\ref{eq:master}) for global disk evolution, which is governed by the radial profiles of $\alpha_S$, $\alpha_W$, and $\lambda$. Following our discussions from Sections \ref{sec:micro} to \ref{sec:fulldisk}, a zeroth-order picture of disk evolution emerges. 
The disk can be divided into three regions with distinct prescriptions of parameters:
The innermost region is dominated by effective viscous transport with $\alpha_S\sim0.1$, which transitions to become wind-dominated with $\alpha_W\sim10^{-3}$ to $10^{-2}$ in the inner and outer disk region with $\lambda\lesssim4$. The inner and outer regions is also expected to be weakly turbulent with $\alpha_S\lesssim10^{-3}$, and it can be boosted to $\alpha_S\lesssim10^{-2}$ in the inner disk by the Hall effect when the background field is aligned with rotation. In addition, solving for disk temperature is also necessary to locate the dead zone inner boundary. However, even to zeroth order, there are still several key missing ingredients, as we elaborate below.

\subsection{Magnetic Flux Transport}\label{ssec:Bfluxtrans}

The established physical picture of disk accretion and evolution crucially depends on the presence and distribution of poloidal magnetic flux threading the disk in setting the radial profile of $\alpha_W$ (and partly $\alpha_S$). What determines this distribution in disks and over time? This is the result of magnetic flux transport.

Magnetic flux transport can be considered as a two-stage process. Starting from an initial condition, (a disk threaded by a vertical field), the system first evolves toward a quasi-steady state over a few local orbital timescales, establishing a locally quasi-equilibrium field configuration and magnetic flux distribution (e.g., Figures \ref{fig:diskinner} and \ref{fig:diskouter}). The second stage involves secular evolution of the quasi-equilibrium field configuration. The key question concerns whether the disk is able to retain or even ``attract" magnetic flux to drive faster accretion, or will the disk lose magnetic flux and slow down its evolution.

Stage one flux transport is generally understood under the advection-diffusion framework. 
In the prototype model of \citet{Lubow_etal94a}, a passive poloidal field through a razor-thin disk is advected inward by the effective viscously-driven accretion flow ($v_{\rm acc} \sim \nu/R$) and diffuses outward due to Ohmic-like (turbulent) resistivity. The latter is given by $v_{\rm diff}=|cE_{\phi,O}/B_z|\sim\eta |B_r^s/HB_z|$, where $E_{\phi, O}$ is the $\phi-$component of the Ohmic electric field with diffusivity $\eta$, $B_r^s$ is the radial field at the disk surface. 
In equilibrium, poloidal field is strongly dragged inward ($|B_r^s|\gtrsim|B_z|$) when $\mathcal{D}\equiv(R/H)(\eta/\nu)\lesssim1$, and becomes more vertical for $\mathcal{D}\gtrsim1$. Incorporating disk vertical structure, the effective advection speed and diffusivity should be their vertical averages weighted by $\eta^{-1}$ \citep{OgilvieLivio01}. As the surface layer can be more electrically-conducting with faster accretion, it can lead to much faster radial advection, and this effect has been incorporated to construct models for global magnetic flux distribution \citep{Okuzumi_etal14,GuiletOgilvie14}.
Generalizations to considering the Hall-effect and ambipolar diffusion are straightforward, which introduce additional Hall-drift and ambipolar-drift (i.e., electron-ion and ion-neutral drift, see Section \ref{sssec:nimhd}) that can also be treated as advection \cite[e.g.][]{BaiStone17}. Effectively, the field configurations discussed in Section \ref{sec:fulldisk} can be considered as established quasi-equilibrium in stage one.

The secular flux transport in stage two remains poorly understood, and represents a major obstacle from understanding long-term disk evolution. Theoretically, \citet{GuiletOgilvie12,GuiletOgilvie13} developed a local model for magnetic flux transport based on asymptotic expansion on top of a strong vertical field, which was later generalized to incorporate all non-ideal MHD effects \citep{LeungOgilvie19}. With linearized dynamical equations and given vertical boundary conditions, the rate of secular flux transport becomes an eigenvalue of the problem \cite[see also][]{Konigl_etal10}. These models offer insights into the general trend of flux transport, but can hardly match the full complexity of microphysics and nonlinear dynamics that simulations can offer.
Computationally, \citet{BaiStone17} found that in 2D simulations with the Hall effect and ambipolar diffusion, magnetic flux consistently migrates outward in quasi-steady state,\footnote{On the other hand, in stage one, the Hall effect tends to attract/expel magnetic flux in the aligned/anti-aligned cases from an hourglass shaped initial field.} with rates increasing strongly with net vertical field (up to $\sim1\%$ of Keplerian speed). At outer disk conditions, the 1D self-similar solutions with ambipolar diffusion by \citet{Lesur21} confirmed the scalings but found a rate $\sim10$ times smaller. Similarly slower rate of outward flux transport was also found in \citet{CuiBai21} with resolved MRI turbulence. Toward the inner disk and with more realistic diffusivity profiles for all three non-ideal MHD effects, \citet{Bai17} found that flux evolution shows more complex behaviors after stage one and no clear trend was identified within the duration of the simulations.

The ambiguities and uncertainties call for major efforts to advance our understandings of (stage two) magnetic flux transport. As establishing magnetic flux distribution in stage one is a pre-requisite, numerical approaches to properly capture realistic gas dynamics is essential. Current results imply that in laminar disks, the diffusivity profile can be the key controlling factor for flux transport, where subtle differences in diffusivity may lead to vastly different rates. Moreover, as an intrinsically global problem, the inner and outer radial boundary conditions (magnetospheric accretion and disk outer truncation) likely also play a crucial role. We therefore encourage in-depth investigations toward these directions.

\subsection{The Early Disk Phase}\label{ssec:early}

A coherent disk evolution picture must consider initial conditions, which links to the process of disk formation and early Class 0/I stages. Observationally, there are fewer disks in Class 0/I phases due to their shorter lifespan. They are also more difficult to observe due to their embedded nature and require kinematic information to separate the rotationally-supported disk from the envelope (see the ALMA eDisk program, \citealp{Ohashi_etal23}). Theoretically, while most physical processes discussed in this review remain applicable, early disks are dynamically evolving with rapid acquisition of mass and angular momentum from the infalling envelope.
We identify two major and inter-related questions for future investigation.

First, initial disk formation and evolution, which connects to the broader context of star formation. Magnetic fields play a crucial role controlling disk formation \citep{Li_etal14}, where the parent protostellar core must lose a substantial fraction of its initial magnetic flux to form the disk. Otherwise, the twisted magnetic fields drive strong magnetic braking that efficiently removes angular momentum from the collapsing envelope without forming a disk, known as ``magnetic braking catastrophe" \cite[e.g.][]{MellonLi08}. This problem is largely resolved by considering non-ideal MHD effects and background turbulence that allow most magnetic flux to diffuse/drift away, with an emerging picture where disks initially form small, and grow over time \cite[see review by][]{Tsukamoto_etal23}. 
As dust size distribution is crucial in setting magnetic diffusivities \cite[e.g.][]{Zhao_etal16,Marchand_etal20,Lebreuilly_etal23}, recent simulations (although excluding the Hall effect) reveal two typical outcomes. With sub-micron sized grains at solar abundance, \citet{XuKunz21a,XunKunz21b} and \citet{Mauxion_etal24} found that Class 0/I disks are gravitationally self-regulated with $Q\sim1$, leading to a surface density profile of $\Sigma\propto R^{-2}$. There is also a disk wind, but it is weak and sub-dominant as the wind region lacks ionization and suffers from strong ambipolar diffusion. Magnetic flux distribution is approximately uniform due to efficient magnetic diffusion, with $B_z\sim$ a few mG. On the other hand, \citet{Tsukamoto_etal23b} incorporated a model of grain growth (to $>10\mu$m), yielding less massive and non-self-gravitating disks, with a power-law distribution of magnetic flux that is generally much stronger. Further considering the Hall effect, its polarity dependence can lead to a bimodality in initial disk size, and may even reverse rotation direction \citep{Tsukamoto_etal15,Wurster_etal16,Zhao_etal20a}. Moreover, incorporating background turbulence leads to the formation of numerous ``gravomagneto sheetlets" as a primary channel to transport mass, angular momentum and magnetic flux to the disk \citep{Tu_etal24}. It has recently also become possible to conduct simulations directly from turbulent molecular clouds, and follow individual collapsing cores toward disk formation
\citep{Kuffmeier_etal17,YangFederrath25,Mayer_etal25}.
Overall, the problem of initial disk formation is highly complex, and it specifically calls for better understanding of its interplay with background turbulent environment and grain growth. 

Second, the origin of accretion outbursts, particularly FU Ori outbursts (FUors). FUors are characterized by sudden rise in luminosity by 3-6 magnitudes and sustained duration over decades or longer. During outbursts, accretion rates are high enough to crush the magnetosphere, and the disk outshines the protostar. FUors have been considered to play a crucial role in protostellar mass assembly, though it is not yet firmly established if all disks experience such events \citep{Fischer_etal23}. Theoretically, the outburst state is a nearly-ideal realization of effective viscously-driven accretion (ideal MHD and no magnetosphere), and with strong effective viscous heating, the MRI-active zone extends to large distances. \citet{Zhu_etal20} found from radiation MHD simulations with net vertical field that the disk primarily undergoes surface layer accretion from the MRI. A disk wind is launched from high altitudes, but it is only subdominant in driving accretion. However, more elusive concerns the outburst trigger, and several categories of (simplified) models have been proposed \citep{Audard_etal14}. These include a combination of MRI and GI activation \cite[e.g.][]{Armitage_etal01,Zhu_etal09a,Bae_etal14}, migration of GI fragments followed by tidal disruption and accretion \cite[e.g.][]{VorobyovzBasu05,VorobyovBasu15,NayakshinLodata12}, and stellar flybys/binarity \citep{Pfalzer08,ForganRice10,Borchert_etal22}. These models predict a range of burst amplitudes, light curves, and kinematic signatures \citep{Vorobyov_etal21}. Currently, spatially resolved observations of FUors are limited, but there is evidence of GI \citep{Liu_etal16} or strong interaction \citep{Perez_etal20,Dong_etal22}, and such disks are both massive and compact \citep{Cieza_etal18,Liu_etal19}. Overall, it is essential to go beyond simplified models to probe the mechanism and universality behind accretion outbursts.

\subsection{Environmental Effects}\label{ssec:environment}

As stars are born in clusters, protoplanetary disks do not live in isolation. Interaction with environment leads to additional sink/source terms and affect the boundary conditions that may profoundly regulate disk evolution. We highlight two aspects of environmental effects with major impact.

First, the external radiation environment in stellar clusters. FUV radiation from nearby massive stars can drive external photoevaporation (pure mass loss), which has been extensively investigated 
\ifbool{supplement}
{(Supplemental Text Section 2.2)}
{(Section \ref{sssec:extphoto})}.
While this effect is only marginally important in nearby star-forming regions \cite[e.g.][]{Winter_etal20}, the FUV flux can be (much) stronger in more typical galactic star-forming environment \cite[e.g.][]{LeeHopkins20}, leading to significant reduction of protoplanetary disk lifetime. We note that tidal truncation from stellar flybys can also be important in dense stellar environment \citep{Cuello_etal23}, but it was found to be subdominant compared to external photoevaporation in general \citep{Winter_etal18}. 
However, under wind-driven accretion, it is unclear if the region near disk outer truncation can be adequately described by the master equation (\ref{eq:master}): this region becomes wrapped by an extended poloidal field loop as discussed in Section \ref{sssec:truncation}. We anticipate that magnetic fields and external radiation act together, and mass loss is also associated with loss and redistribution of angular momentum, but the detailed consequences remain to be explored.

Second, infall and streamers throughout disk evolution. Over recent years, there has been mounting evidence that star formation process is highly anisotropic at all scales \citep{Pineda_etal23}, and infall in many YSOs proceeds in the form of coherent narrow structures called ``streamers". Importantly, streamers have been detected not only in embedded disks \cite[e.g.][]{Yen_etal19,Pineda_etal20}, but also increasingly toward Class II disks known as ``late infall" \cite[e.g.][]{Huang_etal20,Huang_etal22,Hanawa_etal24}. Streamers typically follow free-fall trajectories \citep{Gupta_etal24}, with infall rates, at least in some cases, being comparable to stellar accretion rates, thus continuously delivering mass and angular momentum. If universal, late infall could fundamentally alter the path of disk evolution.
Observationally, it is yet to establish solid statistics on the fraction of disks connected to streamers, but recent surveys in scattered light found up to one-third Class II disks show evidence of interaction with environment \citep{Gupta_etal23,Garufi_etal24,Ginski_etal24}. There is also indirect evidence of higher stellar accretion rates in the denser center of the Lupus star forming region compared to the distributed population \citep{Winter_etal24b}. Theoretically, one promising explanation is Bondi-Hoyle-Lyttleton type accretion from the turbulent interstellar medium, as inferred from hydrodynamic and MHD simulations of supersonic turbulence \citep{Padoan_etal05,Kuffmeier_etal23,Winter_etal24a,Padoan_etal25,HuhnDullemond25}. Altogether, the common assumption of isolated disk evolution may give way to a multi-scale picture of disk-environmental connection, calling for future investigations that integrate more detailed disk physics.

\begin{figure*}[h]
\centering
\includegraphics[width=0.9\textwidth]{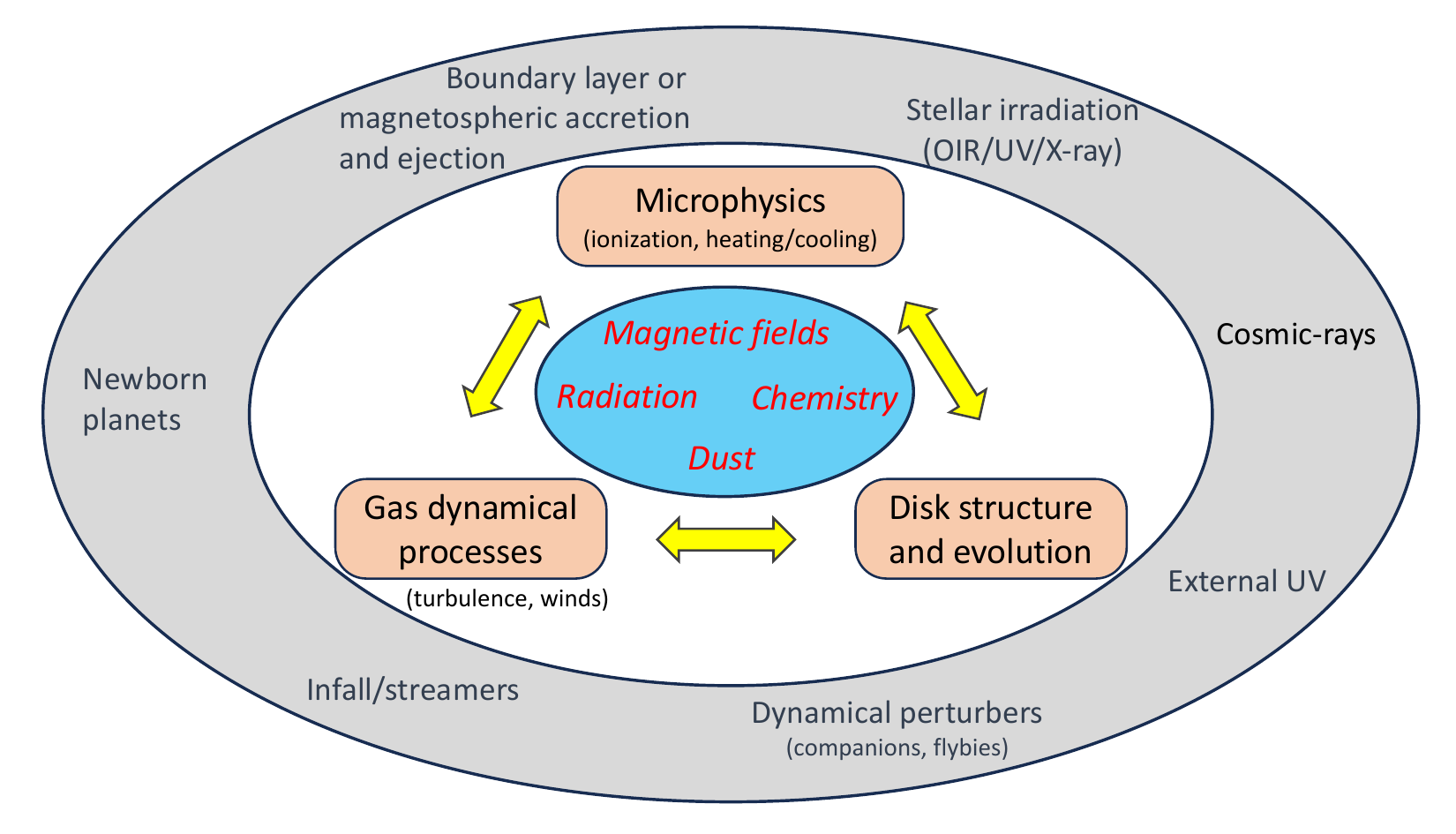}
\caption{The protoplanetary disk ecosystem. Within the disk, the three-level hierarchy mutually affects each other through the coupling with magnetic fields and radiation, which further depends on dust and chemistry. In the meantime, the system constantly interact with external bodies and environments including its host star/companion, forming planets, cosmic-rays and cluster environments.
\label{fig:eco}}
\end{figure*}

\subsection{Protoplanetary Disk as an Ecosystem}\label{ssec:eco}

As our understanding of protoplanetary disks deepens, it is increasingly clear that most physical processes are interdependent of each other, as illustrated in Figure \ref{fig:eco}. 
We have organized this review under the three-level hierarchy, where the microphysics governs gas dynamical processes, which shapes disk structure and drives disk evolution, and this in turn modulates disk ionization and temperature. However, the actual interdependencies are far richer, as we briefly outline below.

Internally, disk physical processes are mediated by magnetic fields, radiation, dust and potentially chemistry, leading to complex interplay among all levels of the hierarchy. Besides magnetic fields and radiation as highlighted in this review, the dust size distribution is another key nexus: it largely controls ionization chemistry and opacity (setting the coupling with magnetic fields and radiation), but is itself shaped by disk structure, turbulence (via radial drift, vertical settling, collision speeds; \citealp{Birnstiel24}), modulated by its composition and material properties. The latter are governed by temperature-sensitive condensation and sublimation processes \cite[e.g.][]{WangL_etal26}, which also release latent heat with dynamical consequences \citep{Owen20,Wang_etal25}. Moreover, dust can be lifted in outflows, which in turn attenuates the radiation which helps drive the outflows \citep{RodenkirchDullemond22,Paine_etal25}. Dust can further drive the streaming instability \citep{YoudinGoodman05} which interplays with other instabilities \citep{Schafer_etal20,XuBai22a,HuangBai25b}. While most studies to date assume a fixed dust size distribution, there can be a number of feedback loops considering all such mutual interactions. 

Externally, the disk is subject to boundary conditions (e.g., boundary layer or magnetospheric accretion/ejection, outer environmental truncation), radiation input (from the star, companions, or nearby massive stars), mass infall from streamers, and dynamical perturbation from companions or stellar flybys. Their influences permeate all three hierarchical levels. We also list newborn planets, whose formation is intimately connected to nearly all physical processes in the disk, and they feedback to the disk not only dynamically \citep{KleyNelson12}, but also thermally \citep{BenitezLlambay_etal15}, magnetically \citep{AoyamaBai23} and chemically \citep{Jiang_etal23}. Another major yet poorly-studied ingredient is cosmic-rays. They control disk ionization, but must propagate through the disk envelope/outflow and atmosphere to reach the disk, calling for more self-consistent treatment of cosmic-ray transport \citep{Nishio_etal25}.

Altogether, protoplanetary disks may be considered as an ecosystem in which a hierarchy of physical processes interact with each other and their star-and-planet-formation environment, where materials, angular momentum and energy flow.
While simultaneously incorporating a large set of physical ingredients is a formidable task at present times, such efforts have become routine in the field of galaxy formation \citep{CrainvandeVoort23}, despite of the necessity to properly calibrate subgrid models. A key advantage in protoplanetary disks is that the level of scale separation is much less extreme, allowing for more solid progress to be made from first principles, which we anticipate to be a promising research avenue in the foreseeable future.

\begin{summary}[SUMMARY POINTS]
\begin{enumerate}
\item The microphysics in protoplanetary disks is not spatially homogeneous. They are largely governed by the coupling between gas and magnetic fields, described by three non-ideal MHD effects set by ionization chemistry, and the coupling between gas and radiation, set by heating and cooling processes. The disk can be approximately divided into three radial sectors accordingly (Figure \ref{fig:division}).
\item Under disk conditions, there are a variety of physical processes that lead to angular momentum transport, including radial transport by turbulence via the MRI, three hydrodynamic instabilities (VSI, COS, ZVI) and GI, as well as vertical transport by magnetized disk winds, which requires the presence of net vertical magnetic field. There are also non-turbulent means of radial transport by spiral density waves and laminar Maxwell stress.
\item The disk innermost sub-AU region is the most complex. Residing inside the (not-well-understood) dead zone inner boundary with temperature $\gtrsim10^3$K, the gas is well coupled to magnetic field. Angular momentum transport is likely MRI-dominated with strong effective viscous heating. Further inside is the inner rim with dust sublimation and disk truncation, where the gas exchanges angular momentum with the star via magnetospheric accretion/ejection and jet launching.
\item The inner disk is governed by all three non-ideal MHD effects with temperature mainly set by stellar irradiation. The MRI turbulence is largely suppressed, and accretion is primarily driven by magneto-thermal disk winds with mass loss comparable to accretion rate.
The disk flow structure, and the wind itself, can be asymmetric depending on the polarity of net vertical field, often exhibiting one-sided surface-layer accretion. 
\item The outer disk ($>$ a few $10$s of AU) is expected to be weakly turbulent thanks to the MRI and hydrodynamic instabilities (VSI), whereas disk accretion is likely dominated by MHD wind. Poloidal magnetic field tends to concentrate into flux sheets, leading to substructure formation. More massive disks can be subject to GI, potentially enhancing radial transport and driving a GI dynamo.
\item The bulk physical picture is by and large supported by observations. Global disk evolution models have been developed to incorporate both radial and vertical transport, but they are subject to major uncertainties concerning magnetic flux evolution and (outer) boundary conditions.
\end{enumerate}
\end{summary}

\section*{DISCLOSURE STATEMENT}
The author is not aware of any affiliations, memberships, funding, or financial holdings that
might be perceived as affecting the objectivity of this review. 

\section*{ACKNOWLEDGMENTS}
I am grateful to Hongping Deng, Oliver Gressel, Greg Herczeg, Henrik Latter, Feng Long, Shoji Mori, Riouhei Nakatani, Eve Ostriker, Shinsuke Takasao, Richard Teague, Lile Wang and Zhaohuan Zhu for helpful discussions and comments. I also thank Xinyu Zheng and Shoji Mori for helping make Figures 2, 3, and 6, and Can Cui for sharing data for Figure 7.
This work is supported by the National Science Foundation of China under grant No. 12325304, 12233004, 12342501.

\bibliographystyle{ar-style2}
\ifbool{supplement}
{\input{ms_main.bbl}}
{
}

\end{document}